\RequirePackage{fix-cm}

\documentclass[aps,pre,twocolumn,superscriptaddress,10pt,nofootinbib,amssymb]{revtex4-2}

\usepackage{subcaption}
\usepackage{amsfonts, amssymb, amsmath}
\usepackage{bm}
\usepackage{graphics}
\usepackage{graphicx}

\usepackage{wasysym}
\usepackage{natbib}
\usepackage{mathtools}
\usepackage{bbold}
\usepackage{makecell}

\usepackage[utf8x]{inputenc}
\usepackage[T1]{fontenc}
\usepackage{paralist}
\usepackage{multirow}

\usepackage[writekey=false,name={}]{notes2bib}

\usepackage[capbesideposition={right,bottom}]{floatrow}

\usepackage[usenames,dvipsnames,table]{xcolor}

\usepackage{color}
\definecolor{darkblue}{rgb}{0,0,0.6}
\definecolor{darkred}{rgb}{0.6,0,0}
\definecolor{blue_R}{RGB}{19, 126, 109}
\definecolor{blue_L}{rgb}{1.0, 0.33, 0.64}
\definecolor{blue_2}{rgb}{0, 0, 1}

\usepackage{tikz}
\usetikzlibrary{decorations.pathreplacing}
\usepackage{hyperref}

\hypersetup{
bookmarksopen=true,
pdftitle=Cascading through the Hierarchy, 
pdfauthor= B. Ladewig et al., 
pdftoolbar=false, 
pdfstartview={FitH},		
pdfmenubar=true,			
pdfhighlight=/O,			
colorlinks=true,			
urlcolor=darkblue,
citecolor=darkblue,		
linkcolor=MidnightBlue,	
}

\usepackage[normalem]{ulem}
\usepackage{comment}

\usepackage{pifont}

\newcommand{\variancevalues}[1]{\kappa_{{#1}}}

\usepackage{abbrevs}
\usepackage{etoolbox} 
\newabbrev\RG{renormalization group (RG)}[RG]

\makeatletter 
\renewcommand\maybe@space@{

  \maybe@ictrue 
  \expandafter   \@tfor
    \expandafter \reserved@a
    \expandafter :
    \expandafter =
                 \nospacelist
                 \do \t@st@ic
 
  \ifmaybe@ic 
    \space
  \fi
}
\makeatother

\usepackage{lmodern}

\begin{document}

\title{Cascading Through the Hierarchy: Regularizer-Induced Feature Detection as Phase Transitions in Deep Linear Neural Networks}

\author{B. Ladewig}
\email[corresponding author: ]{bjoern.ladewig@uni-potsdam.de}
\affiliation{Institute of Physics and Astronomy, University of Potsdam}
\author{I. T. Ersoy}
\affiliation{Institute of Physics and Astronomy, University of Potsdam}
\author{K. Wiesner}
\affiliation{Institute of Physics and Astronomy, University of Potsdam}

\begin{abstract}

A scientific theory of deep learning, comprising learning dynamics and statistical properties of learned models, is rapidly gaining attention. One of the corner stones of this development are analytically solvable toy models, allowing for the fully tractable analysis of the learning dynamics. Here we analytically investigate such a toy model using the regularization strength as a tunable external parameter - akin to external fields in statistical physics. In previous studies, (i) an onset of learning transition was predicted analytically and (ii) it was phenomenologically/numerically established that tuning the regularization strength can result in a \emph{cascade} of phase transitions. The number of those transitions was linked to the geometry of the loss landscape determined by the model complexity. Setting up a rigorous framework underpinning the previous numerical observations, our investigation reveals a precise connection between those cascades of phase transitions, learnable features and the underlying geometry. We provide analytic predictions of these phase transitions as well as tractable order parameters related to learned features. At the level of the minimal model, we connect this macroscopic perspective (that can be condensed into an effective description) to the microscopic perspective in terms of the geometry of the loss landscape characterized by the Hessian spectrum. Thus, the presented model provides a platform to explore and sharpen advances made in the scientific theory of deep learning rooted in statistical physics concepts.

\end{abstract}

\maketitle

\tableofcontents

\vspace{0.5cm}
\hrule
\vspace{0.5cm}

\section{Introduction}
At their core, statistical physics as well as deep learning are concerned with extracting relevant features or representations (e.g. in the form of an effective theory of only a few degrees of freedom) of an underlying probability distribution in order to make problem-dependent predictions about macroscopic behaviour \cite{Transtrum2015,Mehta2019introduction,Bahri2020reviewdeeplearning,Roberts2022,Seroussi2023}.

In supervised deep learning, those features are encoded \emph{microscopically} (geometrically) in a high dimensional non-convex loss landscape formed by the training data in a non-trivial and task-dependent way. In order to extract those features, a learning process of individual models is realized by learning rules e.g. in the form of (stochastic) learning \emph{dynamics} on this landscape. Gradient-based learning dynamics in particular is therefore shaped by geometric properties of the loss landscape like saddle points, local minima and (dynamical) invariant manifolds \cite{Saxe2014exactsolutionsnonlineardynamics,Jacot2022,Menon2025,Zhang2026,Kunin2026} that in turn are related to the macroscopic features to be learned. 
 
Loss landscapes in case of deep neural networks are \emph{hierarchically} organized, due to the architecture of deep neural networks\footnote{where, e.g., critical points and the corresponding input-output relations of networks with lower width (but same depth) are encoded in larger-width networks.} \cite{Fukumizu2000,Zhang2021Embedding} and in particular model-dependent (parameter) symmetries (see, e.g. \cite{Zhao2026symmetry}). The role of parameter symmetries for the hierarchical structures was hypothesized as a key element for unification of theories of deep learning in \cite{Ziyin2025parametersymmetries}. Symmetry-related degeneracies of saddle points etc., result in many-to-one relations between different trained (microscopic) models and \emph{macroscopic} input-output relations \cite{Bahri2020reviewdeeplearning,Mehta2019introduction}. This raises the question to what extend the levels of such a hierarchy can be related to the concept of \textbf{phases} in statistical physics \cite{Goldenfeld1992,Ziyin2025parametersymmetries,Watanabe2025}.

\noindent Addressing this question, the (hierarchical) loss landscape and potential `phases' can be explored by different means: first through (gradient-based) training \emph{dynamics} featuring, e.g., saddle-to-saddle dynamics \cite{Saxe2014exactsolutionsnonlineardynamics,Domine2025,Zhang2026}; second, through the \emph{static/stationary} behaviour by using an \textbf{external field} (in physics terms) that can stabilize qualitatively different phases as put forward in \cite{Ziyin2022,Ziyin2023,Ziyin2024discretesymmetries} and extended by phenomenological observations in \cite{Ersoy2025,Ersoy2025_2}. A physics analogy would be a magnetic field coupled to the magnetization (the \emph{order parameter}), revealing e.g. ferro- and paramagnetic phases. In the context of deep learning, such an external field can be realized by repurposing a ($L_2$-)regularizer as a tunable field of strength $\beta$. 
Taking the physics analogy one step further, we can ask which \textbf{`macroscopic' degrees of freedom/order parameters} can be identified and addressed by using external fields.

Much as in physical systems, the interplay of competing mechanisms, here the often model-complexity-increasing dynamics \cite{Zhang2026,Kunin2026} due to the data-dependent loss landscape and the large-weight penalizing regularization (favoring trivial models), can result in qualitative changes of the system properties in the form of \textbf{phase transitions}. These transitions can be quite generally connected to (broken) parameter symmetries \cite{Ziyin2024discretesymmetries,Ziyin2025parametersymmetries}. Geometrically, for $\beta \to \infty$, the loss landscape becomes convex with a global minimum corresponding to a trivial model (connected to the glass/topological trivialization transitions \cite{Winter2025}). Lowering $\beta$ gives rise to a phase transition, the onset of learning (analytically established for 1d output models in \cite{Ziyin2023}). For higher-dimensional input-output models, a cascade of \emph{multiple} transitions, either second or first order, between different accuracy regimes has been observed phenomenologically \cite{Ersoy2025,Ersoy2025_2}. In \cite{Ersoy2025} the number of transitions was linked to the complexity of the underlying data distribution which forms the loss landscape. Geometrically, the 'phases' have been linked to basins with extended 'loss' barriers \cite{Ersoy2025_2}.\\

Joining `universal' mechanistic and statistical aspects of loss landscape properties, dynamical behaviour and the statistics of ensembles of trained models into a scientific theory of deep learning is an emerging field \cite{ringel2025applicationsstatisticalfieldtheory,Roberts2022,Watanabe2025,Ziyin2024discretesymmetries,Simon2026,Ziyin2025parametersymmetries}. Similar to statistical physics, exactly solvable minimal models provide a platform in making the learning mechanics tractable. One such platform is provided by deep linear models (DLN) that combine a non-convex loss landscape \cite{Baldi1989,Fukumizu1998,Kawaguchi2016,Saddles_DeepLinear,Wendin2025} (with non-trivial Hessian spectra \cite{Singh_DLNHessian,Singh2024closedformsolution}) with non-linear (saddle-to-saddle) dynamics \cite{Fukumizu1998,Saxe2014exactsolutionsnonlineardynamics,Domine2025,Zhang2026} and degenerate critical points lending itself to the introduction of a notion of entropy \cite{Menon2025,Lindsey2025,Chen2025entropicregularizationdeeplinear,Menon2026}. 

On general grounds, a lot of this structure is dictated by symmetries: for deep linear networks important symmetries are inter-layer rotation symmetries (to be defined later). In  \cite{Ziyin2024discretesymmetries} it was shown that in combination with a sufficiently large $L_2$-regularization those symmetries (i) favor lower rank solutions\footnote{$L_2$-regularization for a 1 hidden layer network is related to nuclear norm usage in matrix optimization \cite{cai2010singular} (see also e.g.: \cite{rennie2005fast,scarvelis2024nuclear}), resulting in, e.g., \emph{singular value shrinkage} - a special case of the phase transitions we will discuss in the following. Furthermore, forms of $L_2$-regularization are linked to alternative approaches like data augmentation and hard-wiring (in the context of equivariant and invariant models) \cite{Duan2025} identifying interconnections between these approaches.} \footnote{A corresponding order parameter $O^T \vec{\theta}$ with $\vec{\theta}$ the parameter set and $O$ consisting of columns of orthogonal vectors related to the rotation symmetry was suggested in \cite{Ziyin2024discretesymmetries}.} and (ii) give rise to a block structure in the Hessian.

Combining and extending those ideas, we provide the first (to the best of our knowledge) in-depth investigation of a toy model that features a regularizer-induced \emph{cascade of static phase transitions} while being largely analytically tractable: regularized deep linear models with arbitrary input and output dimensions as well as arbitrary depth at zero temperature/no noise (extending \cite{Ziyin2023}; weak regularization was investigated in \cite{Wang2024}). Here, analytically tractable refers to the following microscopic and macroscopic aspects: (i) data-dependent phase transition points can be analytically predicted; (ii) closed form expressions for complete Hessian spectrum (at the end of training) can be derived; (iii) phases and transitions can be directly linked to geometric structures of the loss landscape; and (iv) an \emph{effective theory} for macroscopic degrees of freedom can be derived. Thereby we link the 'top-down' perspective of the symmetry-predicted presence of phase transitions to the `bottom-up' perspective of a hierarchical organization of the loss landscape \emph{geometry} with and without regularization. We analytically address the following questions:

\begin{enumerate}
\item[\textbf{(P)}] \textbf{phases:} What are the qualitative properties of phases accessible by tuning the regularization strength? What is learned by transitioning to another phase?
\item[\textbf{(OP)}] \textbf{order parameters:} What are the macroscopic degrees of freedom (set of order parameters)? What is a suitable effective description similar to a Landau free energy picture? 
\item[\textbf{(MM)}] \textbf{micro to macro:} How do the geometric properties of the regularized loss landscape from a microscopic point of view connect to the macroscopic input-output relation at the end of training? 
\end{enumerate}

\noindent \textbf{Summary of the Key Findings}

\begin{figure*}
\centering
\includegraphics[width=0.95\textwidth]{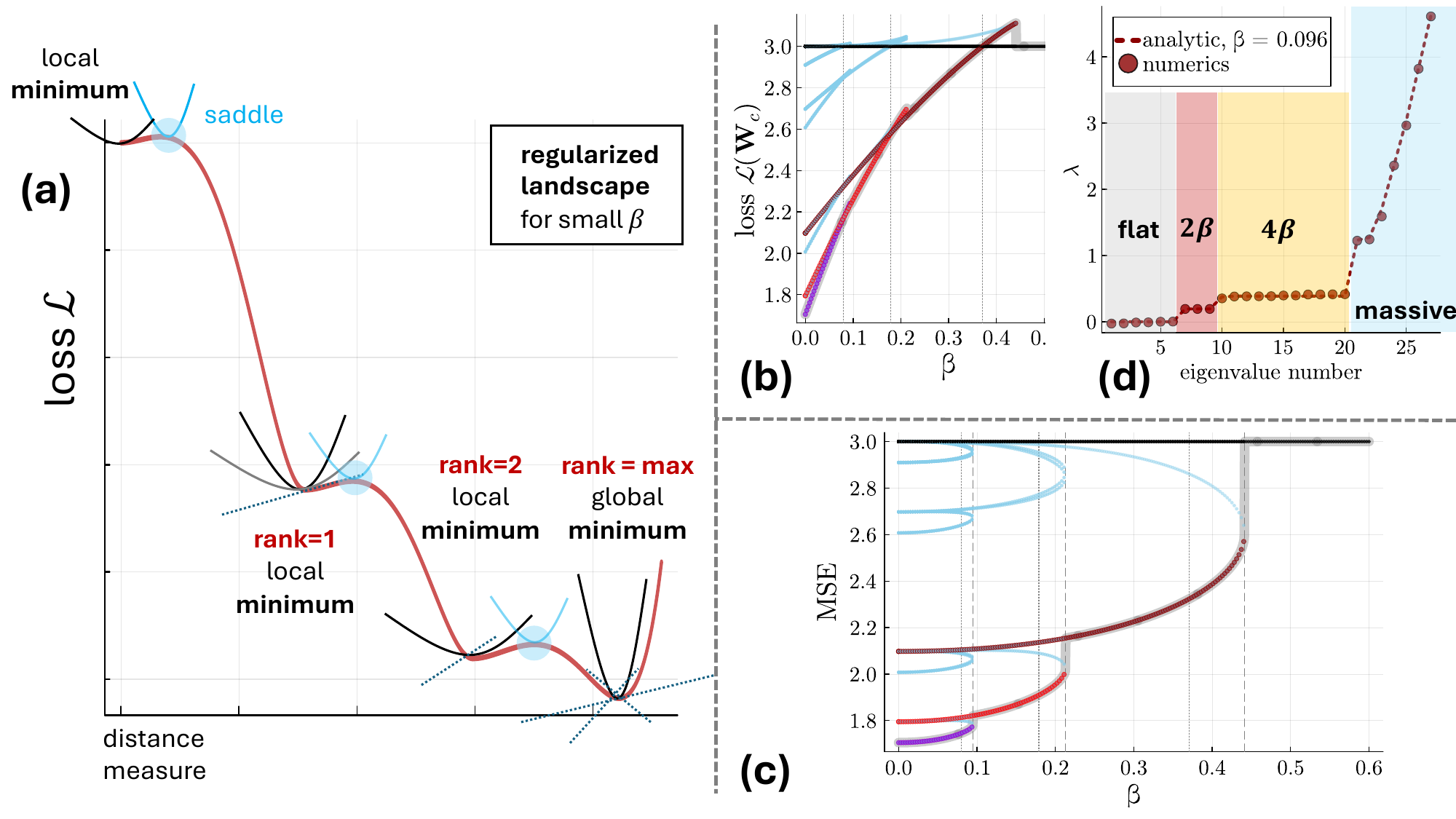}
\caption{\textbf{(a)} Qualitative summary of the main features of the regularized loss landscape for $>1$ hidden layers and small regularization strength; \textbf{(b)} level crossing (marked by dotted vertical lines) at the level of the loss inducing first order transitions; \textbf{(c)} bifurcation (indicated by dashed vertical lines) at the level of the mean squared error; \textbf{(d)} example of the Hessian spectrum for 2 hidden layers. For (b-d): $\eta_1=0.95, \eta_2=0.55,\eta_3=0.3$.}
\label{fig:IntroOverview}
\end{figure*}

\noindent Combining different of the above mentioned dynamical and static perspectives, we provide an in-depth analysis of the critical point structure for analytically tractable cases (see condition \eqref{eq:compatibilitycondition}) for deep linear networks with arbitrary $L_2$ regularization strength. Besides identifying the critical points, we analyze the Hessian spectrum as well as the mode connectivity/degeneracy of the (local) minima and connect this with exact dynamical solutions \emph{in the regularized case}. All analytic investigations are complemented and supported by numerical experiments. The analysis reveals a, though loose, connection to the information bottleneck of Gaussian data \cite{Chechik2003}. Finally, we give some phenomenologically/numerically evidence that the findings in the linear case can be \emph{to some extend} be extended to (deep) non-linear networks. In summary:

\begin{enumerate}
\item[\textbf{(P)}] Phases correspond to (basins around) \emph{rank-restricted} optima of the regularized loss landscape. Each transition is connected with the \emph{learning} of a 'singular direction' of the (co)variance of the underlying data distribution. The onset of learning of individual singular directions is associated with the increase of the rank of the model matrices $W^{(i)}$ as well as the macrostate $\boldsymbol{W}$ with $\vec{y}_\text{output} = \boldsymbol{W} \vec{x}_\text{input}$.
\item[\textbf{(OP)}] The order parameters are the set of singular values $\boldsymbol{\sigma}$ of the macrostate $\boldsymbol{W}$ (defining the input-output relation). The regularized loss can under certain assumptions be reduced to $\mathcal{L}=\mathcal{L}(\boldsymbol{\sigma})$ with a form reminiscent of a Landau free energy for first and second order phase transitions (including \cite{Ziyin2023} as a special case).
\item[\textbf{(MM)}] Stable phases are based on regularizer-induced (local) minima in the microscopic regularized loss landscape (see Fig.~\ref{fig:IntroOverview}). First-order transitions are based on (typical) level crossing\footnote{See, e.g., \cite{Goldenfeld1992} for a discussion of first order phase transitions.} at the level of 'energies' (the loss). Those minima come with qualitatively different Hessian spectra, which can be analytically derived and be qualitatively understood in terms of three contributions: (i) massive data-dependent directions (associated with perturbations of the macrostate $\boldsymbol{W}$), (ii) $0$-balancedness basins (to be defined in \eqref{eq:0balancedcondition}; \cite{Arora2018,Saddles_DeepLinear}) and (iii) flat directions due to symmetries.

\end{enumerate}

Colloquially speaking, the guiding idea of our investigation is two-fold: (i) The loss landscape (without regularization) for the parameter space consists of level sets that separate different accuracy phases, accompanied by reduced ranks (summarized in Sec.~\ref{sec:UnregularizedSection}). (ii) Including a weak ($L_2$) regularizer, on the one hand, turns intermediate accuracy regimes dynamically \emph{stable}, allowing for the analysis of models in those regimes at the end of training (see Sec.~\ref{Sec:RegularizedLandscape}). On the other hand, by increasing the regularizer strength, the most accurate phases become successively unstable at different transition strengths $\beta_c$. The ordering of these transitions is directly related to the underlying data structure. Even though adding a regularizer can serve multiple purposes in deep learning, we advocate the $L_2$-regularizer as a simple means to analyze the loss landscape and the encoded features directly, acting as a kind of bottleneck.

\subsection{Model Setup \& Data}
We are considering a simple model class that already features a non-trivial (non-convex) loss landscape: $n$ layer deep feed-forward networks with linear activation functions with a mean squared error (MSE) \cite{Baldi1989,Kawaguchi2016} and $L_2$ (weight decay) regularization\footnote{$|| \circ ||_2$ is the matrix Frobenius norm (which is, e.g., invariant under left and or right unitary transformations).}:
\begin{align}
&\mathcal{L}(W^{(n)},\dots, W^{(1)}| \{ \vec{x}_\nu, \vec{y}_\nu\}_{\nu=1}^{N_\text{data}}) \\  
&=\underbrace{\mathbb{E}_{\text{training data}} || W^{(n)} \hdots W^{(1)} \vec{x}_\nu- \vec{y}_\nu||^2}_{=:\mathcal{E}}+ \beta \sum_{i=1}^n ||W^{(i)}||_F^2, \nonumber
\end{align}
based on a set of training data $\{(\vec{x}_\nu, \vec{y}_\nu)\}_{\nu=1}^{N_\text{data}}$. The input $\vec{x}$ (output $\vec{y}$) are $d_\text{in}$ ($d_\text{out}$) dimensional: $\vec{x} \in \mathbb{R}^{d_0=d_{in}}$ ($\vec{y} \in \mathbb{R}^{d_n=d_{out}}$). For $n$ layers, the parameter matrices $W^{(i)}$ are of the form $W^{(i)} \in  \mathbb{R}^{d_i \times d_{i-1}}$ and we choose the inner dimensions to be $d_i = d_\text{hidden} \ge d_\text{in},d_\text{out}$ if not stated otherwise (no architectural bottleneck). Therefore, we are dealing with a parameter/model space of dimension $d_\text{parameter} = (d_\text{in}+d_\text{out})\cdot d_\text{hidden} + (n-2)d_\text{hidden}^2$. The goal is to identify the qualitative changes of the loss landscape for different regularization strengths $\beta$. To address this questions, it is convenient to reformulate the loss in the following form:
\begin{align}
&\begin{aligned}
&\mathcal{L}(W^{(n)},\dots, W^{(1)}| \{ \vec{x}_\nu, \vec{y}_\nu\}_{\nu=1}^{N_\text{data}}) = \text{tr}\left[ \boldsymbol{W} \Sigma^{(e)}_{xx} \boldsymbol{W}^T\right] \\ & - 2\text{tr}\left[ \Sigma^{(e)}_{xy}\boldsymbol{W}\right] + \text{tr}\left[ \Sigma^{(e)}_{yy} \right] + \beta \sum_{i=1}^n || W^{(i)}||_F^2,
\label{eq:Lossintraceform} 
\end{aligned}\\
&\text{macrostate:} \, \boldsymbol{W} := W^{(n)} \cdot \dots \cdot W^{(1)}, \\
&\Sigma_{ab}^{(e)} := \frac{1}{N_\text{data}} \sum_{\nu=1}^{N_\text{data}} \vec{a}_\nu \cdot \vec{b}_\nu^T \quad a,b \in \{x,y\}. \label{eq:EmpiricalCovariance}
\end{align}
In case of linear activation functions, the loss is (only) sensitive to the empirical uncentered (co)variances $\Sigma_{xx}^{(e)}$ and $\Sigma_{yx}^{(e)}$ of the data distribution. In the following we assume that the mean is zero for simplicity and that $\Sigma^{(e)}_{xx}$ is invertible and $\Sigma^{(e)}_{yx}$ has full rank. 
If not stated differently, we will be working in the limit of a large set of training data ($N_\text{data} \to \infty$) such that the empirical (co)variance matrices well approximate the underlying covariance matrixes: $\Sigma^{(e)} \to \Sigma$. To keep the discussion as explicit as possible, a specific task can be the (partial) learning of a Gaussian distribution.\\

\noindent \textbf{Explicit Model:} Let $p(\vec{x},\vec{y})$ be a multivariate Gaussian distribution $\mathcal{N}(\vec{0},\Sigma)$ with vanishing mean $\vec{\mu}=\vec{0}$ and (co)variance matrix:
\begin{align}
\Sigma = \begin{pmatrix} \Sigma_{xx} & \Sigma_{xy} \\ \Sigma_{yx}  & \Sigma_{yy} \end{pmatrix}
\end{align}
with $\Sigma_{xx}=\Sigma_{xx}^T$, $\Sigma_{yy} = \Sigma_{yy}^T$ and $\Sigma_{yx} = \Sigma_{xy}^T$. Given samples $\vec{x}_\nu$, the learning task is to predict $\vec{y}_\nu$, where the exact conditional probability distribution $p(\vec{y}|\vec{x}_\nu)$ is Gaussian as well with a mean $\bar{\mu} = \Sigma_{yx}\Sigma_{xx}^{-1}\vec{x}_\nu$. \\

\subsection{Symmetries \label{Sec:Symmetries}}

The structure of the unregularized as well as regularized loss is strongly shaped by the involved symmetries of the loss $\mathcal{L}$: transformations in the parameter space that leave the loss invariant. They influence the dynamics as well as the static (end of training) behaviour: they can enforce conservation laws and solutions may or may not be invariant under certain symmetry transformations (see, e.g., \cite{Zhao2026symmetry,Zhao2023} for an overview). 
The \emph{unregularized} loss $\mathcal{E}(W^{(1)},\dots, W^{(n)})$ is invariant under combinations of general lineare transformations of invertible matrices $\textcolor{orange}{G_i} \in GL(d_i)$ \cite{Baldi1989}: 
\begin{align}
\begin{rcases}
&W^{(i+1)} \to \tilde{W}^{(i+1)}=W^{(i+1)} \textcolor{orange}{G_i}\\
&W^{(i)} \to \tilde{W}^{(i)}=\textcolor{orange}{G_i^{-1}}W^{(i)}
\end{rcases} W^{(i+1)}W^{(i)} = \tilde{W}^{(i+1)}\tilde{W}^{(i)}.
\label{eq:GLSymmetry}
\end{align}
Similar to an external magnetic field, the role of the $L_2$ regularizer at the technical level is an explicit breaking\footnote{Note that this symmetry is broken even in the presence of an infinitesimally small regularization strength.} of those symmetries down to \emph{orthogonal} transformations at the level of $\mathcal{L}$:
\begin{widetext}
\begin{align}
&\mathcal{L}(W^{(n)},\dots, W^{(1)}| \{ \vec{x}_\nu, \vec{y}_\nu\})= \mathcal{L}(\textcolor{blue}{O_n}W^{(n)} \textcolor{orange}{O_{n-1}^T}, \textcolor{orange}{O_{n-1}}W^{(n-1)}\textcolor{orange}{O_{n-2}} , \dots , \textcolor{orange}{O_{1}^T}W^{(1)}\textcolor{blue}{O_0^T} | \{\textcolor{blue}{O_0}\vec{x}_\nu, \textcolor{blue}{O_n} \vec{y}_\nu \} ) \nonumber
\end{align}
\end{widetext}
with $O_i$ elements of the orthogonal group $O(d_i)$. This can be seen from the representation of the loss in \eqref{eq:Lossintraceform}, which is manifestly invariant under the above transformations. 
The \textcolor{orange}{inner symmetry transformations} always leave the loss invariant as $\boldsymbol{W}$ stays unchanged (and the Frobenius norm is invariant under orthogonal transformations). The \textcolor{blue}{outer transformations}  $\textcolor{blue}{O_n}$ and $\textcolor{blue}{O_0}$ describe simultaneous rotations of the model and the data and can be used to rotate into a coordinate system which turns the (co)variance matrices (and therefore the couplings) as simple as possible: $\tilde{\Sigma}^{(e)}_{xx}:=  \textcolor{blue}{O_0} \Sigma^{(e)}_{xx} \textcolor{blue}{O_0^T}$ and $\tilde{\Sigma}^{(e)}_{xy}:=  \textcolor{blue}{O_0} \Sigma^{(e)}_{xy} \textcolor{blue}{O_n^T}$.\\

\subsection{Outline}
The setup we are considering consists of (i) a model architecture; (ii) data (distribution); (iii) training dynamics. In case of (deep) linear networks, a model is determined by the set of the weights of $W^{(1)}, W^{(2)}, \dots$ - corresponding to a point in model space (a \textbf{microstate}). We refer to this as the \textbf{microscopic perspective} and the set of all weights as a microstate $\vec{\theta}$. If we are only interested in the \emph{output} of models, the only relevant quantity is the overall product $\boldsymbol{W} = W^{(n)} \dots W^{(1)}$, capturing the performance of the network. Taking $\boldsymbol{W}$ as the main quantity corresponds to the \textbf{macroscopic perspective}\footnote{Or function space perspective, see, e.g.: \cite{Li2018functionspace,Trager2020functionspacelinear,ringel2025applicationsstatisticalfieldtheory}}. The terminology already implies that many microstates belong to the same macrostate (see also \cite{Menon2025}). Depending on the context a more microscopic or more macroscopic description can be helpful. \\

After initialization, the training dynamics (discrete or continuous; deterministic or stochastic) of choice results in the motion inside the model/parameter space. For simplicity and clarity, we are focusing on the case of gradient flow with an infinitesimal learning rate without added noise nor implicit noise due a small batch size if not stated differently. The gradient flow dynamics is $\partial_t W^{(j)} = -\nabla_{W^{(j)}} L$ and equilibrates eventually into critical points $-\nabla_{W^{(j)}} L =0$ (in particular: (local) minima). Therefore, those critical points are important for two reasons: (i) minima specify the properties of fully trained models (which have equilibrated); (ii) saddle points are important to understand the \emph{dynamics} before finally reaching a stable (local) minimum. 
The presence of a regularizer affects the critical point structure non-trivially, allowing for (zero-temperature) phase transitions in the sense of a qualitative change in the trained models.

\section{Basin Structure in the Absence of Regularization \label{sec:UnregularizedSection}}

Our main objective is to connect the data-formed structure of the unregularized loss landscape with the phases and phase transitions observable in the presence of $L_2$ regularization. The unregularized loss landscape for (deep) linear networks was analyzed in detail in, e.g.,  \cite{Baldi1989,Fukumizu1998,Saddles_DeepLinear,Wendin2025}. For our discussion, the important point is the nested basin structure in the unregularized loss landscape, formed by a global minimum (of the macrostate $\boldsymbol{W}^*$) and many saddle points\footnote{Here, $\nabla$ collects all partial derivatives of all variables $\vec{\theta}$ (all entries of the weight matrices). }: $\nabla \mathcal{E} \stackrel{!}{=} \vec{0}$. In the following, we summarize the known important features that are relevant for the discussion \emph{including} a regularizer. 

\noindent \textbf{Macroperspective:} Under the given assumptions, the optimal macroscopic solution that minimizes the mean squared error term is $\boldsymbol{W}^*= \Sigma_{yx}\Sigma_{xx}^{-1}$. This is the only \textbf{macro}-minimum of the loss landscape. In the presence of hidden layers, the loss landscape also features \emph{saddle points} forming levels of the basin. All critical points of the loss landscape for \emph{arbitrary deep linear networks} correspond to macrostates of the form:

\begin{align}
\boldsymbol{W}^* = W^{(n)} \cdot \dots W^{(1)} = \hat{P}_\mathcal{S} \Sigma_{yx}\Sigma_{xx}^{-1} ,
\label{eq:CriticalMacroStateNoRegularization}
\end{align}
where $\hat{P}_\mathcal{S}$ is a projector. It is constructed from a set of eigenvectors of $\tilde{\Sigma} := \Sigma_{yx} \Sigma_{xx}^{-1} \Sigma_{xy}$ corresponding to the set of eigenvalues $\lambda_i$ specified by the index set $\mathcal{S}$ \cite{Baldi1989,Saddles_DeepLinear}.
 
An important subset of these saddles are those, where we project onto an ordered subset of the $r$ largest eigenvalues of $\tilde{\Sigma}$: those saddles also have the property that they minimize the loss with a rank constraint of $r$ \cite{Bah2022,Saddles_DeepLinear}:
\begin{align}
\boldsymbol{W}^* = \text{arg min}_{\boldsymbol{W}, \text{rank$(\boldsymbol{W})\le r$}} || \boldsymbol{W} \Sigma_{xx}- \Sigma_{yx}||.
\end{align}
These saddle points already give rise to an hierarchical organization of the unregularized loss landscape where the loss for a given critical point is given by \cite{Baldi1989,Saddles_DeepLinear}
\begin{align}
\mathcal{E}(\boldsymbol{W}^*) = \text{tr}[\Sigma_{yy}] - \sum_{i \in \mathcal{S}} \lambda_i,
\end{align}
where the sum runs over all the eigenvalues of the set $\mathcal{S}$. \\

\noindent \textbf{Microperspective:} For more than 1 hidden layer, the critical points come in three flavours \cite{Saddles_DeepLinear}: minima (possibly degenerate), strict saddles (where the Hessian features at least one negative direction) and higher-order/non-strict\footnote{They are not local minima, but negative directions only emerge at higher order derivatives.} saddles. For one hidden layer, non-strict saddles do not exist. This qualitative difference is expected from a power-counting perspective: for 1 hidden layer, the unregularized loss $\mathcal{E}$ is formed by the sum of quartic and quadratic terms of the form $a_{ijlk}\theta_i \theta_j \theta_l \theta_k + b_{ij}\theta_i \theta_j$. Only for deeper linear networks with the lowest order terms being at least cubic like $b_{ijk}\theta_i \theta_j \theta_k$ etc. non-strict/higher-order saddles become possible. \\

\begin{figure*}
\centering
\includegraphics[width=0.9\textwidth]{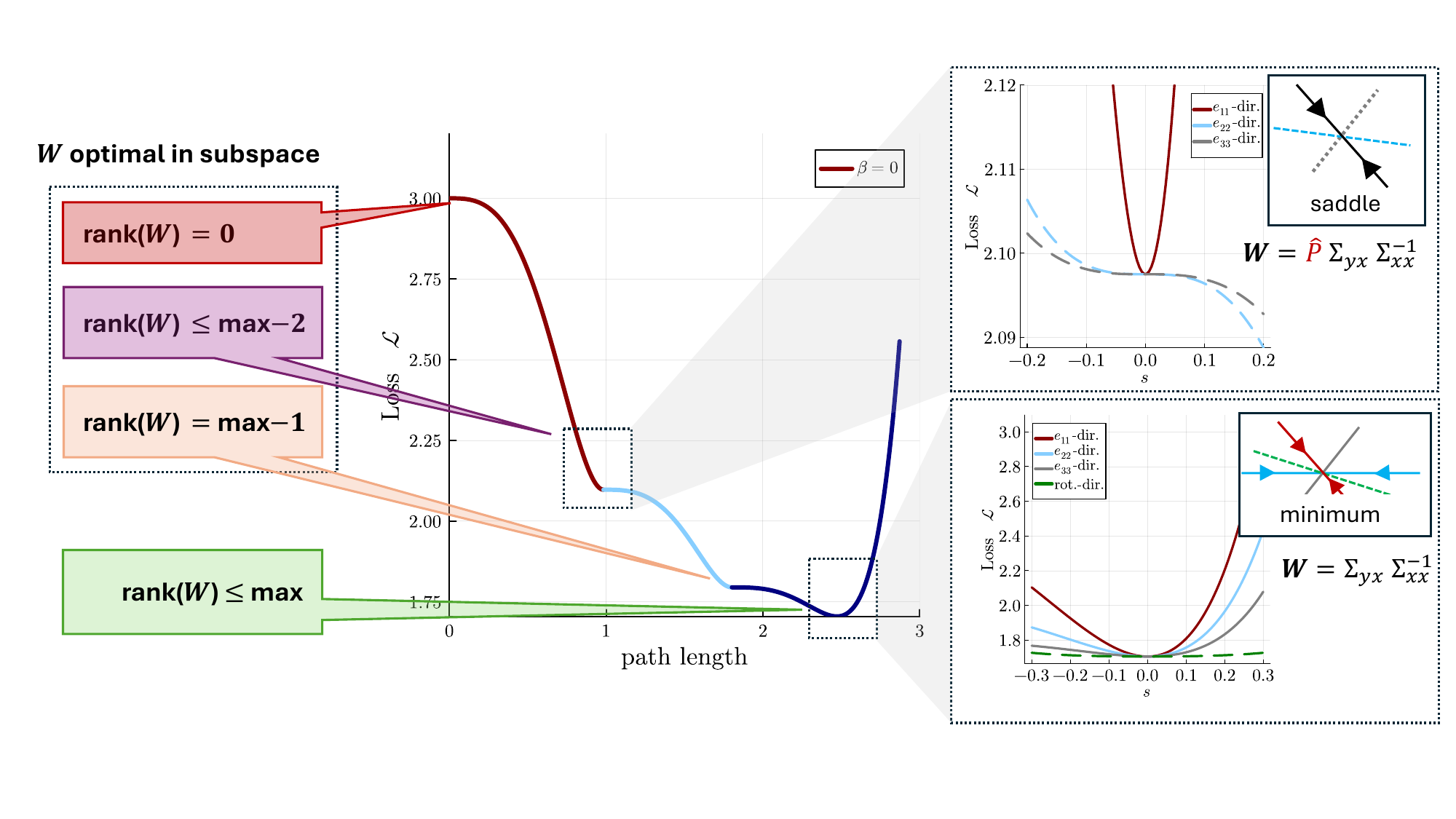}
\caption{Path through the loss landscape, based on Hessian eigendirections revealing a basin structure in the unregularized $(\beta \to 0)$ loss landscape for 2 hidden layers featuring one global minimum and multiple saddle points \eqref{eq:CriticalMacroStateNoRegularization} that partly correspond to rank-restricted optima. With: $\Sigma_{xx}=\mathbb{1}_{3\times3}$, $\Sigma_{yx} =\text{diag}(0.95, 0.55,0.3)$.}
\label{fig:UnregularizedBasin}
\end{figure*}

\noindent \textbf{Basin Structure:} Even though the parameter space is high-dimensional, the basin structure of the loss landscape can be visualized to some extend based on meaningful paths through the parameter space, see Fig.~\ref{fig:UnregularizedBasin}. For special cases, this is possible due the known structure of the Hessian eigenvectors (assuming an infinitesimal regularization strength) as well as the known (asymptotic) paths of gradient flow dynamics.

To do so, we consider for now a 2 hidden layer network with a simple data scenario: $d_\text{in} = d_\text{out}$, $\Sigma_{xx}=\mathbb{1}$ and $\Sigma_{yx} = \text{diag}(\eta_1, \eta_2 , \dots , \eta_{r_\text{max}})$ being diagonal with distinct, decreasing singular values $\eta_j$. In this setting, critical macrostates \eqref{eq:CriticalMacroStateNoRegularization} are diagonal. Starting from the trivial state $\vec{\theta}=\vec{0}$ (a non-strict saddle) and assuming an infinitesimal regularization strength\footnote{For an infinitesimal $L_2$-regularizer, the microstates all have the same set of non-zero singular values.}, we first plot the loss along the leading decay direction determined by $\eta_{j=1}$: $(W^{(1)}(s),W^{(2)}(s),W^{(3)}(s)) = s\cdot (e_{jj}, e_{jj},e_{jj})$ where $e_{jj}$ is the diagonal matrix with the only non-zero entry being $1$ at at position $j$. This path reaches a saddle point at path length $s^*=\eta_{1}^{\frac13}$, corresponding to a rank$=1$ optimal solution. The path is continued by starting from this saddle and following the direction dictated by the next subleading singular value $\eta_{2}$. Repeating this procedure until the full rank minimum is reached, we obtain Fig.~\ref{fig:UnregularizedBasin} in case of $r_\text{max}=3$ (for further details see App.~\ref{App:SymmetryExample}). The emerging picture is an extremely reduced, simplified and non-unique\footnote{Note that there are many different choices for paths: the choice made is related to the following discussion.} visualization of the full loss landscape, based on the ordering of the singular values of the underlying covariance matrix. Constructed in this way, the visible plateaus correspond to saddle points (note that the degeneracy of saddles and minima is not represented).

\section{Regularizer-Stabilized Basins \& Phase Transitions \label{Sec:RegularizedLandscape}}
\begin{figure}[h]
\centering
    \begin{subfigure}[b]{0.43\textwidth}
    \centering
    \includegraphics[width=1.1\textwidth]{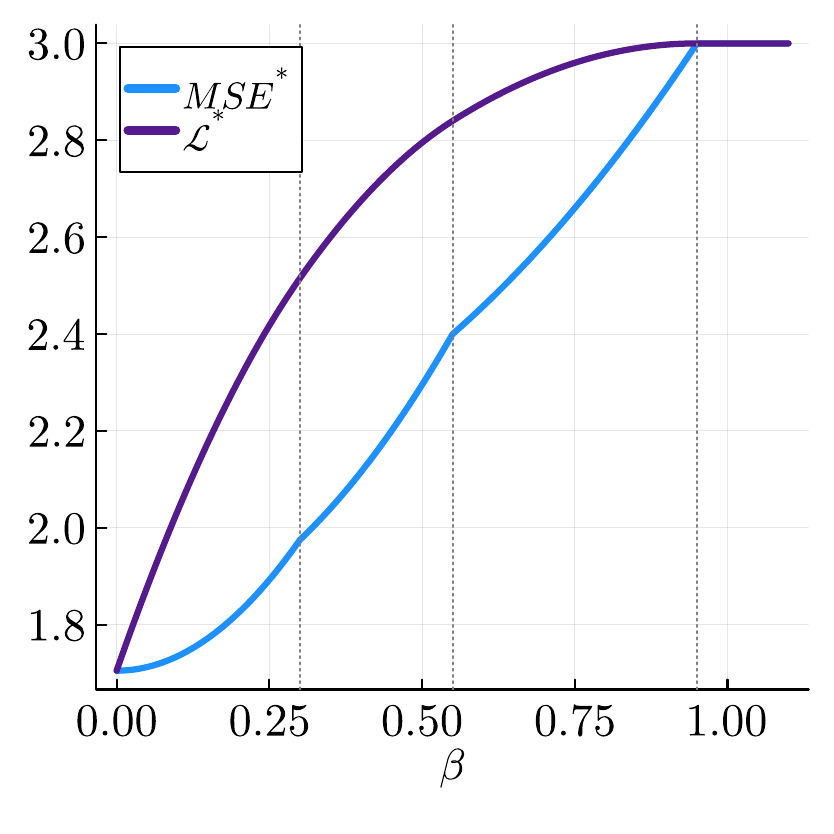}
    \caption{}
    \end{subfigure}
\quad
\centering
    \begin{subfigure}[b]{0.44\textwidth}
    \centering
    \includegraphics[width=1.1\textwidth]{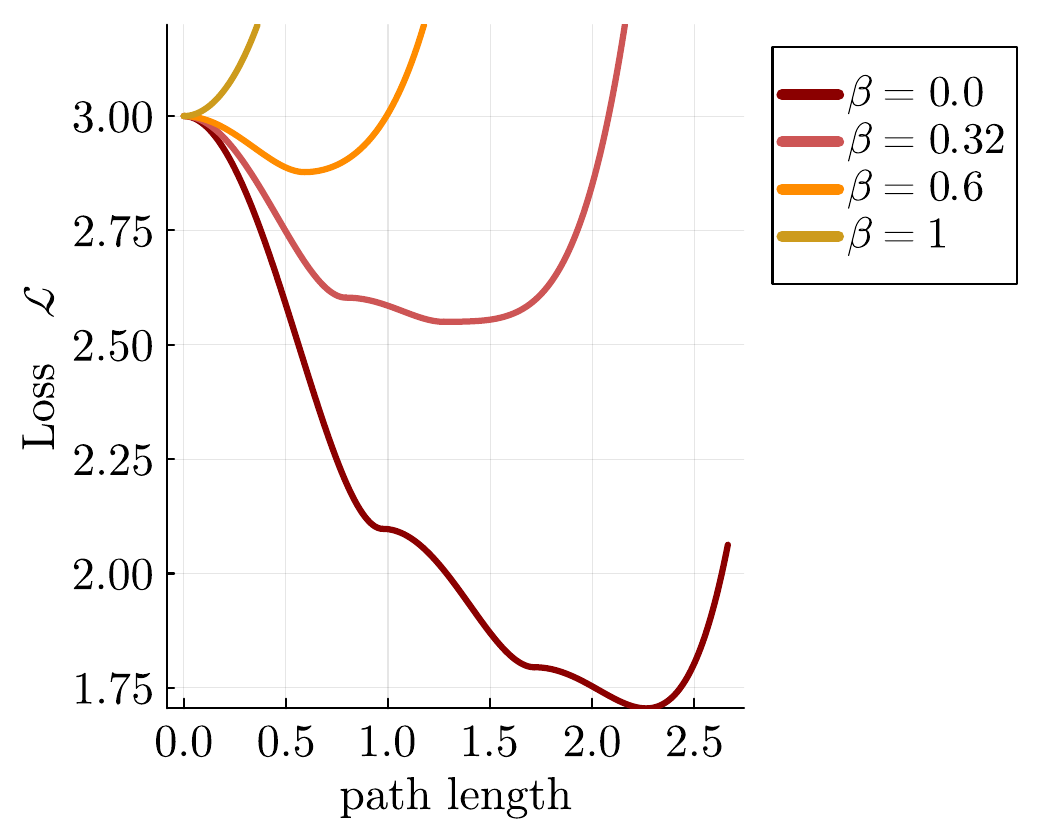}
    \caption{}
    \end{subfigure}
\quad
    \begin{subfigure}[b]{0.44\textwidth}
    \centering
    \includegraphics[width=1.1\textwidth]{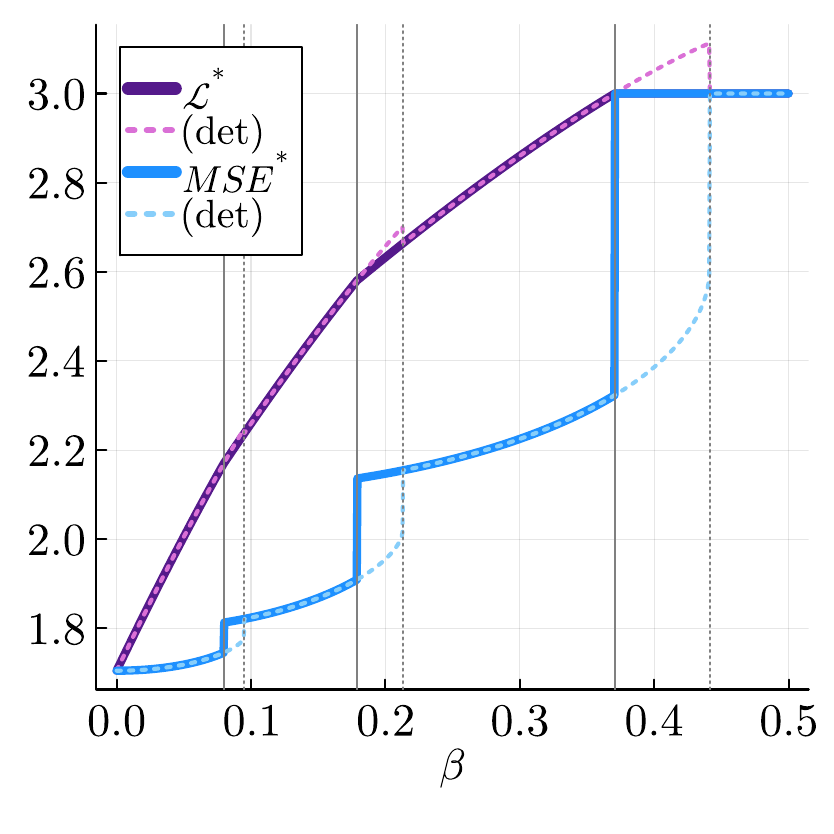}
    \caption{}
    \end{subfigure}
\quad
    \begin{subfigure}[b]{0.43\textwidth}
    \centering
    \includegraphics[width=1.1\textwidth]{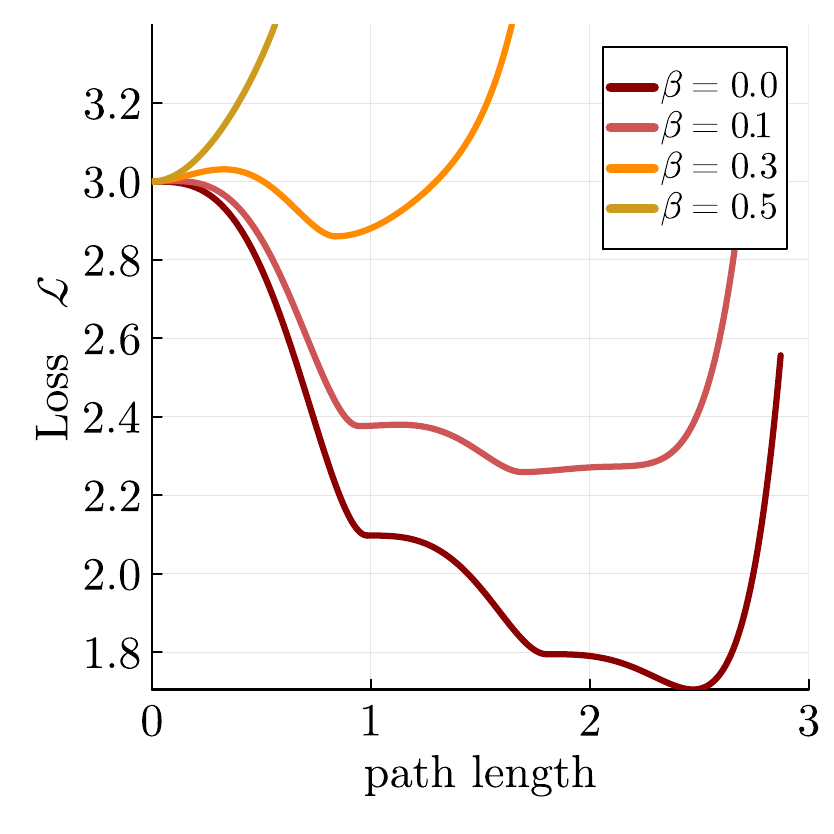}
    \caption{}
    \end{subfigure}
\caption{\textbf{(a, c)} Minimal loss $\mathcal{L}^*$ and MSE for 1 hidden layer (top) and 2 hidden layers (bottom). Colored dashed parts correspond to the loss/MSE for a reversed annealing protocol. \textbf{(b,d)} Basin structure in the regularized loss landscape for deep linear networks with 1 hidden layer (top) - featuring one minimum; and 2 hidden layers (bottom) - featuring multiple minima - for different regularization strengths $\beta$. Vertical lines indicate critical regularization strengths (in case of $1$st order transitions there are two, see main text).}
\label{fig:RegularizedLandscape}
\end{figure}

The conceptual idea behind tuning the $L_2$-regularizer is to access and dynamically stabilize models in qualitatively distinct accuracy regimes/`phases'. The geometric intuition of the $L_2$-regularizer is to add a quadratic potential to the loss landscape, penalizing large parameter values. The qualitative effect of superimposing this additional potential is the emergence of an overall wash-board like regularized loss landscape with potentially multiple local (macro) minima, see Fig.~\ref{fig:RegularizedLandscape}, causing 1st (and 2nd) order phase transitions\footnote{Note that adaptions of weight decay like cautious weight decay \cite{Liang2026cautiousweightdecay,Chen2026cautiousweightdecay}, though similar, do not feature a distorted landscape by the regularization (the critical points are the ones of the unregularized loss landscape).}. \\

\noindent \textbf{Overview:} In case of linear activation functions, we show the existence of those minima and analyze their implication. In the following, we summarize the main findings. In the next sections, we discuss the 1 hidden and $>1$ hidden layer case in more detail. As we are interested in critical points of the loss landscape as possible stationary points of some training dynamics, we start with a dynamical point of view before discussing the loss without reference to the dynamics. For gradient flow as an example, the dynamics can be qualitatively separated into three parts (based on \cite{Arora2018,Lindsey2025}): \\

\noindent \textbf{$0$-balancing dynamics:} The regularization leads to the exponential decay of the following difference (see also, e.g., \cite{Lindsey2025,Wang2024}):
\begin{align}
&W^{(i+1)}{}^T(t) W^{(i+1)}(t) - W^{(i)}(t) W^{(i)}{}^T(t) \label{eq:DecayConservedQuantity}\\
&= \left(W^{(i+1)}{}^T(0) W^{(i+1)}(0) - W^{(i)}(0) W^{(i)}{}^T(0)\right)e^{-2\beta t} \nonumber 
\end{align}
resulting in the  \emph{$0$-balanced condition} (see, e.g., \cite{Arora2018,Saddles_DeepLinear}) for the microstates for $t\to \infty$:
\begin{align}
\text{$0$-balanced condition:} \quad W^{(i+1)}{}^T W^{(i+1)} = W^{(i)} W^{(i)}{}^T.
\label{eq:0balancedcondition}
\end{align}
This condition implies that (for long times) all weight matrices share the same non-zero singular values and the product $\boldsymbol{W} = W^{(n)}\cdots W^{(1)}$ has the (singular value decomposed) form:
\begin{align}
\boldsymbol{W} = \textcolor{blue}{L}S^{(n)}\textcolor{orange}{V_n^T V_n} S^{(n-1)} \textcolor{orange}{V_{n-1}^TV_{n-1}} \cdots S^{(1)} \textcolor{blue}{R^T} = \sum_{i=1}^{r_\text{max}}\sigma_i^n \, \textcolor{blue}{\vec{l}_i} \cdot \textcolor{blue}{\vec{r}_i^T},
\label{eq:SVDMacroStateGeneric}
\end{align}
where $L$ and $R$ are orthogonal matrices of dimension $d_\text{out} \times d_\text{out}$ and $d_\text{in} \times d_\text{in}$ respectively. $S^{(j)}$ are `diagonal' matrices containing the singular values $\sigma_i$ of the matrices $W^{(i)}$. $\textcolor{blue}{\vec{l}_i}$ and $\textcolor{blue}{\vec{r}_i}$ are the left and right singular vectors corresponding to the non-zero singular values. \\

\noindent\textbf{Dynamics in this $0$-balanced subspace}\footnote{compare also to `learning dynamics' in \cite{Lindsey2025}.}: Once the system has entered the $0$-balanced subset, the dynamics can be described either in terms of the macroscopic $\boldsymbol{W}$ alone as 
[adapting \cite{Arora2018,Menon2025}]:
\begin{align}
\frac{d}{dt}\boldsymbol{W}(t) = -2 \left( \sum_{k=0}^{n-1} (\boldsymbol{W}\boldsymbol{W}^T)^{\frac{k}{n}} \hat{N}_{\boldsymbol{W}} (\boldsymbol{W}^T \boldsymbol{W})^{\frac{n-1-k}{n}} + n\beta \boldsymbol{W}\right)
\label{eq:MacroStateEvolutionSingle}
\end{align}
with the disadvantage that the right hand side involves roots of matrices. Alternatively, \emph{three} dependent macroscopic objects can be used: $\{\textcolor{blue_L}{\mathfrak{W}_L},\boldsymbol{W}, \textcolor{blue_R}{\mathfrak{W}_R}\}$ with $\textcolor{blue_L}{\mathfrak{W}_L:=(\boldsymbol{W}\boldsymbol{W}^T)^{\frac{1}{n}} = W^{(n)}W^{(n)}{}^T}$ and $\textcolor{blue_R}{\mathfrak{W}_R:=\boldsymbol{W}^T\boldsymbol{W})^{\frac{1}{n}}=W^{(1)}{}^TW^{(1)}}$ for arbitrary depths (well-known for the exactly solvable 1 hidden layer case \cite{Fukumizu1998}; see also \cite{Arora2018,Menon2026,Lindsey2025} for arbitrary depth):
\begin{align}
\begin{aligned}
&\frac{d}{dt}\textcolor{blue_L}{\mathfrak{W}_L} = -2(\hat{N}_{\boldsymbol{W}} \boldsymbol{W}^T + \boldsymbol{W} \hat{N}_{\boldsymbol{W}}^T +2\beta \textcolor{blue_L}{\mathfrak{W}_L}), \\
&\frac{d}{dt}\boldsymbol{W} = - 2\left(\sum_{i=0}^{n-1}  (\textcolor{blue_L}{\mathfrak{W}_L})^{i} \hat{N}_{\boldsymbol{W}} (\textcolor{blue_R}{\mathfrak{W}_R})^{n-i-1} +n\beta \boldsymbol{W} \right),\\
&\frac{d}{dt}\textcolor{blue_R}{\mathfrak{W}_R} = - 2(\hat{N}_{\boldsymbol{W}}^T \boldsymbol{W} +\boldsymbol{W}^T \hat{N}_{\boldsymbol{W}} +2\beta \textcolor{blue_R}{\mathfrak{W}_R}),
\label{eq:MacroDynamics}
\end{aligned}
\end{align}
with $\hat{N}_{\boldsymbol{W}} := \boldsymbol{W}\Sigma_{xx} - \Sigma_{yx}$. The macro critical points $\boldsymbol{W}_c$ can be directly obtained from the stationary points of this macro dynamics \eqref{eq:MacroStateEvolutionSingle} or \eqref{eq:MacroDynamics} with minimal loss solutions for different $\beta$ shown in Fig.~\ref{fig:RegularizedLandscape}. At the technical level, the three quantities carry information about the right singular vectors, the left singular vectors and the actual function $\boldsymbol{W}$. More physically, they are related to objects like the Neural Tangent Kernel (see, e.g., \cite{Domine2025} for 1 hidden layer case). \\

\noindent\textbf{Special case - dynamics of the singular values \emph{alone}:} 

For a special set of 'aligned' (co)variances (defined in \eqref{eq:compatibilitycondition}) and the condition that the singular vectors of the macrostate are already aligned with the stationary state, the dynamics can be restricted to the singular values $\sigma_j(t)$ that evolve under gradient flow of an effective (loss) potential:

\begin{align}
\begin{aligned}
&\mathcal{L}(\sigma_1, \dots, \sigma_{r_\text{max}}) = \sum_{j=1}^{r_\text{max}} L_j(\sigma_j)\textcolor{gray}{+\text{tr}[\Sigma_{yy}]}, \\
&L_j(\sigma_j) := \variancevalues{j}\sigma_j^{2n} - 2\eta_j \sigma_j^n+ \beta n \sigma_j^2, \\
&\frac{d}{dt}\sigma_j = - \frac{1}{n} \partial_{\sigma_j} L_j(\sigma_j).
\end{aligned}
\end{align}
Here, $\variancevalues{j}$ and $\eta_j$ are singular values of $\Sigma_{xx}$ and $\Sigma_{yx}$ for same singular vector. This expression can equivalently be seen as an effective loss in the subspace of diagonal (singular value) matrices for fixed singular vectors, aligned with $\Sigma_{xx}$ and $\Sigma_{yx}$.

\noindent Under the alignment condition for the (co)variance matrices, the critical points can be determined exactly. The effective loss $\mathcal{L}(\sigma_1,\dots, \sigma_{r_\text{max}})$ splits into independent parts $L_j$ for each singular value, giving rise to a picture reminiscent of Landau free energies for second and first order phase transitions \emph{for the set of \textbf{order parameters} $\boldsymbol{\sigma} = (\sigma_1,\dots, \sigma_{r_\text{max}})$} consisting of the singular values\footnote{Adapting the notation in \cite{Saddles_DeepLinear}, we refer to the index set of finite singular values as with $\mathcal{S}$. A special role is played by the ordered set from $1$ to $r$, denoted as $[[1,r]]$.}:

\begin{itemize}
\item \textbf{$1$ hidden layer:} Each $L_j$ has a unique minimum: for $\beta>\beta_j^* = \eta_j$, the minimum is at $\sigma_j=0$ and wanders off from zero continuously for $\beta < \beta_j^*$, see Fig.~\ref{fig:effectivepotential_and_orderparameters_1hidden}. For a given $\beta$, only a subset of the $\sigma_j$'s is finite at the minimum of $\mathcal{L}(\boldsymbol{\sigma})$ determining the rank of the minimal loss solution. Depending on $\beta$, the minimal loss macro model is rank-reduced (rank $r$).
This picture of decoupled 1d problems can as well be understood from the perspective of the loss landscape as shown in Fig.~\ref{fig:RegularizedLandscape}.

\item \textbf{$>1$ hidden layers:} Each $L_j$ has a (local) minimum at $\sigma_j=0$ and a (local) minimum at $\sigma_j=\sigma_j^*$ (for sufficiently small $\beta$), see Fig.~\ref{fig:ExamplesEffectiveDescription2hidden}. Again, the number of finite $\sigma_j$'s determines the rank of the solution.

\end{itemize}
This effective picture can be compared to the reduced representation of the loss landscape along cuts in parameter space as shown in Fig.~\ref{fig:RegularizedLandscape}(b) (for 1 hidden layer) and Fig.~\ref{fig:RegularizedLandscape}(d) (for more hidden layers).

\subsection{Cascade of Second Order Transitions: 1 hidden Layer Networks}
\begin{figure*}
\includegraphics[width=0.9\textwidth]{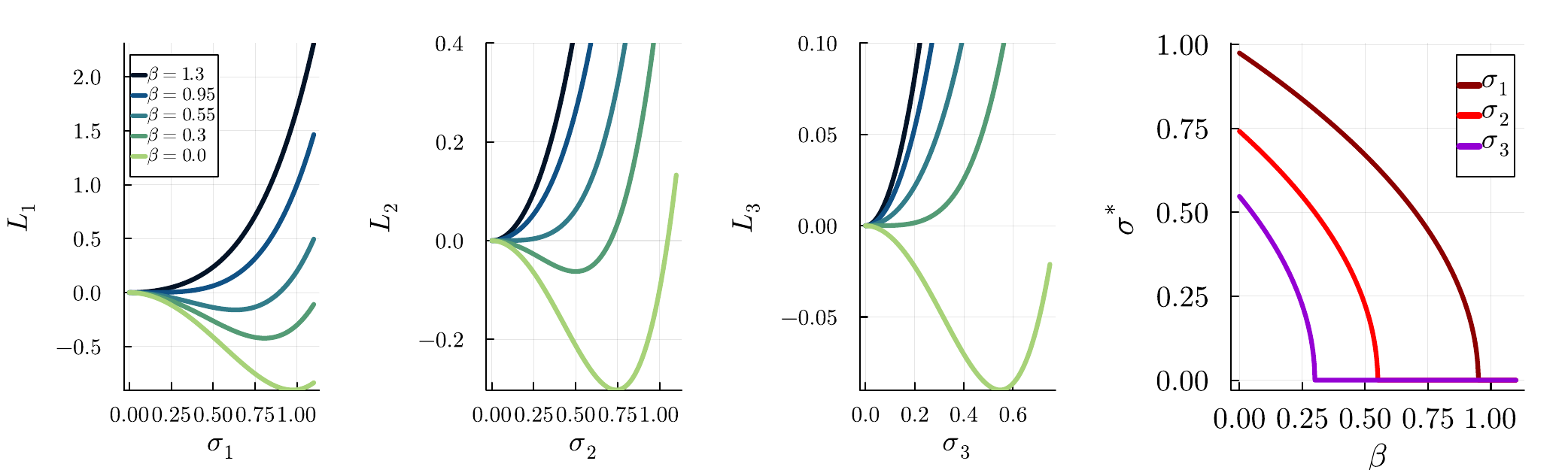}
\caption{Effective loss landscapes (1 hidden layer) for $\eta_1=0.95, \eta_2=0.55,\eta_3=0.3$; the right most plot shows the minima $\sigma_i^*$ of the effective potentials \eqref{eq:EffecrtivePotential1hidden} as a function of $\beta$, displaying the continuous onset at critical regularization strengths $\beta_i^*$.}
\label{fig:effectivepotential_and_orderparameters_1hidden}
\end{figure*}

For 1 hidden layer, decreasing the regularization strength results in a cascade of (continuous) second order transitions. Under the alignment assumption of the data distribution \eqref{eq:compatibilitycondition}), the order parameter are effectively described by $\mathcal{L}(\boldsymbol{\sigma}) = \sum_{j=1}^{r_\text{max}} L_j(\sigma_j)$ with the individual quartic potentials:
\begin{align}
L_j(\sigma_j) = \variancevalues{j} \sigma_j^4 -2(\eta_j-\beta) \sigma_j^2
\label{eq:EffecrtivePotential1hidden}
\end{align}
shown in Fig.~\ref{fig:effectivepotential_and_orderparameters_1hidden}. This is the typical form of a Landau free energy for second order transitions with a continuous and successive onset of the order parameters: $\sigma_j \sim \sqrt{\beta_j^*-\beta}$ with a critical exponent\footnote{'$\beta$'-exponent in the jargon of critical exponents.} of $1/2$ (in accordance with \cite{Ziyin2023}), see Fig.~\ref{fig:effectivepotential_and_orderparameters_1hidden}. Here, $\beta_j^*=\eta_j$ is the critical regularization strength corresponding to the onset of learning of the $j$th singular direction of $\Sigma_{yx}$ (under the alignment assumption of the data distribution \eqref{eq:compatibilitycondition}). This cascade of second order transitions at the level of the loss and MSE is shown in Fig.~\ref{fig:RegularizedLandscape}(a). Note that by lowering the regularization strength, the singular directions of $\Sigma_{yx}$ are learned in descending order: the larger $\eta_j$, the larger $\beta_j^*$ or differently put: the more robust against un-learning. \\

\noindent \textbf{Details:} The critical points can be exactly determined for $\Sigma_{xx}$ and $\Sigma_{yx}$ being compatible in the following sense ($\textcolor{blue}{R} \in O(d_\text{in}), \textcolor{blue}{L} \in O(d_\text{out})$): (this special case was already noted in, e.g., \cite{Saxe2014exactsolutionsnonlineardynamics} for the unregularized setting)
\begin{align}
&\text{alignment condition:} \nonumber \\
&\begin{rcases}
\Sigma_{xx} = \textcolor{blue}{R} S_{xx} \textcolor{blue}{R^T} \\
\Sigma_{yx} = \textcolor{blue}{L} S_{yx} \textcolor{blue}{R^T} \\
\end{rcases}
\quad \Sigma_{yx}\Sigma_{xx}^{-1} = \textcolor{blue}{L} S_{yx} S_{xx}^{-1} \textcolor{blue}{R^T}
\label{eq:compatibilitycondition}
\end{align}
where the $S$-matrices are diagonal, collecting the singular values. Therefore, the singular value decomposition of $\Sigma_{yx}\Sigma_{xx}^{-1}$ is inherited from the ones of $\Sigma_{xx}$ and $\Sigma_{yx}$. In this case, critical solutions take the form:

\begin{align}
W^{(2)} = \textcolor{blue}{L} S^{(2)} \textcolor{orange}{V^T}, \quad W^{(1)} = \textcolor{orange}{V} S^{(1)} \textcolor{blue}{R^T}
\end{align}
with $\textcolor{orange}{V} \in \mathcal{O}(d_\text{hidden})$. The optimal singular values of $W^{(1)}$ and $W^{(2)}$ are given by\footnote{Note the correspondence to the singular value shrinkage discussed in, e.g., \cite{cai2010singular}.}: $\sigma_i^*{}^2  = \text{max}\left((\eta_i -\beta)d_i^{-1},0\right)$, where $\eta_i$ are the singular values of $\Sigma_{yx}$ and $d_i$ are the eigenvalues of $\Sigma_{xx}$. The corresponding loss is 
\begin{align}
\mathcal{L}(\boldsymbol{W}^*) = -\sum_{i \in \mathcal{S}} \underbrace{(\eta_i -\beta)^2 \variancevalues{i}^{-1}}_{=:L_i(\sigma_i^*)}  \textcolor{gray}{+ \text{tr}[\Sigma_{yy}]}.
\end{align}
As soon as $\beta < \eta_i$, an additional singular value becomes non-zero, see Fig.~\ref{fig:effectivepotential_and_orderparameters_1hidden}, resulting in an increased rank (rank transition). Therefore, a cascade of second order "phase transitions" for each eigensubspace of the regression matrix $\Sigma_{yx}\Sigma_{xx}^{-1}$ results in accordance with the effective landscape picture. The corresponding critical regularization strengths are $\beta_j^* = \eta_j$. See Tab.~\ref{Tab:1hiddenLayerOverview} for an overview.

\begin{table*}
\centering
\begin{tabular}{l|l|ll}
\textbf{regularization strength} & \multicolumn{3}{c}{\textbf{optimal macrosolution} $\boldsymbol{W}^*$}  \\ \hline
$\beta > \eta_1$ & $\mathcal{S} = \{\}$ & $\boldsymbol{W}^*_{r=0}=\boldsymbol{0}$ & \multirow{5}{*}{$\begin{rcases} & \\ & \\ & \\ & \\ & \end{rcases} \boldsymbol{W}_r^*=P_{\mathcal{S}}(\Sigma_{yx}-\beta \textcolor{blue}{L}\mathbb{1}_{out,in}\textcolor{blue}{R^T}) \Sigma_{xx}^{-1}$} \\
$\eta_1>\beta > \eta_2$ & $\mathcal{S} = \{1\}$ & $\boldsymbol{W}^*_{r=1}=\sigma_1(\beta) \textcolor{blue}{\vec{l}_1 \cdot \vec{r}_1^T}$ \\ 
$\eta_2>\beta > \eta_3$ & $\mathcal{S} = \{1,2\}$ & $\boldsymbol{W}^*_{r=2}= \sigma_1(\beta) \textcolor{blue}{\vec{l}_1 \cdot \vec{r}_1^T}+ \sigma_2(\beta) \textcolor{blue}{\vec{l}_2 \cdot \vec{r}_2^T}$  \\
\vdots & \vdots & \\
$\beta < \eta_{r_\text{max}}$ & $\mathcal{S} = \{1,\dots , r_{\text{max}}\}$ & $\boldsymbol{W}^*_{r=r_\text{max}} = \sum_{i=1}^{r_\text{max}} \sigma_i(\beta) \textcolor{blue}{\vec{l}_i \cdot \vec{r}_i^T}$ 
\end{tabular}
\caption{Overview of rank transitions for 1 hidden layer (under the alignment condition) with $\eta_1 >\eta_2 > ...$. $P_{\mathcal{S}}=\sum_{i\in \mathcal{S}} \textcolor{blue}{\vec{l}_i \vec{l}_i^T}$ denotes a projection onto the subspace defined by $\mathcal{S}$ (see App.~\ref{App:CriticalPoints}).}
\label{Tab:1hiddenLayerOverview}
\end{table*}

\noindent Summarizing, the effective loss $\mathcal{L}(\boldsymbol{\sigma}) = \sum_{j=1}^{r_\text{max}} L_j(\sigma_j)$ describes the main properties of macrostates $\boldsymbol{W}$ in terms of its singular values. 

A critical macrostate $\boldsymbol{W}^*$ corresponds to a set of singular values $\boldsymbol{\sigma}$ such that $\partial_{\sigma_j} L_j(\sigma_j)=0$. The macrostate of minimal loss is determined by $\boldsymbol{\sigma} = (\sigma_1^*, \dots, \sigma_{r_{\text{max}}(\beta)}^*, 0,\dots)$ such that $\sigma_j^*$'s are minima of $L_j(\sigma_j)$ and $r_{\text{max}}(\beta)$ corresponds to the index of the smallest singular value of $\Sigma_{yx}$ that is still learnable. At the micro level, this effective description already captures that the full regularized loss landscape has a global ($\beta$-dependent) minimum but no \emph{local} minima. However, the effective description is reduced and does not capture the full properties of the critical points of the loss landscape described by $\boldsymbol{\sigma}$ (like saddle points, degeneracies etc.), which we will discuss in the following.\\

\subsubsection{From the effective Description to the full Loss Landscape} 
\begin{figure*}
\centering
    \begin{subfigure}[b]{0.24\textwidth}
    \centering
    \includegraphics[width=\textwidth]{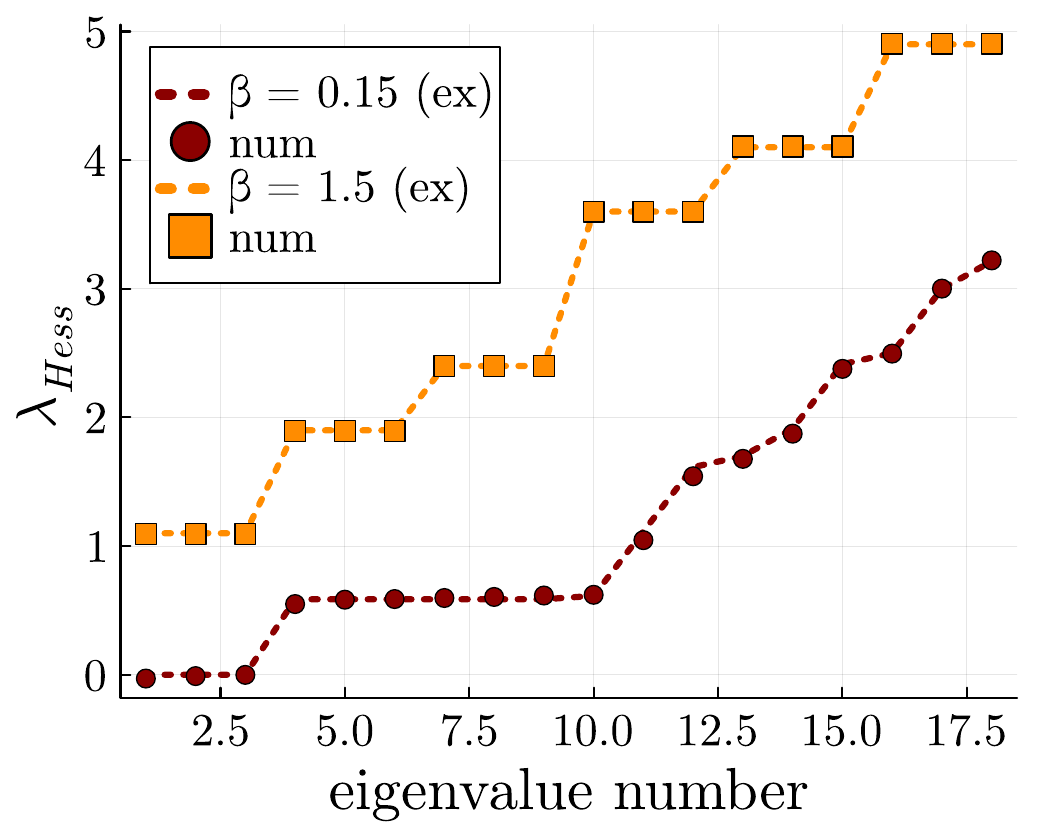}
    \caption{}
    \end{subfigure}
    \begin{subfigure}[b]{0.24\textwidth}
    \centering
    \includegraphics[width=\textwidth]{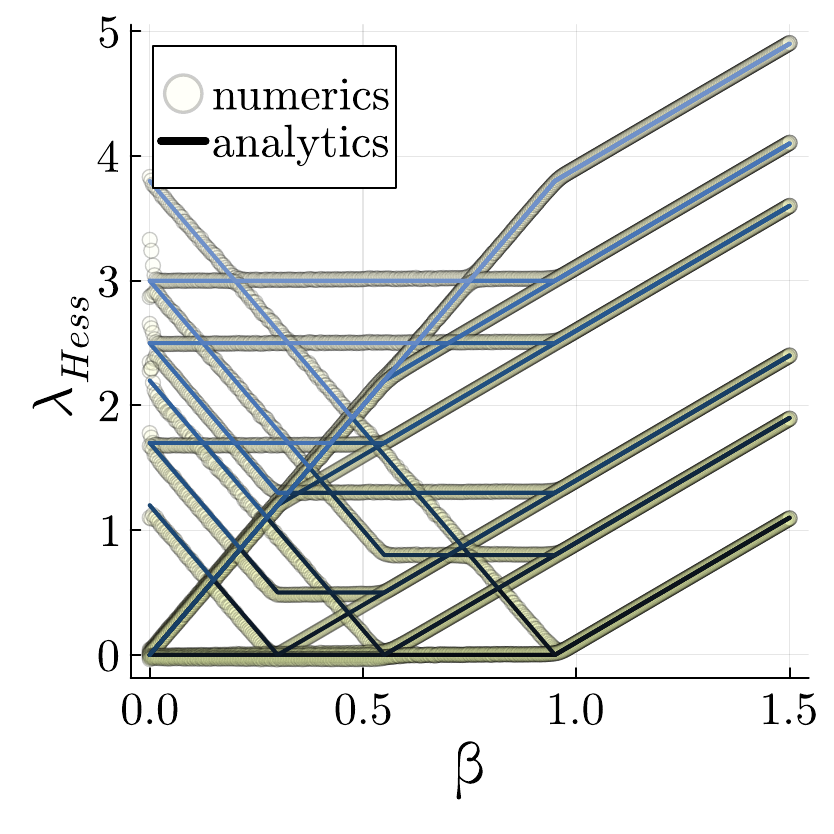}
    \caption{}
    \end{subfigure}
    \vrule
        \begin{subfigure}[b]{0.24\textwidth}
    \centering
    \includegraphics[width=\textwidth]{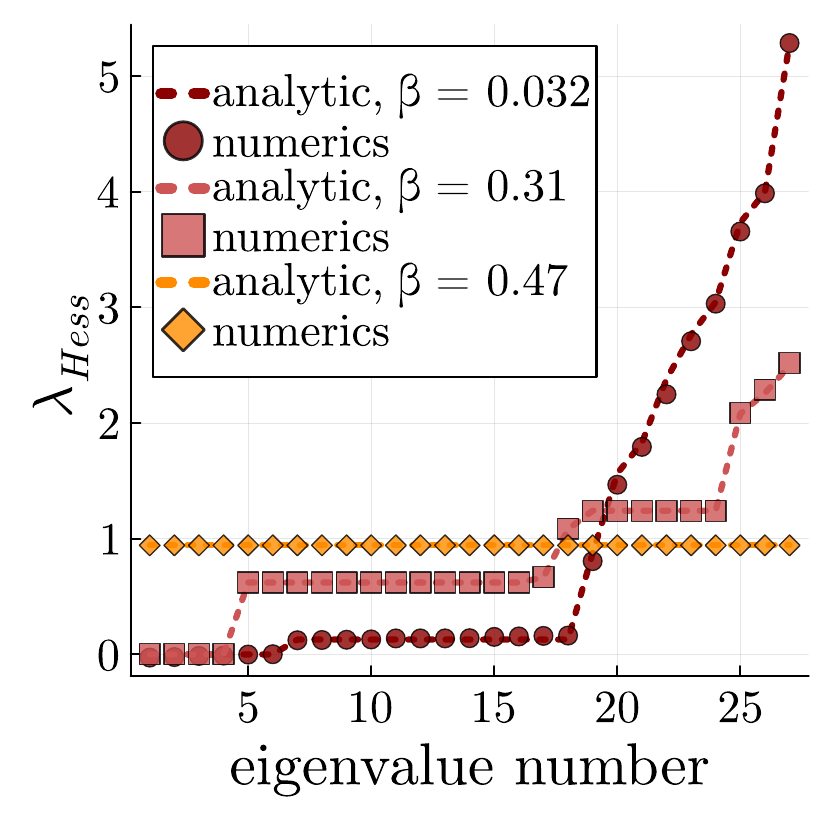}
    \caption{}
    \end{subfigure}
    \begin{subfigure}[b]{0.24\textwidth}
    \centering
    \includegraphics[width=\textwidth]{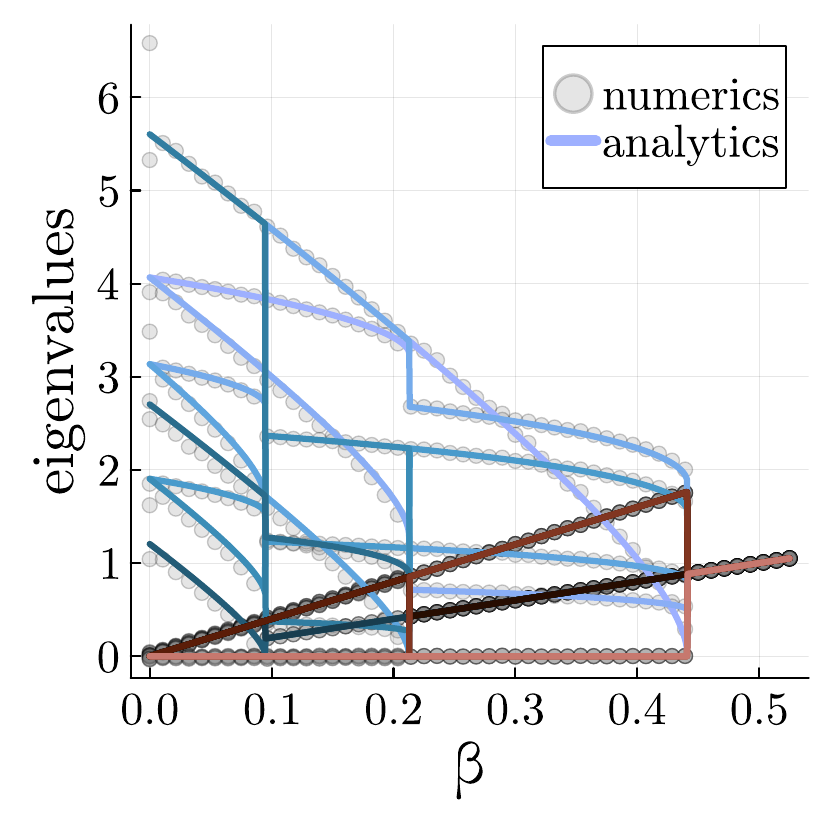}
    \caption{}
    \end{subfigure}   
\caption{Hessian spectra for 1 hidden (a,b) and 2 hidden layers (c,d) with $d_\text{in}=d_\text{out}=d_\text{hidden}=3$ and $\eta_1=0.95,\eta_2=0.55,\eta_3=0.3$ based on numerical simulations (reversed annealing) and analytical prediction: \textbf{(a)} Hessian spectra for 1 hidden layer and different, fixed $\beta$: (i) $r_\text{max}(\beta)=0$ (square); (ii) $r_\text{max}(\beta)=3$ (circle); \textbf{(b)} Hessian spectrum as a function of $\beta$. \textbf{(c)} Hessian spectra for 2 hidden layers and different, fixed $\beta$ for (i) $r_\text{max}(\beta)=0$ (diamond), (ii) $r_\text{max}(\beta)=1$ (square) and (iii) $r_\text{max}(\beta)=r_\text{max}$ (circle); \textbf{(d)} Hessian spectrum as a function of $\beta$.}
\label{fig:HessianSpectrum1hidden}
\end{figure*}

In a first step, we analyze the Hessian spectrum at the critical points in the regularized loss landscape. Mirco critical points $(W^{(2)},W^{(1)})$ belong to macro solutions according to \eqref{eq:SVDMacroStateGeneric}, where the orthogonal matrices $\textcolor{orange}{V_j} \in O(d_j)$ are arbitrary. They have the same set of singular values $\boldsymbol{\sigma}$, but are highly degenerate, resulting in zero eigenvalues in the Hessian spectrum. The spectrum is summarized in Tab.~\ref{tab:HessianSpectrum1hidden} and can be organized into three qualitative regimes: (i) massive data-dependent eigenvalues, (ii) $\beta$-eigenvalues and (iii) zero eigenvalues, see Tab.~\ref{tab:HessianSpectrum1hidden} and Fig.~\ref{fig:HessianSpectrum1hidden}. The $\beta$-eigenvalues stem partly from the explicitly broken symmetry due to the presence of the $L_2$-regularizer. The massive data-dependent eigenvalues instead are tight to the already learned parts of the data distribution. For an order parameter set $\boldsymbol{\sigma}$ \emph{minimizing} $\mathcal{L}(\boldsymbol{\sigma})$, the spectrum only features positive and zero eigenvalues. All zero eigenvalues are associated with the \emph{degeneracy} of the macrostate, as we will discuss below. Therefore, the minimum of the effective description at a given $\beta$ corresponds to the (only) minimum of the full regularized loss landscape. However, a maximum in $\mathcal{L}(\boldsymbol{\sigma})$ corresponds to a saddle point in the loss landscape, since there are always positive eigenvalues in the Hessian spectrum.

\begin{table*}
\centering
\begin{tabular}{l | l | l  }
sector & Hessian eigenvalue & numerosity \\ \hline
\rowcolor{blue!30!white}\multicolumn{3}{l}{\textbf{ \textsf{massive data-dependent sector}} \quad \quad dim. =$r^2+r(d_\text{in}-r)+r(d_\text{out}-r)$} \\ \hline
$P_\mathcal{S}$  &  $2(\sigma_a^2 +\sigma_b^2)$ for $a,b\in \mathcal{S}$  & $\frac12 r(r+1)$  \\
$P_\mathcal{S}$ &  $2(\sigma_a^2 +\sigma_b^2 +2\beta)$ for $a,b\in \mathcal{S}$ & $\frac12 r(r-1)$  \\
mixing  & $2(\eta_a\mp\eta_b)$ for $a\in\mathcal{S},b\notin\mathcal{S}$ & $2r(d_\text{min}-r)$ \\ 
in/out mismatch  & $2\eta_a$ for $a \in \mathcal{S} $ & $r(d_\text{min}-d_\text{max})$ \\\hline 
\rowcolor{lightgray}\multicolumn{3}{l}{\textbf{\textsf{null space sector}} \quad \quad \quad \quad  dim.=$(d_\text{min}-r)(d_\text{hidden}-r)+(d_\text{max}-r)(d_\text{hidden}-r)$} \\ \hline
$P_\mathcal{S}^\perp$ & $2(\mp \eta_b + \beta)$ for $b\notin \mathcal{S}$ & $2(d_\text{min}-r)(d_\text{hidden}-r)$ \\
in/out mismatch & $2\beta$ & $(d_\text{max}-d_\text{min})(d_\text{hidden}-r)$ \\ \hline
\rowcolor{lightgray}\multicolumn{3}{l}{\textbf{\textsf{$\beta$ sector (explicitly broken symmetry)}} \quad \quad dim.=$(d_\text{min}-r)(d_\text{hidden}-r)+(d_\text{max}-r)(d_\text{hidden}-r)$}  \\ \hline
$P_\mathcal{S}$ & $4\beta$ & $\frac12 r(r+1)$   \\
mixing & $4\beta$ & $r(d_\text{hidden}-r)$ \\ \hline
\rowcolor{lightgray}\multicolumn{3}{l}{\textbf{\textsf{zero eigenvalues (from symmetry)}} \quad \quad \quad dim.=$\frac12r(r-1) + r(d_\text{hidden}-r)$} \\ \hline
$P_\mathcal{S}$  & $0$ & $\frac12 r(r-1)$ \\
mixing  & $0$ & $r(d_\text{hidden}-r)$ 
\end{tabular}
\caption{Hessian spectrum at critical points of rank $r$ of the regularized loss landscape for 1 hidden layer. $P_\mathcal{S}$ refers to the projection of the $W^{(i)}$ onto the $r^2$-dim. subspace of finite singular values. See App.~\ref{App:Hessian1hiddenLayer} for more details.}
\label{tab:HessianSpectrum1hidden}
\end{table*}

\noindent \textbf{Degeneracies \& Structure of different Rank Solutions:} The different rank (critical) macrosolutions are accompanied by increasing degeneracies at the microlevel: many different microstates belong to the same macrostate. This degeneracy can be captured by an \emph{order parameter space} for a given phase (see, e.g., \cite{Altland_Simons_2023}): given $(\sigma_1, \dots, \sigma_r, 0,\dots)$, we can pair up each extremal microsolution, belonging to the same macrosolution $\boldsymbol{W}_r^*$ of rank $r$, with an element in this order parameter space. For the trivial state $\boldsymbol{W}_{r=0}^*$, there is only one valid \emph{extremal} microsolution $(W^{(2)}=\boldsymbol{0}, W^{(1)}=\boldsymbol{0})$. The order parameter space contains only a single element. In case of rank $r=1$, we have 
\begin{align}
\boldsymbol{W}_{r=1}^* = \sigma_1^2(\beta) \textcolor{blue}{\vec{l}_1} \cdot \textcolor{blue}{\vec{w}_1^T} = W^{(2)}W^{(1)} = \sigma_1^2(\beta) \textcolor{blue}{\vec{l}_1} \textcolor{orange}{\vec{v}_1^T \vec{v}_1} \textcolor{blue}{\vec{w}_1^T}
\end{align}
Since the singular vectors are normalized, any choice of (normalized) \textcolor{orange}{$\vec{v}_1$} is valid and corresponds to a different microstate. This implies that the (global) minimum is highly degenerate. Since $\vec{v}_1$ is a $d_\text{hidden}$-dim. vector with the constraint $|| \vec{v}_1||=1$, a choice of $\vec{v}_1$ can be associated with an element of the $d_\text{hidden}-1$-dim. sphere $S^{d_\text{hidden}-1}$. Therefore, the dimension of the order parameter space is $d_{\text{hidden}-1}$ and for any of those micro critical points, the Hessian features $d_{\text{hidden}-1}$ flat directions.\\

\begin{figure}[h]
\centering
\includegraphics[width=\textwidth]{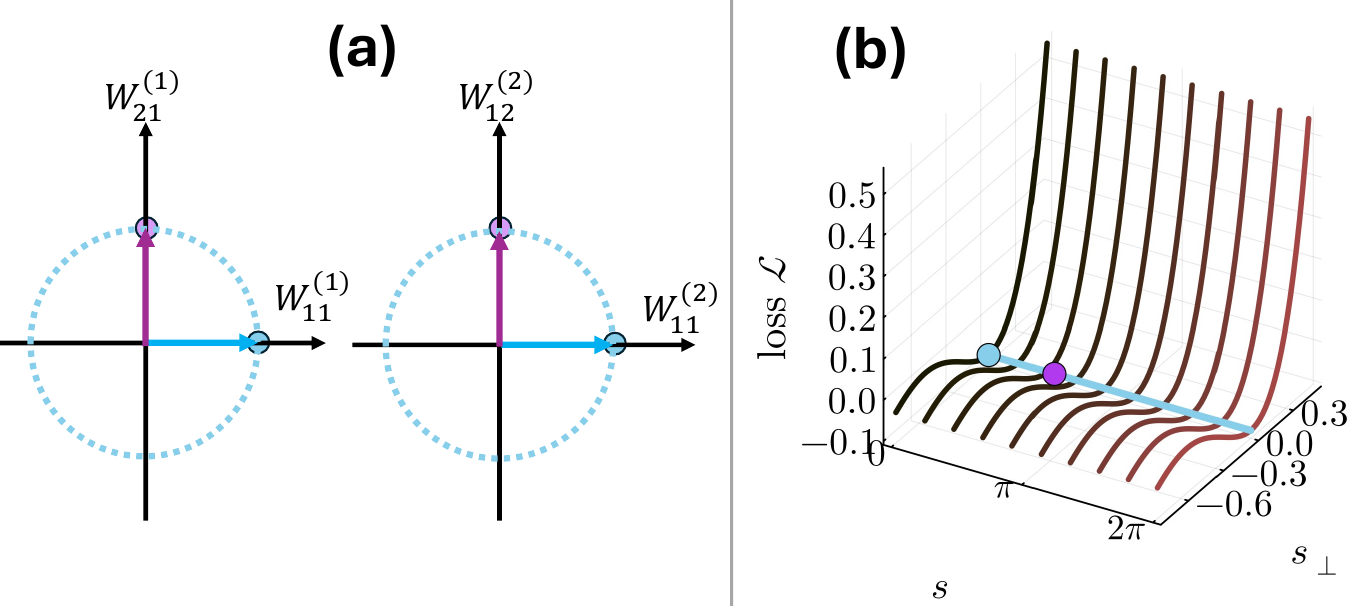}
\caption{\textbf{(a)} In the example ($d_\text{in} = d_\text{out}=1$ and $d_\text{hidden}=2$) minima are degenerate and continuously connected:  any pair $(W^{(2)},W^{(1)})$, examples indicated by points of the same color, belong the same macrostate. These pairs of points lie on a continuous path that can be parametrized by $s$ and can be labeled by a point on an (abstract) $S^1$, represented in \textbf{(b)}. Shown is the local loss along the symmetry direction (constant) and in one orthogonal direction (see also App.~\ref{App:SymmetryExample}).}
\label{fig:OrderParameterSpaceIllustration}
\end{figure}

\noindent \textbf{Example:} The simplest example corresponds to $d_\text{in} = d_\text{out}=1$ and $d_\text{hidden}=2$: In this case, the norm is restricted to $||W^{(i)}||_F^2 = \text{max}(||\Sigma_{yx}||_F-\beta,0)$ and $W^{(1)} = W^{(2)}{}^T \Sigma_{yx}/ ||\Sigma_{yx}||_F$. Therefore, any matrix (vector) $W^{(2)}$ with the correct norm is valid and $W^{(1)}$ is fully determined by the choice of $W^{(2)}$: there is a continuous and closed path in parameter space with the same (minimal) loss. This path can be parametrized by $s\in [0,2\pi]$, labeling the different $W^{(2)}$ states, shown in Fig.~\ref{fig:OrderParameterSpaceIllustration}. In locally perpendicular directions, perturbations (parametrized by $s_\perp$) are massive.

\noindent \textbf{Arbitrary rank:} Following the same logic, a rank$=r$ macrostate $\boldsymbol{W}_r^*$ of the form
\begin{align*}
\boldsymbol{W}_{r}^* = \sum_{j=1}^r \sigma_j^2(\beta) \textcolor{blue}{\vec{l}_j} \cdot \textcolor{blue}{\vec{w}_j^T} = W^{(2)}W^{(1)} = \sum_{j=1}^r \sigma_j^2(\beta) \textcolor{blue}{\vec{l}_j} \textcolor{orange}{\vec{v}_j^T \vec{v}_j} \textcolor{blue}{\vec{w}_j^T}
\end{align*}
has degeneracies due to the choices of normalized \textcolor{orange}{$\vec{v}_j$}. Following, the logic from above, \textcolor{orange}{$\vec{v}_1$} can be chosen arbitrarily under the constraint $||\vec{v}_1||=1$. The next vector \textcolor{orange}{$\vec{v}_2$} additionally has to be orthogonal to \textcolor{orange}{$\vec{v}_1$}: this reduces the available subspace dimension by one. Continuing this argument, the resulting order parameter space can be thought of as $S^{d_\text{hidden}-1} \times S^{d_\text{hidden}-2} \times \dots \times S^{d_\text{hidden}-r}$ with dimension $(d_\text{hidden}-1) + \dots + (d_\text{hidden}-r) =r(d_\text{hidden}-r)+\frac12 r(r-1)$. 
Therefore, the cascade of rank transitions starting from large $\beta$ is accompanied by an increase of degeneracies of the minimum with the most accurate regime featuring the largest order parameter space.\\

\noindent 

\noindent Summarizing, the regularized loss landscape features one global (degenerate) minimum and many (degenerate) saddle points. The cut through the loss landscape, as shown in Fig.~\ref{fig:RegularizedLandscape}(b), therefore depicts those saddles as well as the global minimum. The degeneracy of the minimum depends on $\beta$ which determines $r=r_\text{max}(\beta)$ and is reflected in the Hessian spectrum, Fig.~\ref{fig:HessianSpectrum1hidden}(a). For $\beta > \beta_\text{onset}$ such that $r_\text{max}(\beta)=0$ all directions are massive: the macrostate is not degenerate and the minimum is at the origin. The second example corresponds to a lower regularization strength with $r_\text{max}(\beta)=3$ featuring a degenerate minimum: the spectrum contains three zero eigenvalues - in accordance with the counting above. Another nine eigenvalues $\lambda$ with $0 < \lambda \le 4\beta$ belong to the $\beta$ classes mentioned in Tab.~\ref{tab:HessianSpectrum1hidden}. The remaining eigenvalues belong to the massive data dependent sector.

\subsubsection{Dynamical Aspects of the Regularized Setting \& Critical Slowing Down}
For 1 hidden layer networks, exact solutions of the full gradient flow \emph{dynamics} are possible under some assumption adapting \cite{Fukumizu1998,Braun2022,Domine2025} (see also \cite{Lindsey2025} for a discussion of general aspects of the dynamics under regularization). For times $t\gg\beta^{-1}$, the dynamics takes place in the subspace of $0$-balanced solutions. In this subspace, the evolution of \textbf{macro-quantities} can be solved exactly for (e.g.) $d_\text{in} = d_\text{out}=d_\text{hidden}$ and $\Sigma_{xx}=\mathbb{1}$ (see App.~\ref{App:AlgebraicScaling}, adapting \cite{Fukumizu1998,Braun2022,Domine2025}) with $Q := \begin{pmatrix}W^{(1)}{}^T & W^{(2)} \end{pmatrix}^T$:

\begin{align}
&\begin{pmatrix} \textcolor{blue_R}{\mathfrak{W}_R}(t) & \boldsymbol{W}^T(t) \\  \boldsymbol{W}(t) & \textcolor{blue_L}{\mathfrak{W}_L}(t) \end{pmatrix} = QQ^T(t)\\
&=e^{Ft} Q_0 \left(\mathbb{1} + \frac12 Q_0^T \left(e^{Ft} F^{-1}e^{Ft}-F^{-1}\right)Q_0 \right)^{-1} Q_0^T e^{Ft},\nonumber\\
& F := \begin{pmatrix} -\beta \mathbb{1} & \Sigma_{yx}^T\\ \Sigma_{yx} & -\beta \mathbb{1} \end{pmatrix},  \quad D_F = \begin{pmatrix} S_{xy}-\beta\mathbb{1} & 0\\ 0& -S_{xy}-\beta \mathbb{1}  \end{pmatrix}.
\end{align}
The stationary state is determined by the positive eigenvalues of $F=F(\beta)$. The qualitative change in behaviour by increasing $\beta$ is reflected in the decrease of the number of positive eigenvalues of $F$. Right at the second order transition, zero eigenvalues emerge in $F$ that allow for an algebraic decay for long times (adapting the treatment in \cite{Braun2022}, see Appendix~\ref{App:AlgebraicScaling}). A particularly simple case arises once the initial state already shares the singular directions with the stationary state: in this case, the dynamics is reduced to the singular values alone and the 'critical' singular value evolves as:
\begin{align}
\sigma^c_j(t) = \sqrt{\frac{\sigma_j^2(0)}{1+4t\cdot \sigma_j^2(0)}} \quad \rightarrow \quad \sim \frac{1}{\sqrt{t}},
\end{align}
indicating the critical slowing down typical for second order phase transitions with an exponent\footnote{The exponent is denoted as $\beta/(\nu z)$ in the jargon of critical exponents.} $1/2$ as expected from the critical exponent of the order parameter.

\subsection{Cascade of First Order Transitions: $n-1$-hidden Layer Networks}

\begin{figure*}
\centering
    \begin{subfigure}[b]{0.65\textwidth}
    \centering
    \includegraphics[width=\textwidth]{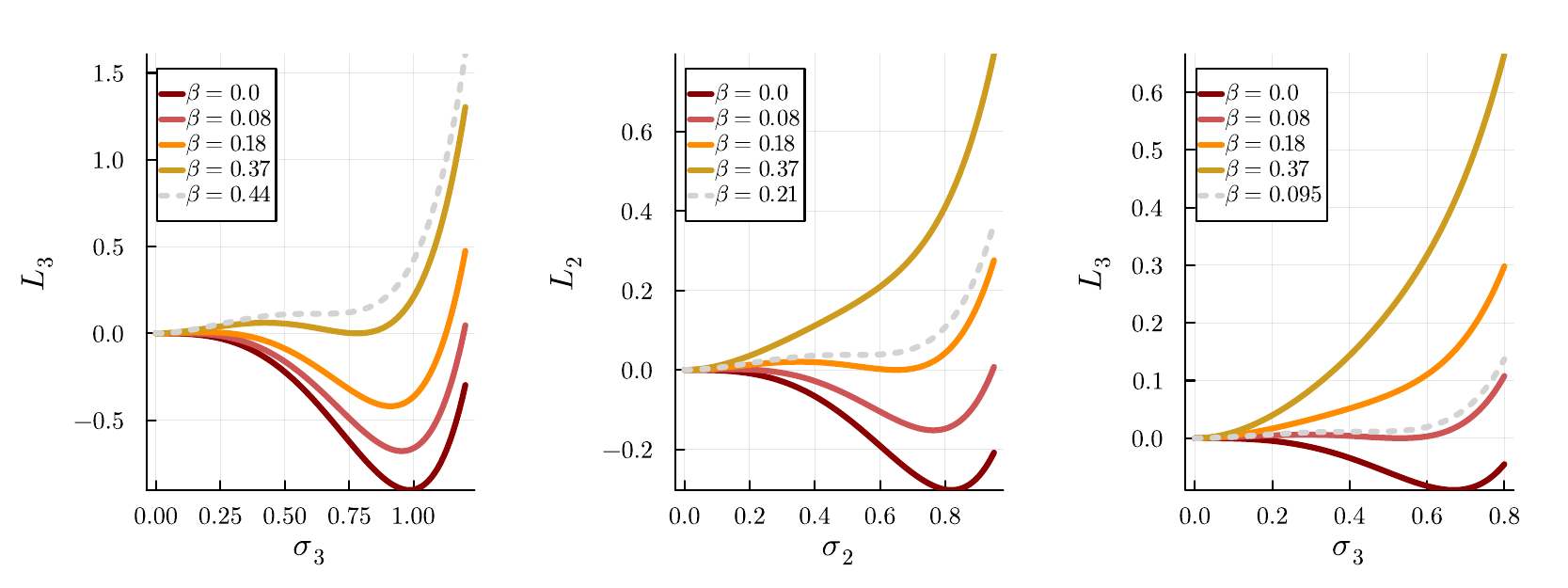}
    \caption{}
    \end{subfigure}
\quad
    \begin{subfigure}[b]{0.3\textwidth}
    \centering
    \includegraphics[width=1.2\textwidth]{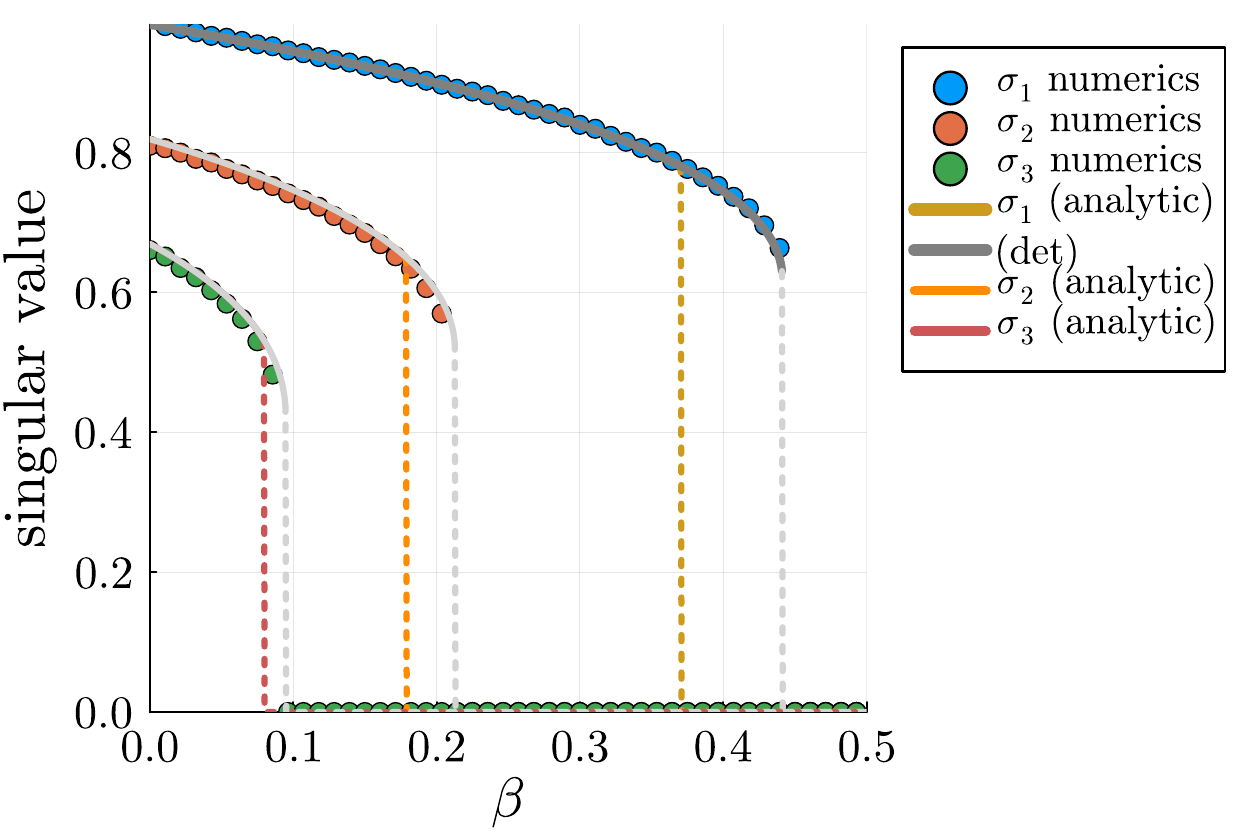}
    \caption{}
    \end{subfigure}
\caption{\textbf{(a)} Effective loss potential of the three singular values $\sigma_j$ for a 2 hidden layer network with $d_\text{in}=d_\text{out}=3$ with $\eta_1=0.95,\eta_2=0.55,\eta_3=0.3$; \textbf{(b)} onset of the singular values for (i) numerical simulation based on (reversed) annealing and (ii) analytical calculation. The gray dotted lines correspond to the $\beta$-value, where a bifurcation takes place; the colored lines indicate the $\beta$-values of level crossings.}
\label{fig:ExamplesEffectiveDescription2hidden}
\end{figure*}

For a network with $n>3$ layers ($>2$ hidden layers), the singular values of $\Sigma_{yx}$ are learned as well in descending order, accompanied by first order transitions. This first-order scenario is more generic: the error-term (here: MSE) features higher-order couplings of the $W^{(j)}$ parameters, whereas the $L_2$-regularizer results in quadratic terms in the loss. Under the alignment assumption \eqref{eq:compatibilitycondition} of the data distribution, each order parameter $\sigma_j$ is effectively described by an order $2n$ polynomial:
\begin{align}
L_j(\sigma_j) = \variancevalues{j} \sigma_j^{2n} -2\eta_j \sigma_j^{n} + n\beta \sigma_j^2.
\end{align}
This is the typical form of a Landau free energy for \emph{first order} transitions, featuring multiple (local) minima. For 2 hidden layers and $d_\text{in} = d_\text{out}=3$ the effective losses for the three singular values are shown in Fig.~\ref{fig:ExamplesEffectiveDescription2hidden}. The trivial state $\sigma_j=0$ is always a minimum. For sufficiently low $\beta <\beta_j^{(det)}$, a non-trivial minimum emerges with ($n \ge 3$ layers):
\begin{align}
\beta^{(det)}_{j,n} = \frac{\eta_j^{2(1-1/n)}}{\variancevalues{j}^{1-2/n}} \cdot \frac{n \cdot (n-2)^{1-2/n}}{(2n-2)^{2(1-1/n)}}  \stackrel{n \gg1}{\rightarrow} \sim \frac14 \frac{\eta_j^2}{\variancevalues{j}}.
\label{eq:DeterministicTransitionBeta_n>2}
\end{align}
This means that for too large regularization, this singular direction of the data distribution cannot be learned. From the perspective of $\mathcal{L}(\boldsymbol{\sigma}) = \sum_{j=1}^{r_\text{max}} L_j$, decreasing $\beta$ results in a number of bifurcations as shown in Fig.~\ref{fig:IntroOverview}, which we will discuss in more detail below.

\noindent Reducing $\beta$ even further, the (non-trivial) local minimum in the effective picture turns into the (global) minimum $L_j(\sigma_j^*) = L_j(0)$ with a critical regularization strength of:
\begin{align}
\beta^*_{j,n} = \frac{\eta_j^{2-2/n}}{\variancevalues{j}^{1-2/n}} \cdot \frac{(n-2)^{1-2/n}}{(n-1)^{2-2/n}} \quad \quad \stackrel{n \gg1}{\rightarrow} \sim \frac{\eta_j^2}{\variancevalues{j}} n^{-1}.
\end{align}
At the level of the full loss, this is indeed a level crossing (see also Fig.~\ref{fig:IntroOverview}(b)), indicating a phase transition point. However, this `static' picture needs to be paired with a dynamical picture. In case of deterministic gradient flow (`zero temperature/noise'), a model that equilibrates into a \emph{local} minimum will be trapped there. Therefore, the behaviour of individual trajectories depends strongly on the initial conditions. For finite noise instead (e.g., due to finite batch size, finite learning rate, added noise etc.), models in a local minimum \emph{can} escape via noise activation (see \cite{Ersoy2026,Wang2024}), but those events are \emph{rare} due to the necessity to overcome a potential barrier (see also \cite{Mori2022}), manifesting as hysteresis as observed in, e.g., \cite{Ziyin2022}. \\

\subsubsection{Form the effective Description to the full Loss Landscape}
In the following, we analyze how this effective description manifests at the level of the full regularized loss landscape for networks with multiple hidden layers. The different critical points of the full loss landscape are labeled by $\boldsymbol{\sigma}$ such that the $\sigma_j$'s are critical points (minima as well as maxima) of the effective potentials $L_j$, giving rise to minima and \emph{saddle points} at the level of the full loss landscape, see Fig.~\ref{fig:IntroOverview}(a) for an overview (for the weak regularization case see also \cite{Wang2024}).\\

\noindent \textbf{Structure of different rank solutions:} Due to similar arguments as in the $1$ hidden layer case, a macrosolution $\boldsymbol{W}_r^*$ of rank $r$ is compatible with many different microstates. For each hidden layer, the argumentation can be repeated, since the normalized \textcolor{orange}{'inner'} singular vectors (see \eqref{eq:SVDMacroStateGeneric}) can be chosen freely, resulting in different mircostates.
Therefore, the order parameter set $\boldsymbol{\sigma}$ with $r$ finite singular values is accompanied by an order parameter space of dimension $d_r = n \cdot r(d_\text{hidden}-r) + \frac{n}{2} r(r-1)$ for an $n$ layer network. \\

\noindent \textbf{Local Structure \& Hessian:} For more than $1$ hidden layer, the Hessian spectrum at a critical point can be cleanly organized into three qualitative regimes: (i) massive data-dependent eigenvalues, (ii) $\beta$-eigenvalues and (iii) zero eigenvalues, see Tab.~\ref{tab:HessianSpectrum2hidden} and Fig.~\ref{fig:HessianSpectrum1hidden}. Based on this, the Hessian at the critical points can be decomposed into the following form (for an in-depth derivation see App.~\ref{App:HessianSpectrum}):
\begin{align}
H_\text{crit} = \textcolor{orange}{H_0 + \textcolor{red!70!black}{H_{\text{null}}^{(\beta)}}+ H_{4\beta}} + \textcolor{blue}{H_{ex} + H_{r^2}}.
\label{eq:HessianGeneralDecomposition}
\end{align}
To better understand the implications, we pair up the micro- and macroperspective: The eigendirections for the first three parts correspond to perturbations of the microsolutions that leave the macrostate $\boldsymbol{W}_r$ invariant (at leading order). 
$\textcolor{orange}{H_0}$ collects directions associated with rotations of the \textcolor{orange}{inner} singular vectors (zero eigenvalue directions). 
In the absence of the $L_2$-regularization, all inner general linear ($GL$) transformations of the form \eqref{eq:GLSymmetry} would be symmetries of the loss, giving rise to different microstates and therefore giving rise to additional flat directions. For finite regularization strength $\beta$, the $GL$ symmetry is broken down to orthogonal symmetries (in particular rotations), see Sec.~\ref{Sec:Symmetries}. At the level of the Hessian, directions in parameter space that are associated with general linear transformations that are \emph{not} orthogonal transformation will become massive with eigenvalues $\lambda=4\beta$. These directions, corresponding to violating $0$-balancedness, form a basin of the $0$-balanced solutions, described by the exponential decay in \eqref{eq:DecayConservedQuantity}. The corresponding block in Hessian is denoted $\textcolor{orange}{H_{4\beta}}$.

\noindent The last two contributions in \eqref{eq:HessianGeneralDecomposition} instead feature massive data-dependent eigendirections and correspond to perturbations that do \emph{change} the macrostate as well. 
In case of $\textcolor{blue}{H_{r^2}}$, these massive directions can be associated with changing the singular values and de-alining the singular vectors (rotations). The eigendirections of $\textcolor{blue}{H_{ex}}$ are associated with 'exchanging' learned singular values/directions. Even though these perturbations change the macrostate, the rank is unchanged. In line with this, the number of massive eigenvalues corresponds exactly to the dimension of the manifold of rank-$r$ matrices in the space of $d_\text{out} \times d_\text{in}$ macromatrices $\boldsymbol{W}$. A special role is played by $\textcolor{red!70!black}{H_{\text{null}}^{(\beta)}}$: those are perturbations in the null sector of the microsolutions $W^{(i)}$ that do increase the rank of the microsolutions. By violating the $0$-balanced condition, the corresponding eigenvalue scales with $\beta$. However, the \emph{macrostate}, being a product of weight matrices, is unchanged by such perturbations of individual $W^{(i)}$. \\

\noindent In summary, the macro as well as micro critical points are again labeled by the order parameter set $\boldsymbol{\sigma}$: all combinations of $\sigma_j$ that are critical points of $L_j$ give rise to (micro) critical points. The set of finite $\sigma_j$'s determines the rank $r$ and is specified by the index set $\mathcal{S}$. However, only if the those $r$ finite singular values are associated with the $r$ \emph{largest} singular values of $\Sigma_{yx}$ (meaning: $\mathcal{S}=[[1,r]]$), the micro critical point is a (local) minimum of the regularized loss landscape. Examples of Hessian spectra at those minima are shown in Fig.~\ref{fig:HessianSpectrum1hidden} for the (i) trivial phase, (ii) $r=1$ and (iii) full rank phase, where all flat directions are associated with the different degeneracies of those solutions. If $\boldsymbol{\sigma}$ instead features an unordered set of locally minimal singular values ($\mathcal{S} \neq [[1,r]]$), the critical point is a \emph{strict saddle}: there is at least one negative eigenvalue in $\textcolor{blue}{H_{ex}}$ (corresponding to the exchange of a suboptimal singular value with a larger one) due to $\eta_b >\eta_a$ (see Tab.~\ref{tab:HessianSpectrum2hidden}). Note that these saddle points are again degenerate with the same degeneracy as a rank$=r$ minimum. All other combinations of critical $\sigma_j$'s give rise to strict saddles as well. This is qualitatively summarized in Fig.~\ref{fig:IntroOverview}(a,c).

\begin{table*}
\centering
\begin{tabular}{l| l| l| l}
\textbf{sector} & \textbf{eigenvalue} & \textbf{direction} & \textbf{numerosity}  \\ \hline
\rowcolor{blue!30!white}\multicolumn{4}{l}{\textbf{ \textsf{massive data-dependent}}} \\ \hline
$P_\mathcal{S}$, diagonal & $6\sigma_a^4-2\beta$, $a\in \mathcal{S}$& $\delta W^{(1)}, \delta W^{(2)}, \delta W^{(3)}\neq0$ & $r$ \\
$P_\mathcal{S}$, off-diagonal & $\lambda^{(ab)}_+$, $\lambda^{(ab)}_-$, $a,b\in \mathcal{S}$ & $\delta W^{(1)}, \delta W^{(2)}, \delta W^{(3)}\neq0$ & $r(r-1)$ \\
exchange & $2\sigma_a(\eta_a\pm \eta_b)$, $a \in \mathcal{S}, b\notin \mathcal{S}$ & $\delta W^{(1)},\delta W^{(3)} \neq0$ & $2r(d_\text{min}-r)$ \\
 & $2\sigma_a\eta_a$, $a \in \mathcal{S}, b>d_\text{out}$ & $\delta W_1\neq0$ & $r(d_\text{in}-d_\text{out})$ \\ \hline
\rowcolor{lightgray}\multicolumn{4}{l}{\textbf{ \textsf{null space sector}}} \\ \hline
$P_\mathcal{S}^\perp$ & $2\beta$ & arbitrary in null space & $(d_\text{in}+d_\text{out}+d_\text{hidden}-3r)(d_\text{hidden}-r)$ \\ \hline
 \rowcolor{lightgray}\multicolumn{4}{l}{\textbf{\textsf{$\beta$-sector (explicitly broken symmetry)}}} \\ \hline
$P_\mathcal{S}$ & $4\beta$ & $\delta W^{(i)},\delta W^{(i+1)} \neq0$ & $2\frac12 r(r+1)$ \\
mixing & $4\beta$ & $\delta W^{(i)},\delta W^{(i+1)} \neq0$ & $2\frac12 r(r+1)$ \\ \hline
\rowcolor{lightgray}\multicolumn{4}{l}{\textbf{\textsf{zero eigenvalues (from symmetries)}}} \\ \hline
$P_\mathcal{S}$ & $0$ &  $\delta W^{(i)},\delta W^{(i+1)} \neq0$ & $2\frac12 r(r-1)$ \\
mixing & $0$ &  $\delta W^{(i)},\delta W^{(i+1)} \neq0$ &  $2r(d_\text{hidden}-r)$ \\
\end{tabular}
\caption{Overview of the Hessian spectrum at a critical point of rank $r$ for 2 hidden layer networks (can be directly generalized to arbitrary depth, see Tab.~\ref{tab:HessianSpectrumgeneral}) for $\Sigma_{xx}=\mathbb{1}$. $\delta W^{(i)}$ are the Hessian matrix-valued eigendirections, collected in vectorized form in $\delta \vec{\theta}$ with $H\delta \vec{\theta} = \lambda \delta \vec{\theta}$. $P_\mathcal{S}$ refers to the projection of the $W^{(i)}$ onto the $r^2$-dim. subspace of finite singular values. $\lambda_\pm$ are defined in \eqref{eq:DefLambdaplusminus2hidden}.}
\label{tab:HessianSpectrum2hidden}
\end{table*}

\subsubsection{Macro-Dynamics Picture} 
This micro-picture at the level of the parameters can be complemented by a \emph{macro-dynamics picture} of $\bar{\boldsymbol{W}}:= (\textcolor{blue_L}{\mathfrak{W}_L}, \boldsymbol{W}, \textcolor{blue_R}{\mathfrak{W}_R})$ based on \eqref{eq:MacroDynamics}. The behaviour of the macrostate close to a critical point $\bar{\boldsymbol{W}}_r^*$ is determined by the Jacobian $\mathcal{J}$ of the macro dynamics - the Jacobian for the unregularized case was discussed in \cite{Menon2026}. Close to a stationary state $\bar{\boldsymbol{W}} = \bar{\boldsymbol{W}}_r^* +\delta\bar{\boldsymbol{W}}$, we expect that (i) changing the singular values and (ii) de-aligning will be massive/stable (data-dependent) perturbations $\delta \bar{\boldsymbol{W}}$ in the macro (function) space. Therefore, massive perturbations $\delta \bar{\boldsymbol{W}}$ should correspond to infinitesimal rotations in the complement of the null spaces of the matrices in $\bar{\boldsymbol{W}}_r^*$, which can be expressed as ($\epsilon \ll 1$; and assuming $\Sigma_{xx}=\mathbb{1}$), see App.~\ref{App:Jacobian2Hidden} for further details:

\begin{align}
&\bar{\boldsymbol{W}}_r^*  \rightarrow \quad  (\textcolor{black}{L_\epsilon}\textcolor{blue_L}{\mathfrak{W}_L} \textcolor{black}{L_\epsilon^T},\textcolor{black}{L_\epsilon} \boldsymbol{W} \textcolor{black}{R_\epsilon^T},\textcolor{black}{R_\epsilon} \textcolor{blue_R}{\mathfrak{W}_R} \textcolor{black}{R_\epsilon^T})^T \approx \bar{\boldsymbol{W}}_r^* + \textcolor{gray}{\epsilon} \delta \bar{\boldsymbol{W}}, \nonumber\\
&\delta \bar{\boldsymbol{W}} = \begin{cases}
 (\sigma_i^2-\sigma_j^2) \textcolor{blue_L}{(\vec{l}_i \vec{l}_j^T + \vec{l}_j \vec{l}_i^T)} \\
(\sigma_i^n -\sigma_j^n)  \textcolor{black}{(\vec{l}_i \vec{r}_j^T + \vec{r}_j\vec{l}_i^T)} \\
(\sigma_i^2 -\sigma_j^2) \textcolor{blue_R}{(\vec{r}_i \vec{r}_j^T+ \vec{r}_j \vec{r}_i^T)} \end{cases},
\label{eq:MassiveDirectionsMacroDynamics}
\end{align}

with infinitesimal orthogonal rotations $R_\epsilon \approx \mathbb{1} + \epsilon (\vec{r}_i \vec{r}_j^T - \vec{r}_j \vec{r}_i^T)$, $L_\epsilon \approx \mathbb{1} + \epsilon (\vec{l}_i \vec{l}_j^T - \vec{l}_j \vec{l}_i^T)$. As we show in the App.~\ref{App:Jacobian2Hidden}, eigendirections $\delta \bar{\boldsymbol{W}}$ of the Jacobian at the critical point in the finite rank sector are indeed given by the blue-marked terms in \eqref{eq:MassiveDirectionsMacroDynamics}. The corresponding eigenvalues are the same as the eigenvalues of micro-$\textcolor{blue}{H_{r^2}}$ belonging to (infinitesimal) rotations and shifts of the singular values (compatible with the $0$-balanced condition), see Tab.~\ref{tab:HessianSpectrum2hidden}. Therefore, the massive data-dependent eigenvalues and eigendirections - encoding information about the already learned directions of the data distribution - are encoded in the micro as well as the macro picture. This is to be expected since those directions at the micro level indeed correspond to changes in the macrostate. \\

\section{Discussion}

In the following, we compare the transition mechanism with the \emph{dynamical} saddle-to-saddle mechanism observed in the unregularized case and discuss some of the assumptions made during the analysis.\\

\subsection{Connection to the unregularized Case}
The guiding idea of our analysis is that the $L_2$-regularizer can be used to detect (geometric) features of the data distribution, being encoded in the (unregularized) loss landscape. Increasing $\beta$ restricts the resolution of those features to the $r_\text{max}(\beta)$ \emph{largest} singular values of $\Sigma_{yx}$ (under the alignment condition). 
Those singular values (and singular vectors) are directly linked to the unregularized case. To make the connection clear, we can interpret the $r_\text{max}(\beta)$ (local) minima (with different ranks $r$) in the regularized case\footnote{with $d_\text{hidden} \ge d_\text{in},d_\text{out}$} as solutions of the \emph{rank-restricted} and regularized learning problem: $\boldsymbol{W}^* = \arg \min_{\boldsymbol{W}, \text{rank } r' \le r} \mathcal{L}_\beta(\boldsymbol{W})$ for $r \le r_\text{max}(\beta)$. For $\beta \to 0$, those solutions become rank-restricted optima of the mean squared error itself (as discussed in Sec.~\ref{sec:UnregularizedSection}), which in turn are \emph{saddle points} of the unregularized loss landscape\footnote{Note that there is a difference between $\beta\to0$ and $\beta=0$, since in the former case the general linear symmetries are explicitly broken.} with rank $r<r_\text{max}(0)$. These saddle point solutions stand out in the set of all saddle points since they correspond to projections $P_\mathcal{S}\Sigma_{yx}\Sigma_{xx}^{-1}$ onto the $r$ \emph{largest} eigenvalues of $\tilde{\Sigma}=\Sigma_{yx}\Sigma_{xx}^{-1}\Sigma_{xy}$ (with $\mathcal{S}=[[1,r]]$), see Sec.~\ref{sec:UnregularizedSection} \cite{Bah2022,Saddles_DeepLinear}. Therefore, the $L_2$-regularization dynamically stabilizes those states that are still optimal under a rank restriction, underpinning that we are observing as many transitions as the rank of $\Sigma_{yx}$ (assuming $\Sigma_{xx}$ is full rank). For an overview of the connection of the critical points in the regularized and unregularized setting see Tab.~\ref{tab:ConnectioRegUnreg}.

\begin{table}[h]
\centering
\begin{tabular}{l|l}
\textbf{regularized} & \textbf{unregularized} $(\beta \to 0)$ \\ \toprule
critical points $0$-balanced & $0$-balanced (and tightend*) \\ \hline
global minimum with  & global minimum with \\ 
rank $r_\text{max}(\beta) \in [0,r_\text{max}]$ & rank $r_\text{max}(0)=r_\text{max}$ \\ \hline
minima with  & higher-order saddles \\
rank $r \le r_\text{max}(\beta)$ & \\
$\arg \min_{\boldsymbol{W}, \text{rank } r' \le r} \mathcal{L}_\beta(\boldsymbol{W})$ & $\arg \min_{\boldsymbol{W}, \text{rank } r' \le r} \mathcal{E}(\boldsymbol{W})$\\ \hline
strict saddle points & strict saddle points \\ 
\botrule
\end{tabular}
\caption{Connection between the critical points of the regularized and unregularized ($\beta\to0$) case (for the unregularized case see \cite{Baldi1989,Saddles_DeepLinear}, * 'tightend' defined in \cite{Saddles_DeepLinear}). The mentioned higher-order saddle corresponds to the projection $P_{\mathcal{S}}\Sigma_{yx}\Sigma_{xx}^{-1}$ for the index set $\mathcal{S}=[[1,r]]$.}
\label{tab:ConnectioRegUnreg}
\end{table}

\noindent \emph{Dynamic vs. static:} The transition scenario we are observing here can also be interpreted as a static version of the dynamical rich-feature learning regime in the absence of regularization \cite{Zhang2026}. Initializing the gradient flow close to the origin, multiple saddle points are met during the dynamics eventually reaching the minimum \cite{Jacot2022}. In this case, \emph{early stopping} can allow to work with models in the vicinity of a saddle (which are lower rank/coarser solutions) (see also \cite{Advani2020}). This is reminiscent of (our) trapping a model in a local minimum via a sufficiently large regularization strength. Both cases are tight to the underlying saddle point and `basin' structure of the unregularized, purely data-dependent loss landscape. \\

\subsection{Beyond the Alignment Condition and $N_\text{data} \to \infty$}

So far, we have discussed special cases where the covariances $\Sigma_{xx}$ and $\Sigma_{yx}$ of the underlying data distribution can be simultaneously diagonalized \eqref{eq:compatibilitycondition}, allowing for an analytic treatment. Besides being a special case, as soon as we are working with a finite number of training samples $N_\text{data}$, this condition is violated at the level of the (stochastic) \emph{emperical} covariances $\Sigma^{(e)}$ \eqref{eq:EmpiricalCovariance}. However, we do expect the logic of those special cases to be valid for generic covariance matrices with earlier numerical investigations \cite{Ersoy2025} for non-aligned covariance matrices showed the same behaviour as discussed before. In the generic case, an exact analytic treatment is not possible to the best of our knowledge. The challenge is that an effective description in terms of independent $\sigma_j$'s is not possible anymore because the optimal left and right singular vectors of $\boldsymbol{W}$ become $\beta$-dependent as well. Nevertheless, the underlying structure of the task is the same: the $L_2$ regularization is expected to result in (local) minima in the regularized loss landscape\footnote{We cannot rule out that more saddle points will result in the non-aligned scenario, but we do not expect more minima to emerge.}, corresponding to models that are optimal under rank constraint with $r \le r_\text{max}(\beta)$. Those in turn are again connected to the saddle points in the unregularized landscape for $\beta \to 0$, where the singular vectors align with $\hat{P}_{\mathcal{S}}\Sigma_{yx}\Sigma_{xx}^{-1}$.\\

\noindent \textbf{Beyond Alignment:} The qualitative change due to non-aligned (co)variances in contrast to the previous discussion is the interdependence of the singular values and singular vectors. In particular, singular vectors of $\boldsymbol{W}^*$ will \emph{rotate} as a function of $\beta$. For 1 hidden layer, the limiting cases can be determined exactly: 
\begin{itemize}
\item the onset of learning is always determined by the largest singular value of $\Sigma_{yx}^{(e)}$ (in accordance with \cite{Ziyin2023}) with $\beta_\text{onset} = \eta_\text{max}$\footnote{This can be seen by solving \eqref{eq:MacroDynamics} for $\beta^*=\eta_1=\eta_\text{max}$.}. At the onset, the singular vectors of $\boldsymbol{W}^*$ are aligned with those of $\Sigma_{yx}^{(e)}$.
\item for $\beta\to0$ the singular vectors of a critical $\boldsymbol{W}^*$ (for a given $\mathcal{S}$) align with $P_{\mathcal{S}}\Sigma_{yx}^{(e)}\Sigma_{xx}^{(e)}{}^{-1}$. 
\end{itemize}
For weak perturbations around the exactly solvable cases, solutions can be obtained perturbatively. For 1 hidden layer networks, an example of the \emph{rotating} (right) singular vectors of $\boldsymbol{W}^*$ as a function of $\beta$ is shown in Fig.~\ref{fig:DiscussionNonAlignedCase}(b), based on directly numerically solving for stationary states \eqref{eq:MacroDynamics} together with a perturbative solution (see App.~\ref{App:Perturbation} for details). In Fig.~\ref{fig:DiscussionNonAlignedCase}(a), the (square root) singular value spectrum of $\boldsymbol{W}^*$ is shown as well (based on directly numerically solving for stationary states \eqref{eq:MacroDynamics} as well as using gradient descent dynamics).\\

\noindent \textbf{Beyond $N_\text{data} \to \infty$:} For different finite $N_\text{data}$ the average of the singular values over $10^3-10^5$ realizations for 1 hidden layer is shown in Fig.~\ref{fig:DiscussionNonAlignedCase}(c), indicating a smeared transition strength, as different finite sets of training data lead to different quantitative predictions of the transition points. An example for a non-aligned case for 2 hidden layers is shown in Fig.~\label{fig:DiscussionNonAlignedCase}(d), featuring the same phenomenology.

\begin{figure*}
\centering
    \begin{subfigure}[b]{0.23\textwidth}
    \centering
    \includegraphics[width=\textwidth]{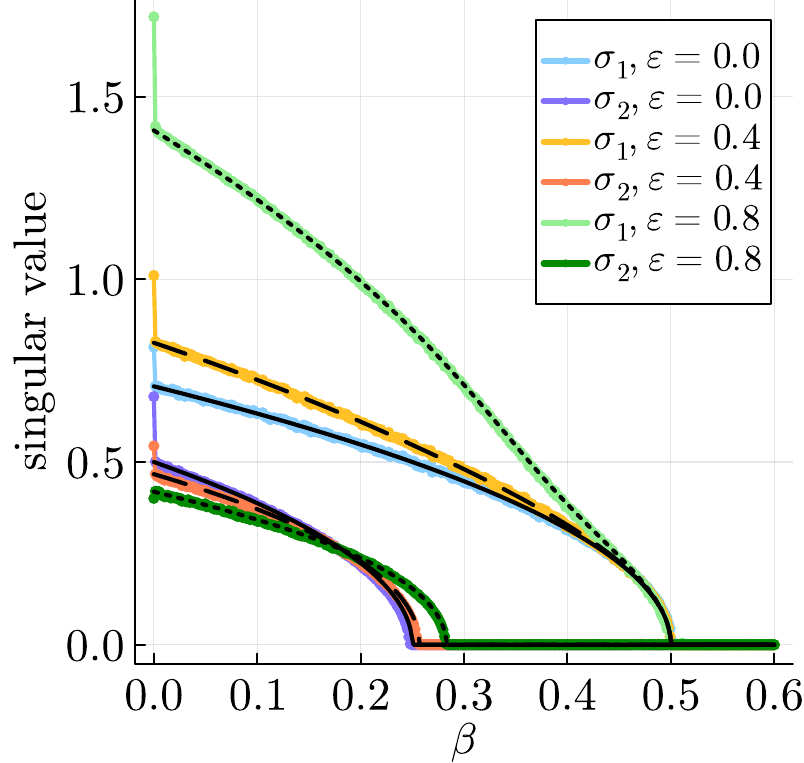}
    \caption{\label{fig:SVnonalgined}}
    \end{subfigure}
\quad
    \begin{subfigure}[b]{0.23\textwidth}
    \centering
    \includegraphics[width=\textwidth]{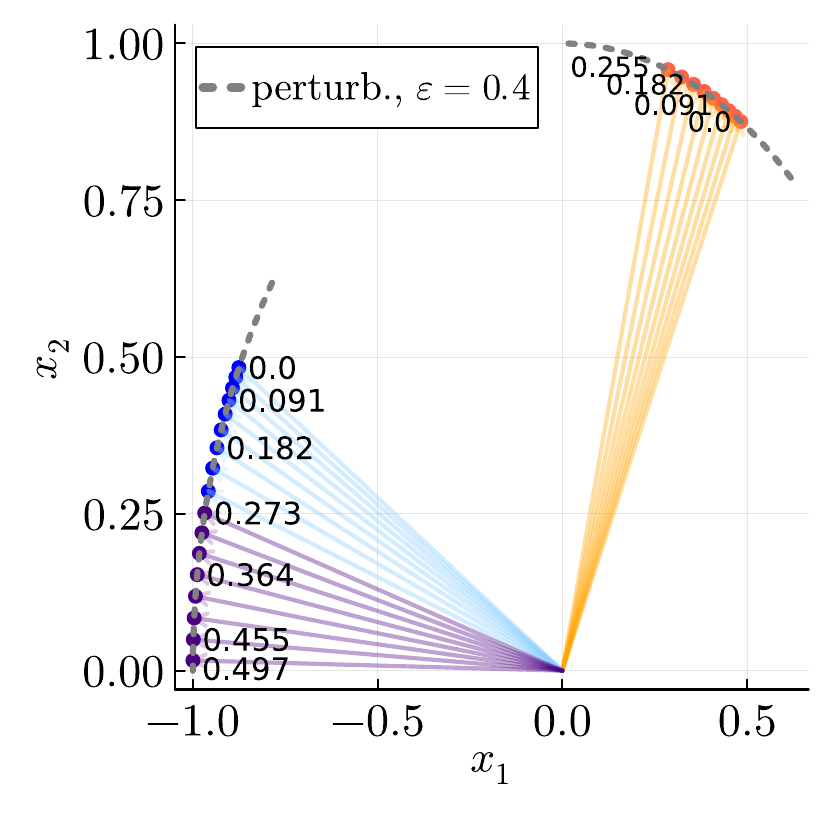}
    \caption{}
    \end{subfigure}
    \quad
    \begin{subfigure}[b]{0.23\textwidth}
    \centering
    \includegraphics[width=\textwidth]{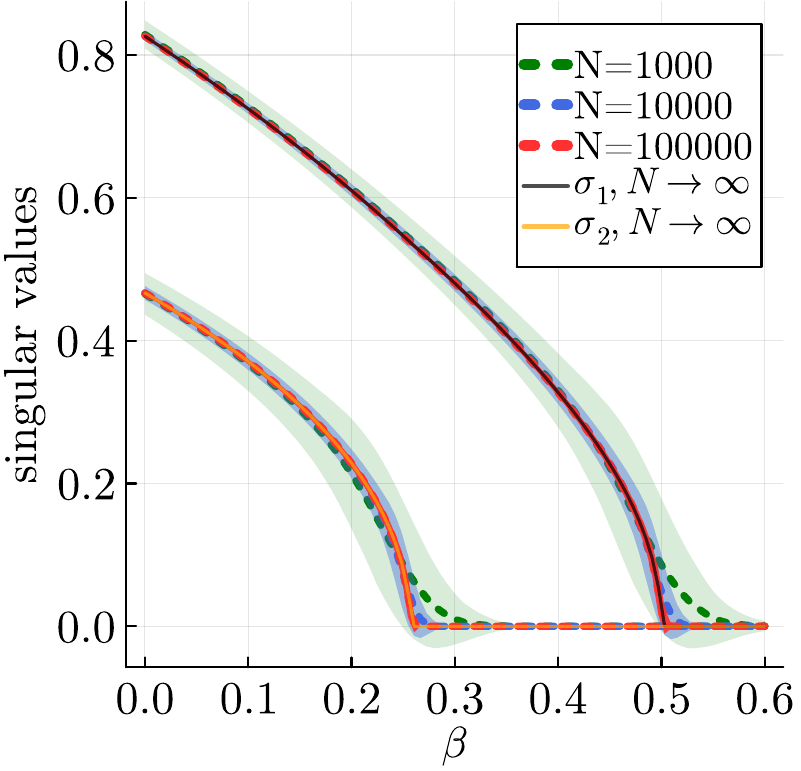}
    \caption{}
    \end{subfigure}
     \quad
    \begin{subfigure}[b]{0.23\textwidth}
    \centering
    \includegraphics[width=1.1\textwidth]{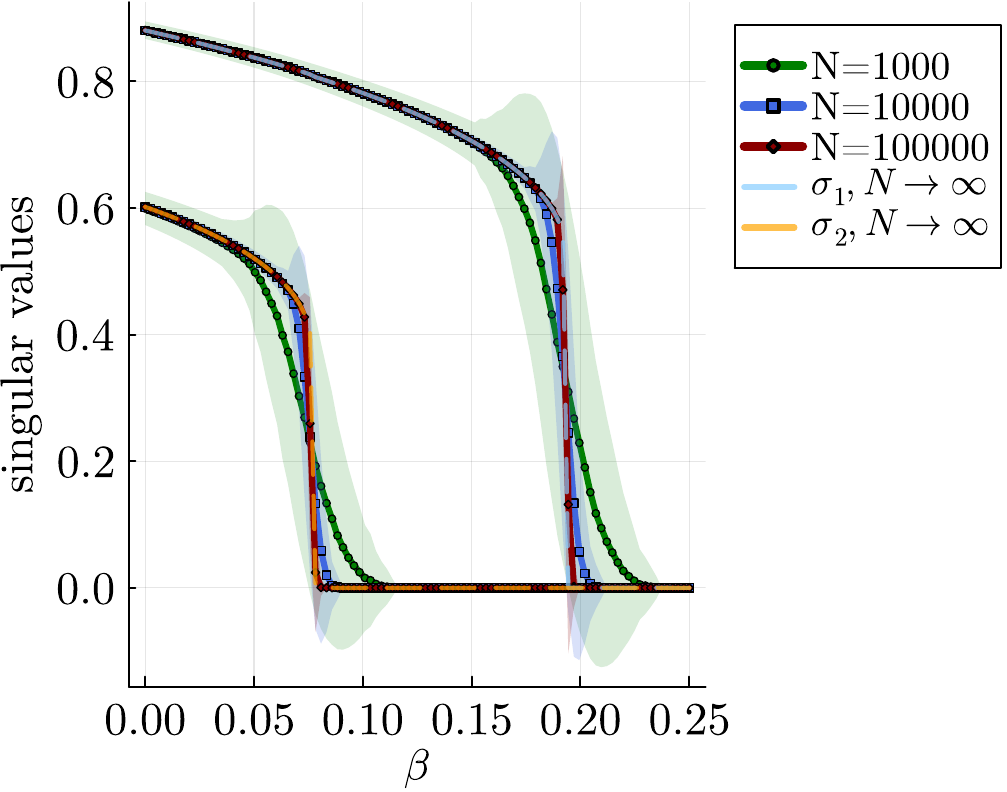}
    \caption{}
    \end{subfigure}
\caption{\textbf{(a)} Singular values and \textbf{(b)} right singular vectors $\vec{r}_i = (x_1,x_2)^T$ for non-aligned (co)variance matrices for a 1 hidden layer network with $d_\text{in} = d_\text{out}=2$, $\Sigma_{yx}=\text{diag}(0.5,0.25)$ and $\Sigma_{xx} = \mathbb{1} + \epsilon \begin{pmatrix} 0 & 1 \\ 1 & 0 \end{pmatrix}$. \textbf{(c,d)} examples of the dependence on sample size $N_\text{data}$ ($\epsilon=0.4$) with the dashed lines corresponding to the averaged singular values over different samples for 1 hidden and 2 hidden layer networks.}
\label{fig:DiscussionNonAlignedCase}
\end{figure*}

\subsection{Comparison to Non-Linear Cases}
\begin{figure*}
\centering
   \begin{subfigure}[b]{0.23\textwidth}
    \centering
    \includegraphics[width=1.1\textwidth]{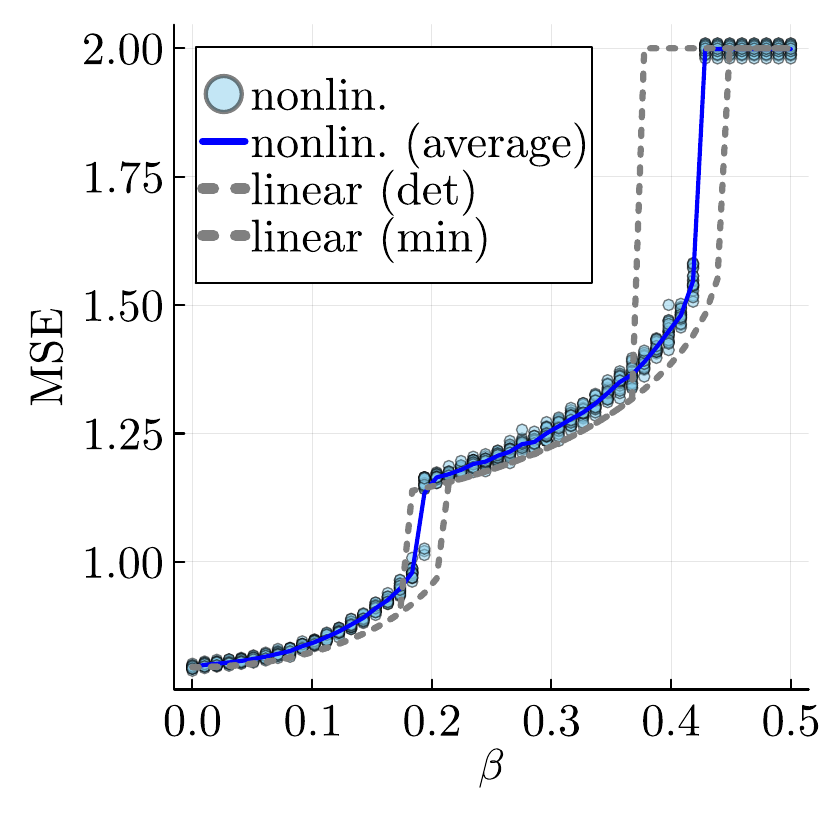}
    \caption{Gaussian data}
    \end{subfigure}
    \quad
    \begin{subfigure}[b]{0.23\textwidth}
    \centering
    \includegraphics[width=1.1\textwidth]{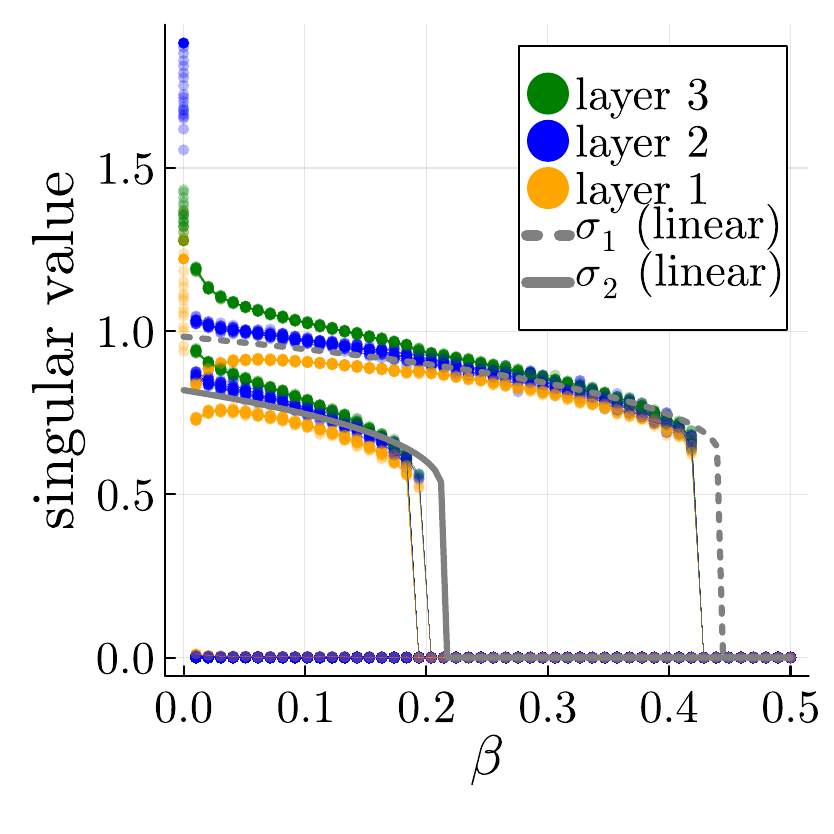}
    \caption{Gaussian data}
    \end{subfigure}
\quad 
    \begin{subfigure}[b]{0.23\textwidth}
    \centering
    \includegraphics[width=1.1\textwidth]{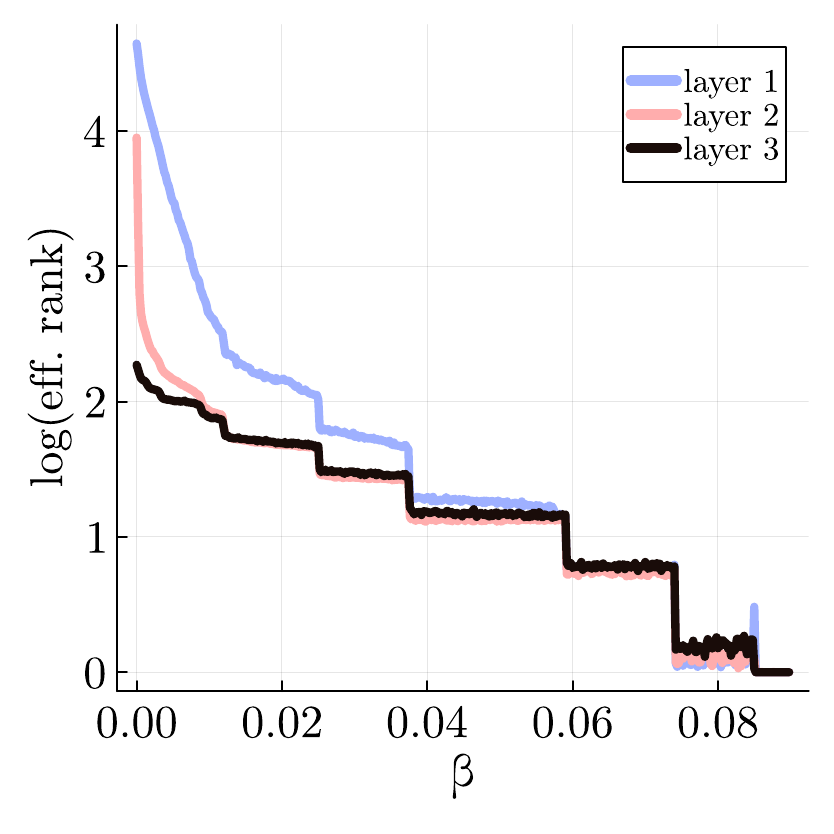}
    \caption{MNIST data}
    \end{subfigure}
\quad
    \begin{subfigure}[b]{0.23\textwidth}
    \centering
    \includegraphics[width=1.1\textwidth]{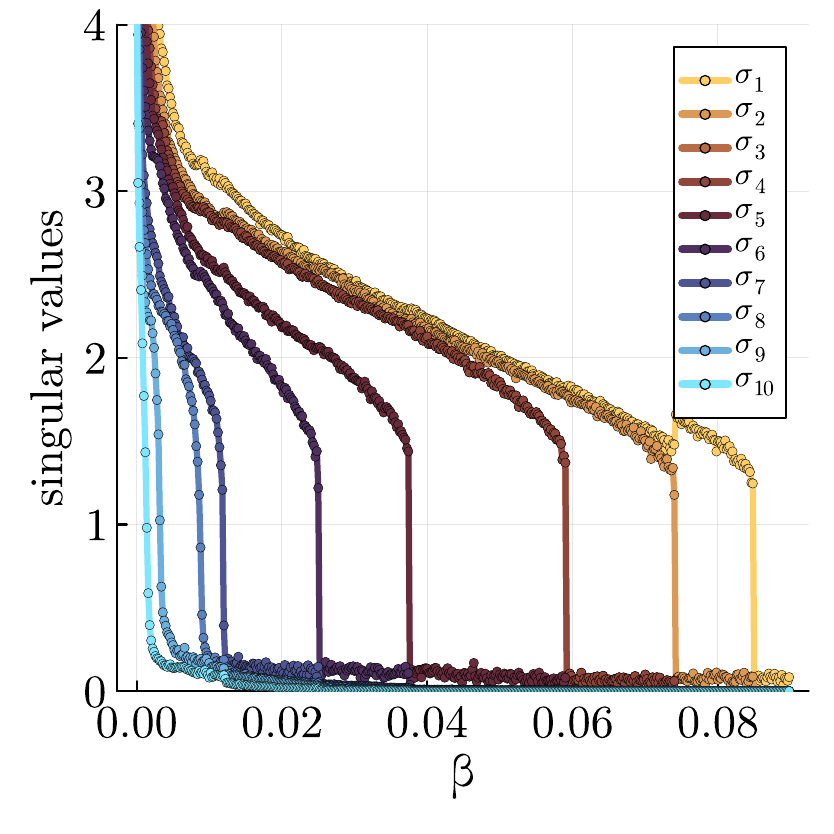}
    \caption{MNIST data}
    \end{subfigure}
    \caption{\textbf{(a,b)} MSE and singular values of a 2 hidden layer network with $\tanh$-activation function ($d_\text{hidden}=20$) for 20 trajectories; linear case as reference. \textbf{(c,d)} Effective rank and singular values of a 2 hidden layer network ($d_\text{hidden}=128$) with sigmoid activation function for MNIST (extracted from the experiments in \cite{Ersoy2025_2}).}
\label{fig:MNISTsvsandrank}
\end{figure*}

Remarkably, the $L_2$-phenomenology of the deep \emph{linear} networks bears strong similarity with models based on non-linear activation functions and also other cost functions (comparing to \cite{Ersoy2025,Ersoy2025_2} and \cite{Ziyin2022}). The similarity holds (i) at the level of the error/accuracy (see Fig.~\ref{fig:MNISTsvsandrank}) as well as [to some extend] (ii) the newly defined order parameter set itself. In case of $n$ layer deep linear models, the order parameter set $\boldsymbol{\sigma}$ is the set of the ($n$th roots of the) singular values of the macrostate or, directly linked, the singular values of the individual micro matrices $W^{(j)}$. For non-linear networks, we also observe numerically that the singular values of the weight matrices (at the individual layers) are indicators of the transition, see Fig.~\ref{fig:MNISTsvsandrank}. In the first case, the data set consists of Gaussian data with a non-linear $n=3$-network based on a $\tanh$-activation function: individual trajectories display the discontinuous behaviour as a function of $\beta$ with comparable onsets of singular values at each layer. Note as well the quantitative similarity to the linear case (dashed lines), which is to be expected as in the Gaussian data case linear activation functions are sufficient for an optimal description. \\

The second case is based on the model data from \cite{Ersoy2025_2} for annealing experiments (trained on the MNIST data set), also related to the topological trivialization transitions observed in \cite{Winter2025}. The singular values extracted from the models at the end of training show a similar behaviour: onset of different singular values corresponding to different transitions at the accuracy level. In this non-linear setting, in contrast to the linear case, there is no formal reason that enforces the rank or even the singular values of the weight matrices of \emph{different} layers to be comparable. Using the effective rank as defined in \cite{Olivier2007effectiverank} (see also \cite{Yunis2024rankminimization}), we however observe that \emph{for sufficiently large} regularization the effective rank is comparable at all layers with distinct jumps at the transitions. Therefore, these phase transitions can be roughly characterized as transitions between different effective ranks. This is in line with the \emph{linear origin theorem} in \cite{Ziyin2022}: at the onset of learning $\beta_\text{onset}$ (at the upper end of regularization strengths), non-linear networks are well described by linear ones (where the $0$-balanced condition enforces the same rank at all layers). However, for lower $\beta$ (here around $<0.02$) the effective ranks deviate more strongly and the singular values as such do not display sudden onsets. We interpret this regime of low regularization strength as the \emph{effectively} non-linear regime, where a description in terms of a linear model qualitatively fails.

\section{Summary \& Outlook}

One of the central pillars for the development of a scientific theory of deep learning are tractable minimal models, in the spirit of the Ising model or the harmonic oscillator in statistical physics \cite{Simon2026}. Those models provide a unique platform to establish rigorous connections between statistical physics and the theory of deep learning, complementing more abstract frameworks like singular learning theory \cite{Watanabe2025}.

We introduced such a minimal model in the form of a deep linear network with regularization, arbitrary depth as well as arbitrary input and output dimensions, \emph{repurposing the regularization strength as an external field strength} as put forward in \cite{Ziyin2023,Ziyin2024discretesymmetries,Ziyin2025parametersymmetries}. This largely analytically tractable toy model complements existing research in this field e.g. \cite{Wang2024,Ziyin2022,Ziyin2023,Ziyin2024discretesymmetries,Lindsey2025}, and in particular the phenomenologically observed cascade of transitions in \cite{Ersoy2025,Ersoy2025_2}, linked to the model complexity. In the toy model presented here, multiple successive transition points, Hessian spectra and an effective description based on an order parameter \emph{set} have been exactly determined. Tuning the regularization strength reveals hierarchically organized phases and phase transitions in the form of successively learned singular directions of the covariance matrix, accompanied by jumps in the ranks (in line with \cite{Ziyin2024discretesymmetries}). In terms of statistical physics, different phases can be classified by a set of order parameters (singular values of the learned macrostate $\boldsymbol{W}$), including as a special case the order parameter introduced in \cite{Ziyin2023} (see also \cite{EFTcollectiveDL2024} for \emph{coupled} neural networks). In terms of learning, the order parameters correspond to learned features of the data distribution. At the microscopic level, those phases and phase transitions are related to the Hessian spectrum of the loss landscape, providing a direct link between the macroscopic and microscopic description.

\noindent \textbf{At the microlevel}, to be learned features of the data distribution in deep linear networks (in the form of singular directions of the (co)variance matrices) are geometrically encoded in the form of saddle points at the level of the loss $\mathcal{L}$ \cite{Baldi1989,Fukumizu1998,Kawaguchi2016,Saddles_DeepLinear,Wendin2025} in the unregularized case. The $L_2$ regularizer turns certain saddle points (corresponding to rank-restricted optima) into stable (local) minima, which are therefore more easily accessible at the end of training. Geometrically, the number of massive data-dependent modes of the Hessian (at those minima) is directly linked to the rank $r$ of the learned model and the Hessian consists of four blocks: (i) data-dependent massive block; (ii) $\beta$-block associated with the explicitly broken $GL$-symmetry; (iii) zero-block due to symmetries and (iv) null-block corresponding to perturbations in the null space of the (mirco) solutions.

\noindent \textbf{At the level of learning}, a finite $L_2$ regularization strength $\beta$ restricts the learnable singular directions of the data (co)variances $\Sigma_{yx}$ to the largest $r_\text{max}(\beta)$ ones (given the alignment condition \eqref{eq:compatibilitycondition}), resulting in a cascade of rank transitions by tuning $\beta$. At vanishing regularization, the optimal macrostate $\boldsymbol{W}^*$ has rank $r_\text{max}(0) = \text{rank}(\Sigma_{yx})$, which cascades down to $r_\text{max}(\beta >\beta_\text{onset})=0$ beyond the onset of learning (see also \cite{Ziyin2023}). In terms of glass physics, this was referred to as a topological trivialization transition \cite{Winter2025}.

\noindent \textbf{At the macroscopic (statistical) level}, our arguments are valid at zero temperature/zero noise, as for zero noise, phase transitions in the strict sense do not require a thermodynamic limit in the system size \cite{Goldenfeld1992}. The numerical experiments that support our theory, are based on low noise gradient descent dynamics with finite learning rates and batch sizes, indicating a certain robustness against noise. It is a question for future investigation whether these observed transitions can be true thermodynamic phase transitions (in the limit of large model sizes as investigated in, e.g., \cite{Li2021DLNbackpropagating} at the level of the Gibbs distribution). \\

\noindent Putting our observations in perspective, two analogies can be drawn:
\begin{itemize}
\item The structure of the static transitions in the linear case, where successively singular vectors of the (co)variance [from large to small] are learned, is reminiscent of the structural transitions in the information bottleneck (IB) approach for Gaussian variables \cite{Chechik2003}. Here, the tuning parameter (also denoted $\beta$) is the tradeoff between compression and relevance. In analogy, the regularization strength acts as an effective compression parameter\footnote{However, in the IB setting, the main geometric object is the matrix $\Sigma_{x|y} \Sigma_{xx}^{-1} = \mathbb{1}- \Sigma_{xy} \Sigma_{yy}^{-1} \Sigma_{yx} \Sigma_{xx}^{-1}\Sigma_{xy}$ in contrast to (only) $\Sigma_{yx}$ and $\Sigma_{xx}$ in the linear setting.
} or bottleneck.

\item The regularization/external field induced \emph{static} transitions can be interpreted as a \emph{static} analog of the saddle-to-saddle \emph{dynamics} in the unregularized setting as discussed in \cite{Jacot2022,Domine2025} and generalized in \cite{Zhang2026,Kunin2026}. Similarly, cascades of (dynamical) phase transitions (corresponding the progressive learning of features) have been observed \cite{Bachtis2025cascade} in the context of restricted Boltzmann machines. 
\end{itemize}

We have primarily focused on minimal models that can be analyzed analytically to establish notions like order parameters, connecting to the theoretical considerations in \cite{Ziyin2024discretesymmetries}. Broadening the scope of our investigation, we provided numerical evidence that phase transitions \emph{as (effective) rank transitions} can also be observed for non-linear activation functions to some extend. These findings fit to observations of rank collapse and tighter alignment of singular vectors in \cite{Yunis2024rankminimization} (see also \cite{Kuzborskij2025lowrankbiasweightdecay}). The limitation of this description for non-linear networks is expected since the presence of different symmetries in those networks suggests the definition of different order parameters. \\

\noindent Extending the phase transition and statistical physics (analogy), the observed cascade of phase transitions can also be rephrased as symmetry-breaking transitions. In follow-up work, we will embed the observed phenomenology and effective description into an extended symmetry framework, related to the ideas put forward in \cite{Ziyin2025parametersymmetries}. \\

\noindent \textit{Acknowledgements} -- We thank Andrés Fernando Cardozo Licha for providing us with the singular values of the experiments in \cite{Ersoy2025_2}. We also made use of Julia packages \cite{Pal2024nonlinearsolve}.

\bibliography{Bib_Linear_Networks}

\begin{widetext}
\addcontentsline{toc}{section}{\textbf{Appendix}}
\appendix

\section{Loss Landscape, Equations of Motion and Critical Points}

For a deep forward network with $n$ layers ($n-1$ hidden layers), the equations of motion for gradient flow dynamics are $-\frac{d}{dt}W^{(i)} = \nabla_{W^{(i)}} \mathcal{L}$, which read:

\begin{align}
\begin{aligned}
&-\frac{d}{dt}W^{(1)} = 2(W^{(n)} \dots W^{(2)})^T \underbrace{(\boldsymbol{W}\Sigma_{xx} - \Sigma_{yx})}_{=:\hat{N}_{\boldsymbol{W}}} + 2\beta W^{(1)}, \\
&-\frac{d}{dt}W^{(i)} = 2(W^{(n)} \dots W^{(i+1)})^T \hat{N}_{\boldsymbol{W}} (W^{(i-1)}\dots W^{(1)})^T + 2\beta W^{(i)}, \\
&-\frac{d}{dt}W^{(n)} = 2\hat{N}_{\boldsymbol{W}} (W^{(n-1)}\dots W^{(1)})^T+ 2\beta W^{(n)}.
\end{aligned}
\end{align}
The equations of motions can be simplified \emph{in the $0$-balanced sector}: the macro objects are $\textcolor{blue_R}{\mathfrak{W}_R} = (\boldsymbol{W}^T \boldsymbol{W})^{\frac{1}{n}}$, $\textcolor{blue_L}{\mathfrak{W}_L}= (\boldsymbol{W} \boldsymbol{W}^T)^{\frac{1}{n}}$ and $\boldsymbol{W} = W^{(n)}\cdot \dots \cdot W^{(1)}$. The closed set of equations of motion \emph{in the $0$-balanced sector} for these macro objects are (as a reminder: $\hat{N}_{\boldsymbol{W}}:= \boldsymbol{W} \Sigma_{xx}-\Sigma_{yx}$): 

\begin{align}
\begin{aligned}
&\frac12 \frac{d}{dt}\textcolor{blue_R}{\mathfrak{W}_R} = - (\hat{N}_{\boldsymbol{W}}^T \boldsymbol{W} +\boldsymbol{W}^T \hat{N}_{\boldsymbol{W}} +2\beta \textcolor{blue_R}{\mathfrak{W}_R}) := \boldsymbol{g}_{\textcolor{blue_R}{\mathfrak{W}_R}},\\
&\frac12 \frac{d}{dt}\boldsymbol{W} = -( \hat{N}_{\boldsymbol{W}}(\textcolor{blue_R}{\mathfrak{W}_R})^{n-1} + \dots + (\textcolor{blue_L}{\mathfrak{W}_L})^{i+1} \hat{N}_{\boldsymbol{W}} (\textcolor{blue_R}{\mathfrak{W}_R})^{i-1}  + \dots + (\textcolor{blue_L}{\mathfrak{W}_L})^{n-1}\hat{N}_{\boldsymbol{W}} +n\beta \boldsymbol{W}):= \boldsymbol{g}_{\boldsymbol{W}}, \\
& \frac12 \frac{d}{dt}\textcolor{blue_L}{\mathfrak{W}_L} = -(\hat{N}_{\boldsymbol{W}} \boldsymbol{W}^T + \boldsymbol{W} \hat{N}_{\boldsymbol{W}}^T +2\beta \textcolor{blue_L}{\mathfrak{W}_L}):=\boldsymbol{g}_{\textcolor{blue_L}{\mathfrak{W}_L}}. 
\label{eq:MacroDynamicsAppendix}
\end{aligned}
\end{align}
In case of $n=2$ and $\Sigma_{xx}=\mathbb{1}$ this reduces to the well-known Riccati equation \cite{Fukumizu1998}, see also App.~\ref{App:AlgebraicScaling}. Replacing $\textcolor{blue_R}{\mathfrak{W}_R}$ and $\textcolor{blue_L}{\mathfrak{W}_L}$ in terms of $\boldsymbol{W}$ gives rise to the equation of motion \eqref{eq:MacroStateEvolutionSingle} of $\boldsymbol{W}$ alone (in this subspace).

\subsection{Critical Points \label{App:CriticalPoints}}
The critical points of the loss landscape are determined by $\nabla_{W^{(i)}} \mathcal{L}=0$. This directly gives rise to the $0$-balanced condition for any finite $\beta$ since:
\begin{align}
\begin{rcases}
&2\beta W^{(i)} \textcolor{orange}{W^{(i)}{}^T} \stackrel{!}{=} - (W^{(n)} \dots W^{(i+1)})^T N (\textcolor{orange}{W^{(i)}}\dots W^{(1)})^T \\
&2\beta \textcolor{orange}{W^{(i+1)}{}^T} W^{(i+1)} \stackrel{!}{=} - (W^{(n)} \dots \textcolor{orange}{W^{(i+1)}})^T N (W^{(i)}\dots W^{(1)})^T
\end{rcases}
W^{(i+1)}{}^T W^{(i+1)} \stackrel{!}{=} W^{(i)} W^{(i)}{}^T.
\end{align}
Due to this condition, the macro critical points can be determined by:
\begin{align}
 \left( \sum_{k=0}^{n-1} (\boldsymbol{W}\boldsymbol{W}^T)^{\frac{k}{n}} \hat{N}_{\boldsymbol{W}} (\boldsymbol{W}^T \boldsymbol{W})^{\frac{n-1-k}{n}} + n\beta \boldsymbol{W}\right) \stackrel{!}{=}0.
\label{eq:criticalityconditionfrommacroevolution}
\end{align}

\subsection{Aligned (Co)Variances}
If the alignment condition \eqref{eq:compatibilitycondition} holds, $\Sigma_{xx}$ and $\Sigma_{yx}$ can be 'diagonalized' from the onset. In this frame of reference, \eqref{eq:criticalityconditionfrommacroevolution} can be solved by matrices $\boldsymbol{W}^* = S_{\boldsymbol{W}}$ with singular values $s_i\ge0$:
\begin{align}
\sum_{k=0}^{n-1} S_{\boldsymbol{W}}^{\frac{2k}{n}}( S_{\boldsymbol{W}}\tilde{\Sigma}_{xx} - \tilde{\Sigma}_{yx})S_{\boldsymbol{W}}^{\frac{2(n-1-k)}{n}} + n\beta S_{\boldsymbol{W}} \stackrel{!}{=}0 \quad &\Leftrightarrow \quad s_i^{3-\frac{2}{n}} \variancevalues{i} - s_i^{2-\frac{2}{n}} \eta_i +\beta s_i =0 \quad \forall i \\
&\Leftrightarrow \quad  \sigma_i^n(\sigma_i^{2n-2}\variancevalues{i} - \sigma_i^{n-2} \eta_i + \beta)=0,
\end{align}
where in the last step we used $s_i = \sigma_i^n$ with $\sigma_i$ the singular values of the matrices $W^{(i)}$. Besides $\sigma_i=0$, we can have up to 2 non-trivial solutions for each $i$ (depending on $\beta$) for $n>2$. For a given solution $S_{\boldsymbol{W}}$, further valid solutions are obtained by setting some or all of its finite singular values to zero (corresponding to a projection $\boldsymbol{W}_{r'<r}^* = P^{r'}_{\boldsymbol{W}^*_r} \boldsymbol{W}^*_r$ onto $r'$ finite singular values). In this diagonalized setting, the projection operators are simply $P^{r'}_{\boldsymbol{W}^*_r} = P^{r'} = \sum_{i\in \mathcal{S}_{r'}} |e_i \rangle \langle e_i|$ where $|e_i\rangle$ are the canonical basis vectors and $\mathcal{S}_{r'}$ denotes the index set of finite singular values.\\

\noindent \textbf{Onset of non-trivial solutions:} For $n>2$, the $\beta$-values $\beta^{(det)}_{i,n}$ below which non-trivial solutions for a given index $i$ become possible are determined by requiring \emph{additionally} that the second derivative in the singular values vanishes:
\begin{align}
(2n-1) \variancevalues{i} \sigma_i^{2n-2}-(n-1) \eta_i \sigma_i^{n-2}+\beta^{(det)}_{i,n} \stackrel{!}{=}0
\end{align}
giving rise to \eqref{eq:DeterministicTransitionBeta_n>2} in the main text.

\subsection{Formal Connection to unregularized Case}
The critical point structure and \eqref{eq:criticalityconditionfrommacroevolution} can also be cast into a form reminiscent of the $\beta=0$ case. Assuming that $\boldsymbol{W}_r$ has rank $r$, we can rewrite the condition \eqref{eq:criticalityconditionfrommacroevolution} formally by introducing a shifted \textcolor{blue}{variance matrix}:
\begin{align}
 \left( \sum_{k=0}^{n-1} (\boldsymbol{W}_r\boldsymbol{W}_r^T)^{\frac{k}{n}} \left(\boldsymbol{W}_r\textcolor{blue}{\underbrace{\left(P_R^{(r)}\Sigma_{xx}P_R^{(r)}+\beta (\boldsymbol{W}^T \boldsymbol{W})_r^{\frac{1-n}{n}} \right)}_{=: \hat{\Sigma}^{(r)}_{xx}(\boldsymbol{W}_r,\beta)}} - \underbrace{P_L^{(r)}\Sigma_{yx}P_R^{(r)}}_{:= \Sigma_{yx}^{(r)}}\right) (\boldsymbol{W}_r^T \boldsymbol{W}_r)^{\frac{n-1-k}{n}} \right)\stackrel{!}{=}0,
\label{eq:extremaeffectivesigmaxx},
\end{align}
where $P_R^{(r)}$ ($P_L^{(r)}$) projects onto the subspace spanned by the right (left) singular vectors for finite singular values of $\boldsymbol{W}_r$ and $(\boldsymbol{W}^T \boldsymbol{W})_r^{\frac{1-n}{n}}$ is only evaluated on this finite rank subspace.
These equations have the same form as the $\beta=0$ case with $\hat{\Sigma}_{xx}\rightarrow \hat{\Sigma}_{xx}^{(r)}(\boldsymbol{W}_r,\beta)$. Even though we might not be able to solve these equations in closed form, we can state the formal solutions based on the solution \emph{structure} in the $\beta=0$ case (adapting the arguments in \cite{Baldi1989}): 
\begin{itemize}
\item full rank solutions (for sufficiently low $\beta$), are of the form $\boldsymbol{W}_{r=r_{\text{max}}} = \Sigma_{yx} \hat{\Sigma}_{xx}^{-1}(\boldsymbol{W}^*,\beta)$ with $P_L=\mathbb{1}_{o,o},P_R=\mathbb{1}_{i,i} $. Further rank-reduced solutions are obtained by projections $P_{\boldsymbol{W}^*} \Sigma_{yx} \hat{\Sigma}_{xx}^{-1}(\boldsymbol{W}^*,\beta)$ where the projection operators have the properties $P_{\boldsymbol{W}^*} \tilde{\Sigma}_{\boldsymbol{W}^*} P_{\boldsymbol{W}^*} = P_{\boldsymbol{W}^*} \tilde{\Sigma}_{\boldsymbol{W}^*} = \tilde{\Sigma}_{\boldsymbol{W}^*}P_{\boldsymbol{W}^*}$. They are constructed from the eigenvectors (with eigenvalues $\hat{\lambda_i}^{\text{xx-shift}}$) of
\begin{align}
\hat{\Sigma}^{\text{xx-shift}}_{\boldsymbol{W}^*} := \Sigma_{yx} \hat{\Sigma}_{xx}^{-1}(\boldsymbol{W}^*,\beta) \Sigma_{xy}
\end{align}
similar to the $\beta=0$-case. The loss values at critical points are given by:
\begin{align}
\mathcal{L}^*(\boldsymbol{W}^*,\beta) = -\sum_{i\in \mathcal{S}} \left(\hat{\lambda_i}^{\text{xx-shift}} - (n-1)\beta \sigma_i^*{}^2\right) +\text{const.}.
\end{align}
\item Depending on $\beta$, \eqref{eq:extremaeffectivesigmaxx} will feature solutions with rank $r\le r_{\text{max}(\beta)}$: if a solution $\boldsymbol{W}_{r}^*$ of \eqref{eq:extremaeffectivesigmaxx} was found, further projections to lower-rank solutions are possible according to the same logic: $\boldsymbol{W}_{r'<r}^* = P^{r'}_{\boldsymbol{W}^*_r} \boldsymbol{W}^*_r$ where the projection operator projects onto eigensubspaces of $\tilde{\Sigma}^{(r)} := \Sigma_{yx}^{(r)} \textcolor{blue}{\hat{\Sigma}_{xx}^{(r)}{}^{-1}(\boldsymbol{W}_r^*,\beta)} \Sigma_{yx}^{(r)}{}^{(r)}$ in the $r$-dim. subspace. Note that the inverse is only evaluated on the one this subspace (it is the pseudoinverse).
\end{itemize}

\noindent \textbf{Shifting the Covariance Matrix:} Note that we could have rewritten \eqref{eq:criticalityconditionfrommacroevolution} (for a rank-$r$ solution) equivalently by redefining the \textcolor{orange}{covariance matrix}:
\begin{align}
\left( \sum_{k=0}^{n-1} (\boldsymbol{W}_r\boldsymbol{W}_r^T)^{\frac{k}{n}} \left(\boldsymbol{W}_r P_R^{(r)}\Sigma_{xx}P_R^{(r)} - \textcolor{orange}{\underbrace{\left(P_L^{(r)}\Sigma_{yx} P_R^{(r)}-\beta \boldsymbol{W}_r (\boldsymbol{W}_r^T \boldsymbol{W}_r)^{\frac{1-n}{2}}\right)}_{=:\hat{\Sigma}^{(r)}_{yx}(\boldsymbol{W}_r)}}\right) (\boldsymbol{W}_r^T \boldsymbol{W}_r)^{\frac{n-1-k}{n}} \right)\stackrel{!}{=}0,
\end{align}
with $\hat{\Sigma}^{(r)}_{yx}(\boldsymbol{W}_r):= P_L^{(r)}\Sigma_{yx}P_R^{(r)}-\beta L_r S_r^{2-n} R_r^T$ (for a given singular value decomposition $\boldsymbol{W}_r=L_rS_rR_r^T$) while leaving $\Sigma_{xx}$ unchanged. Here, $S_r^{2-n}$ is meant to be evaluated on the $r$-dimensional subspace of finite singular values only. Projected solutions are found based on projection operators constructed from eigenvectors (to eigenvalues $\hat{\lambda_i}^{\text{xy-shift}}$):
\begin{align}
\tilde{\Sigma}^{\text{xy-shift}}_r := \textcolor{orange}{\hat{\Sigma}_{yx}^{(r)}} \Sigma_{xx}^{(r)}{}^{-1} \textcolor{orange}{\hat{\Sigma}_{yx}^{(r)}}
\label{eq:ShiftedCovarianceMatrix}
\end{align}
with loss
\begin{align}
\mathcal{L}^*(\boldsymbol{W}^*,\beta) = -\sum_{i\in \mathcal{S}} \left(\hat{\lambda_i}^{\text{xy-shift}} - (n-2)\beta \sigma_i^*{}^2\right)+\text{const.}.
\end{align}
Note that the case $n=2$ (1 hidden layer) is special: the critical points can be entirely discussed in terms of \eqref{eq:ShiftedCovarianceMatrix} (once a maximum rank solution was found). This makes the $n=2$-case very similar in structure to the unregularized case.

\subsection{Non-Aligned (Co)Variances: Perturbatively (1 hidden layer) \label{App:Perturbation}}

Based on criticality conditions \eqref{eq:criticalityconditionfrommacroevolution}, we can perturbatively study the situation where $\Sigma_{xx}$ and $\Sigma_{yx}=\text{diag}(\eta_1,...,\eta_{r_\text{max}})$ are not aligned with, e.g.: $\Sigma_{xx} = D + \epsilon \sum_i \alpha_i (e_{ij}+e_{ji})$ with $\epsilon \ll 1$. Here, $D$ is a diagonal matrix with entries $\variancevalues{i}$. At leading order in $\epsilon$, the singular values of $\boldsymbol{W}^*$ are not affected, but the matrices $\textcolor{blue}{L}$ and $\textcolor{blue}{R}$ of singular vectors change as follows:

\begin{align*}
&\textcolor{blue}{R} \approx \mathbb{1} + \epsilon \textcolor{blue!70!white}{R^{(1)}} \approx \exp(\epsilon \textcolor{blue!70!white}{R^{(1)}}), \quad \textcolor{blue}{L} \approx \mathbb{1} + \epsilon  \textcolor{blue!70!white}{L^{(1)}} \approx \exp(\epsilon \textcolor{blue!70!white}{L^{(1)}}), \\
& \textcolor{blue!70!white}{R^{(1)}} = \sum_{\text{$(i,j)$ pairs}} r^{(ij)} (e_{ij}-e_{ji}), \quad r^{(ij)} = \alpha_i \frac{\eta_i \variancevalues{j} \cdot\text{max}(\eta_i-\beta,0)+\eta_j \variancevalues{i} \cdot \text{max}(\eta_j -\beta,0)}{\eta_i^2 \variancevalues{j}^2 -\eta_j^2 \variancevalues{i}^2}, \\
& \textcolor{blue!70!white}{L^{(1)}} = \sum_{\text{$(i,j)$ pairs}} l^{(ij)} (e_{ij}-e_{ji}), \quad l^{(ij)} = \alpha_i \frac{\eta_j \variancevalues{i} \cdot\text{max}{(\eta_i-\beta,0)}+\eta_i \variancevalues{j} \text{max}\cdot(\eta_j -\beta,0)}{\eta_i^2 \variancevalues{j}^2 -\eta_j^2 \variancevalues{i}^2},
\end{align*}
where $R^{(1)}, L^{(1)}$ are anti-symmetric matrices (linear combinations of generators of the orthogonal group). $e_{ij}$ is the matrix with a single $1$ at position $(i,j)$.

\subsection{Paths through Loss Landscape \& Symmetry Example \label{App:SymmetryExample}}

\noindent\textbf{Unregularized Case:} Due to the simplicity of the chosen (co)variances, the geometry of the loss landscape can be visualized based on straight paths through the loss landscape in the directions of the Hessian eigendirections (where we assume an infinitesimal $L_2$ regularization to explicitly break the general linear symmetries). 

\noindent \underline{For 1 hidden layer ($n=2$):} The first path corresponds to the direction dictated by the leading singular value of $\Sigma_{yx}$, which is $\eta_1$ with (right) singular vector $\vec{r}_1=\vec{e}_1$. In parameter space, this translates into the direction $(e_{11},e_{11})$ (or more generally: $(\vec{e}_1 \textcolor{orange}{\vec{v}_1^T},\textcolor{orange}{\vec{v}_1}\vec{e}_1^T)$ for any normalized $\textcolor{orange}{\vec{v}_1}$. The corresponding path is $(W^{(2)},W^{(1)}) = s\cdot (e_{11},e_{11})$ with $s\in[1,(\eta_1)^{\frac{1}{n}}]$ the path length. The corresponding 1-dimensional representation of the loss corresponds to the effective potential for a system that is initialized as (e.g.) $(W^{(2)},W^{(1)}) = \epsilon(e_{11},e_{11})$ with $\epsilon \ll1$ close to the origin that will approach $(W^{(2)},W^{(1)})_{t\to \infty} = (\eta_1)^{\frac{1}{n}}(e_{11},e_{11})$ in the long time limit under gradient flow\footnote{This direction is the (local) direction corresponding to the largest negative eigenvalue of the Hessian.}. This is a saddle point with corresponding macrostate $\boldsymbol{W} =\hat{P}_{\mathcal{S}} \Sigma_{yx}\Sigma_{xx}^{-1}$ with $\mathcal{S}=\{1\}$. In the next step, we consider the path along a negative Hessian direction (which exists for a 1 hidden layer network) of this saddle point, corresponding to the direction given by the first subleading singular value $\eta_2$ of $\Sigma_{yx}$. Starting from $(\eta_1)^{\frac{1}{n}}(e_{11},e_{11})$ in parameter space, we plot the parameter direction $(W^{(2)},W^{(1)}) = \eta_1^{\frac{1}{n}}(e_{11},e_{11}) + (s-\eta_1^{\frac{1}{n}})\cdot (e_{22},e_{22})$ with $s\in[\eta_1^{\frac{1}{n}},\eta_2^{\frac{1}{n}}+\eta_1^{\frac{1}{n}}]$. This path corresponds to the effective potential seen by a system initialized somewhere on this path. For long times, such a system would reach $(W^{(2)},W^{(1)})_{t\to\infty} = \eta_1^{\frac{1}{n}}(e_{11},e_{11}) + \eta_2^{\frac{1}{n}}(e_{22},e_{22})$. This is a saddle point with corresponding macrostate $\boldsymbol{W} =\hat{P}_{\mathcal{S}} \Sigma_{yx}\Sigma_{xx}^{-1}$ with $\mathcal{S}=\{1,2\}$. This construction can be continued until the global minimum is reached (similar to  Fig.~\ref{fig:UnregularizedBasin} in the main text with a path length $s$ on the $x$-axis). The choice of cuts/paths is not unique (there are multiple negative directions at the saddle point). However, the above choice corresponds to the successive traversing of those saddles that correspond to rank-restriced optima, which play a distinct role once a finite $L_2$-regularization is introduced (as they turn into (local) minima). 

\noindent \underline{For more hidden layers ($n>2$):} The same construction can be made for arbitrary depth ($n$ layers). In this case, the follow-up direction after having reached a non-strict saddle corresponds \emph{locally} to a flat direction of the Hessian. The accuracy-decreasing character of this direction can be seen at higher-order derivatives. More directly (for the simple data structure in the main text): plugging in the Ansatz $W^{(i)} = \sum_{k\in \mathcal{S}} \eta_k^{\frac{1}{n}} e_{kk} + f(t) e_{jj}$ with $j\notin \mathcal{S}$ we get a single ordinary differential equation for $f(t)$:
\begin{align*}
\frac{d}{dt}f(t) = -2 f^{n-1}(t)(f^n(t) - \eta_j),
\end{align*}
such that for $f(t)\approx0$, the strongest growth is given by choosing the direction $e_{jj}$ such that $\eta_{j\notin \mathcal{S}}$ is maximal.\\

\noindent \textbf{Regularized Case:} The construction is similar, though the intermediate critical points are local minima for $n>2$ layers. Formerly flat directions (in the direction of the next lower lying critical point, increasing the rank) are now stiff (positive Hessian eigenvalue) due to the finite regularization. \\

\noindent \textbf{Connected Minima:} Minima in this model are highly degenerate. An explicit parametrization of the connected minima in the example discussed in the main text (see Fig.~\ref{fig:OrderParameterSpaceIllustration}) is given by $s\in [0,2\pi]$ ($\sigma:= \sqrt{||\text{max}(\Sigma_{yx}-\beta,0)||} \ge 0$): $W^{(2)*}(s) = \sigma \begin{pmatrix} \cos(s) \\ \sin(s)\end{pmatrix}$, $W^{(1)*}(s) = W^{(2)}{}^T \frac{\Sigma_{yx}}{||\Sigma_{yx}||}$.
Perturbations in orthogonal directions can be included for example as:
\begin{align*}
(W^{(2)*}(s),W^{(1)*}(s)) \rightarrow \left(W^{(2)*} + s_\perp \begin{pmatrix} \cos(s) \\ \sin(s)\end{pmatrix},W^{(1)*}+s_\perp\begin{pmatrix} \cos(s) \\ \sin(s)\end{pmatrix} \right),
\end{align*}
where $s$ labels the minimum and $s_\perp$ parametrizes the deviation from the minimum. The loss as a function of $s_\perp$ is shown in Fig.~\ref{fig:OrderParameterSpaceIllustration}.

\section{Numerical Details}

\begin{table}[h]
\centering
\begin{tabular}{l|l|l}
\textbf{figure} & \textbf{(co)variance data} & \textbf{hyperparameters (numerics)} \\ \toprule
\makecell{Fig.~\ref{fig:IntroOverview} \\ Fig.~\ref{fig:ExamplesEffectiveDescription2hidden} \\ Fig.~\ref{fig:HessianSpectrum1hidden} (c,d)} & $\eta_1=0.95, \eta_2=0.55,\eta_3=0.3$, $\Sigma_{xx}=\mathbb{1}_{3\times 3}$ & \makecell{sgd (single run) with $\Delta t=0.01$ and $2000$ epochs, batch size: $128$\\ $N_\text{training}=10^5$, IA} \\ \hline
Fig.~\ref{fig:UnregularizedBasin} & $\eta_1=0.95, \eta_2=0.55,\eta_3=0.3$, $\Sigma_{xx}=\mathbb{1}_{3\times 3}$ &  \\ \hline
Fig.~\ref{fig:RegularizedLandscape} & $\eta_1=0.95, \eta_2=0.55,\eta_3=0.3$, $\Sigma_{xx}=\mathbb{1}_{3\times 3}$ & \\ \hline
Fig.~\ref{fig:HessianSpectrum1hidden} (a,b) & $\eta_1=0.95, \eta_2=0.55$, $\Sigma_{xx}=\mathbb{1}_{3\times3}$ & \makecell{sgd (single run) with $\Delta t=0.01$ and $50$ epochs,\\ batch size: $128$, $N_\text{training}=10^5$, IA}\\ \hline
Fig.~\ref{fig:DiscussionNonAlignedCase} & $\eta_1=0.5, \eta_2=0.25$, $\Sigma_{xx} = \mathbb{1} + \epsilon \begin{pmatrix}0&1\\1&0\end{pmatrix}$ & \makecell{using package NonlinearSolve \cite{Pal2024nonlinearsolve} solving for stable stationary solutions} \\ \hline
Fig.~\ref{fig:DiscussionNonAlignedCase}(a) & $\eta_1=0.5, \eta_2=0.25$, $\Sigma_{xx} = \mathbb{1} + \epsilon \begin{pmatrix}0&1\\1&0\end{pmatrix}$ & adamw, $\Delta t=0.001$, $200$ epochs, $N_\text{training} =10^5$, batch size: $1000$, IA\\ \hline
Fig.~\ref{fig:MNISTsvsandrank} & $\eta_1=0.95,\eta_2=0.55$, $\Sigma_{xx}=\mathbb{1}_{3\times 3}$  & \makecell{$d_\text{hidden}=20$, adamw, $\Delta t=0.001$, $50$ epochs,$N_\text{training}=10^6$\\batch size: $128$, IA, $20$ runs}
\end{tabular}
\caption{Overview of hyperparameters used in the numerical experiments. Note that we have used $\Sigma_{yy}=\mathbb{1}_{o,o}$ in all cases. IA: inverse annealing protocol (starting from $\beta=0$ and annealing towards larger $\beta$).}
\end{table}

\section{Hessian Spectrum \label{App:HessianSpectrum}}
In the following, we discuss how the Hessian spectrum at the critical points for the 1 hidden layer case and the 2 hidden layer case can be derived in full detail. The case of arbitrary depth is discussed at the end. Formally, the derivation is based on vectorization (see also, e.g., \cite{Singh_DLNHessian}) and the eigenvalue equations:
\begin{align}
&H_{\mathcal{L}} \delta \vec{\theta}_a = \lambda_a \delta \vec{\theta}_a, \quad \left(H_{\mathcal{L}}\right)_{ij} = \left.\frac{\partial^2 \mathcal{L}}{\partial W^{(i)} \partial W^{(j)}}\right|_{\vec{\theta} = \vec{\theta}^*}\\
&\vec{\theta}^* = \text{vec}_c(W^{(1)}{}^*,W^{(2)}{}^*), \quad \delta \vec{\theta} := \text{vec}_c(\delta W^{(1)}, \delta W^{(2)}),
\end{align}
where $\delta W_i$ are the matrix-values deviations from the critical point. In the following, we are working with the alignment condition and (if not stated differently) used the outer transformations to 'diagonalize' the (co)variances. 

\subsection{1 hidden Layer \label{App:Hessian1hiddenLayer}}
Starting point of the analysis are the gradients of the loss (which are matrix valued; $\hat{N}_{\boldsymbol{W}} := W^{(2)}W^{(1)} \Sigma_{xx}-\Sigma_{yx}$):
\begin{align*}
&\frac{\partial \mathcal{L}}{\partial W^{(1)}} = 2\left(W^{(2)}{}^T \hat{N}_{\boldsymbol{W}} + \beta W^{(1)}\right), \quad \frac{\partial \mathcal{L}}{\partial W^{(2)}} =  2\left(\hat{N}_{\boldsymbol{W}}W^{(1)}{}^T  + \beta W^{(2)}\right).
\end{align*}
To analyze the Hessian, we are working with column vectorization, where $\text{vec}_c(W) := (W_{:,1}, W_{:,2},\dots)$, such that the Hessian becomes a matrix. Useful examples are:
\begin{align}
&W^{(1)}{}^* = \sum_i \sigma_i \textcolor{orange}{\vec{v}_i} \textcolor{blue}{\vec{r}_i^T} \quad \rightarrow \text{vec}_c(W^{(1)}{}^*) = \sum_i \sigma_i \textcolor{blue}{\vec{r}_i} \otimes \textcolor{orange}{\vec{v}_i}, \\
&W^{(2)}{}^* = \sum_i \sigma_i \textcolor{blue}{\vec{l}_i} \textcolor{orange}{\vec{v}_i^T} \quad \rightarrow \text{vec}_c(W^{(2)}{}^*) = \sum_i \sigma_i \textcolor{orange}{\vec{v}_i} \otimes \textcolor{blue}{\vec{l}_i}, \\
&\text{vec}_c((W^{(1)},W^{(2)}))^* = \sum_i \sigma_i \begin{pmatrix} \textcolor{blue}{\vec{r}_i} \otimes \textcolor{orange}{\vec{v}_i}  \\ \textcolor{orange}{\vec{v}_i} \otimes \textcolor{blue}{\vec{l}_i} \end{pmatrix} \quad \Leftrightarrow \quad \text{vec}_c((W^{(1)},W^{(2)}{}^T))^* = \sum_i \sigma_i \begin{pmatrix} \textcolor{blue}{\vec{r}_i} \otimes \textcolor{orange}{\vec{v}_i}  \\ \textcolor{blue}{\vec{l}_i} \otimes \textcolor{orange}{\vec{v}_i} \end{pmatrix}.
\end{align}
Working with $\text{vec}_c(\delta W^{(1)}, \delta W^{(2)}{}^{\textcolor{orange}{T}})$ turns the expressions more symmetric (and we will often make use of this notation in the following). For 1 hidden layer (denoting with indices $i,o,h$ the input, output or hidden dimension for \emph{matrices}):
\begin{align}
\begin{rcases}
&\frac{1}{2} \frac{\partial^2 \mathcal{L}}{\partial W^{(1)} \partial W^{(1)}} = \Sigma_{xx} \otimes W^{(2)}{}^T W^{(2)} + \beta \mathbb{1}_i \otimes \mathbb{1}_h \\
& \frac{1}{2}\frac{\partial^2 \mathcal{L}}{\partial W^{(1)} \partial W^{(2)}} = (\hat{N}_{\boldsymbol{W}}^T \otimes \mathbb{1}_{h})\mathcal{K}_{o,h} + \Sigma_{xx}W^{(1)}{}^T \otimes W^{(2)}{}^T \\
& \frac{1}{2} \frac{\partial^2 \mathcal{L}}{\partial W^{(2)} \partial W^{(1)}} = (\mathbb{1}_{h} \otimes \hat{N}_{\boldsymbol{W}})\mathcal{K}_{h,i} + W^{(1)}\Sigma_{xx} \otimes W^{(2)} \\
&  \frac{1}{2}\frac{\partial^2 \mathcal{L}}{\partial W^{(2)} \partial W^{(2)}} =  W^{(1)} \Sigma_{xx} W^{(1)}{}^T \otimes \mathbb{1}_{o} + \beta \mathbb{1}_{h} \otimes \mathbb{1}_{o}
\end{rcases} \quad H_{\mathcal{L}} = 2\begin{pmatrix} \frac{\partial^2 \mathcal{L}}{\partial W^{(1)} \partial W^{(1)}} & \frac{\partial^2 \mathcal{L}}{\partial W^{(1)} \partial W^{(2)}} \\ \frac{\partial^2 \mathcal{L}}{\partial W^{(2)} \partial W^{(1)}} & \frac{\partial^2 \mathcal{L}}{\partial W^{(2)} \partial W^{(2)}} \end{pmatrix},
\label{eq:FormalHessian1hidden}
\end{align}
where $\mathcal{K}$ is a commutation matrix that turns a vectorized matrix into the vectorized transpose of that matrix: $\mathcal{K} \text{vec}_c(W) = \text{vec}_c(W^T)$. Note that $\mathcal{K}_{a,b}$ is a $(d_a \times d_b) \otimes (d_b \times d_a)$ matrix:
\begin{align}
\mathcal{K}_{a,b} = \sum_{i=1}^a \sum_{j=1}^b \textcolor{blue}{|e_i^{(a)}\rangle} \textcolor{orange}{\langle e_j^{(b)}|} \otimes \textcolor{orange}{|e_j^{(b)}\rangle} \textcolor{blue}{\langle e_i^{(a)}|}.
\end{align}
As an example: $\mathcal{K}_{h,i} \text{vec}_c(\textcolor{orange}{\vec{v}_m} \textcolor{blue}{\vec{r}_l^T}) = \mathcal{K}_{h,i} (\textcolor{blue}{\vec{r}_l} \otimes \textcolor{orange}{\vec{v}_m})= \textcolor{orange}{\vec{v}_m} \otimes \textcolor{blue}{\vec{r}_l}= \text{vec}_c(\textcolor{blue}{\vec{r}_l} \textcolor{orange}{\vec{v}_m^T})$.\\

\noindent From now on we will mainly work with $\text{vec}_c(\delta W^{(1)}, \delta W^{(2)}{}^T)$ and the correspondingly transformed Hessian $H\to H'$, such that the expressions become more symmetric. From the very structure of the Hessian matrix we directly observe that eigenvectors $\delta \vec{\theta}' = \text{vec}_c(\delta W^{(1)}, \delta W^{(2)}{}^T)$ are constructed (mainly) from linear combinations of $(\vec{r}_i \otimes \vec{v}_j ,\vec{l}_i \otimes \vec{v}_j)^T$ for different $i,j$ as e.g.:
\begin{align}
&\left(\Sigma_{xx} \otimes W^{(2)}{}^T W^{(2)} + \beta \mathbb{1}_{i} \otimes \mathbb{1}_{h}\right)  \left(\vec{r}_i \otimes \vec{v}_j \right)  =  (\variancevalues{i} \sigma_j^2 + \beta) \left( \vec{r}_i \otimes \vec{v}_j \right),  \\
&\left(\hat{N}^T \otimes \mathbb{1}_{h} + (\Sigma_{xx}W^{(1)}{}^T \otimes W^{(2)}{}^T)\mathcal{K}_{o,h}\right) \left( \vec{l}_i \otimes \vec{v}_j \right)= n_i \left(\vec{r}_i \otimes \vec{v}_j \right)+ \variancevalues{i} \sigma_i \sigma_j  \left(\vec{r}_j \otimes \vec{v}_i\right), \\
&\left((\mathbb{1}_{h} \otimes \hat{N})\mathcal{K}_{h,i} + W^{(1)}\Sigma_{xx} \otimes W^{(2)}\right) \left( \vec{r}_i \otimes \vec{v}_j \right) = n_j \left( \vec{v}_j \otimes \vec{l}_i \right) + \variancevalues{i} \sigma_i \sigma_j \left( \vec{r}_i \otimes \vec{v}_j \right)
\end{align}
with $n_i =-\beta$ for $i\in \mathcal{S}$ and $n_i = \eta_i$ for $i \notin \mathcal{S}$. We can now distinguish the following subspaces (as a reminder: $\mathcal{S}$ denotes the set of indices of \emph{finite} singular values) listed in Tab.~\ref{tab:HessianSpectrum1hiddenAppendix}.

\begin{table}[h]
\centering
\begin{tabular}{l|l|l|l}
\textbf{subspace} & \textbf{condition} & \textbf{eigenvalues} & \textbf{numerosity}\\ \hline
$\left\{\begin{pmatrix} \vec{r}_i \otimes \vec{v}_j  \\ \vec{l}_i \otimes \vec{v}_j \end{pmatrix}, \begin{pmatrix} \vec{r}_j \otimes \vec{v}_i  \\ \vec{l}_j \otimes \vec{v}_i \end{pmatrix} \right\}$ & $i,j \in \mathcal{S}$ & $\lambda = 0$, $\lambda = 2\left(\sigma_i^2 + \sigma_j^2\right)$ &  $\frac12r(r+1)$,$\frac12r(r+1)$ \\ 
$\left\{\begin{pmatrix} \vec{r}_i \otimes \vec{v}_j  \\ -\vec{l}_i \otimes \vec{v}_j \end{pmatrix}, \begin{pmatrix} \vec{r}_j \otimes \vec{v}_i  \\ -\vec{l}_j \otimes \vec{v}_i \end{pmatrix} \right\}$ & $i,j \in \mathcal{S}$ & $\lambda = 4\beta$, $\lambda = 2(\sigma_i^2 + \sigma_j^2 +2 \beta)$ & $\frac12r(r+1)$,$\frac12r(r-1)$ \\  \hline
$ \vec{v}^\pm_{ij} =\begin{pmatrix}  \vec{r}_i \otimes \vec{v}_j  \\ \pm \vec{l}_i \otimes \vec{v}_j \end{pmatrix}$ & $i \notin \mathcal{S}$, $j\notin \mathcal{S}$ & $\lambda_+ = 2(-\eta_i + \beta)$ and $\lambda_- = 2(\eta_i + \beta)$ & $(r_\text{max}-r)(d_\text{hidden}-r)$ \\
& $i \in \mathcal{S}$, $j\notin \mathcal{S}$ & $\lambda_+ =0$ and $\lambda_- = 4\beta$ & $r(d_\text{hidden}-r)$ \\
&$i \notin \mathcal{S}$, $j\in \mathcal{S}$ & $\lambda_\pm = 2(\eta_j \pm \eta_i)$ & $r(r_\text{max}-r)$ \\ \hline
$\vec{v}^\pm_{ij} =\begin{pmatrix}  \vec{r}_i \otimes \vec{v}_j \\ 0 \end{pmatrix}$ & $i > d_\text{out}$, $j\in \mathcal{S}$ & $\lambda= 2(\sigma_j^2 +\beta)$ & $(d_\text{in}-r_\text{max})r$ \\
& $i > d_\text{out}$, $j\notin \mathcal{S}$ & $\lambda= 2\beta$ & $(d_\text{in}-r_\text{max})(d_\text{hidden}-r)$
\end{tabular}
\caption{Eigenvectors and eigenvalues of the Hessian at a critical point (for $\Sigma_{xx}=\mathbb{1}$) for 1 hidden layer network. In our case $r_\text{max}=d_\text{out}$.}
\label{tab:HessianSpectrum1hiddenAppendix}
\end{table}
The set of vectors mentioned first in Tab.~\ref{tab:HessianSpectrum1hiddenAppendix} form an invariant subspace under $H'$, such that $H'$ can effectively be formulated as a $2\times2$ matrix in those subspaces (for $\Sigma_{xx}=\mathbb{1}$ for simplicity):
\begin{align*}
\boldsymbol{(A):}\quad &\frac12 H' \begin{pmatrix} \vec{r}_i \otimes \vec{v}_j  \\ \vec{l}_i \otimes \vec{v}_j \end{pmatrix} = \sigma_j^2 \begin{pmatrix} \vec{r}_i \otimes \vec{v}_j  \\ \vec{l}_i \otimes \vec{v}_j \end{pmatrix} + \sigma_i \sigma_j \begin{pmatrix} \vec{r}_j \otimes \vec{v}_i  \\ \vec{l}_j \otimes \vec{v}_i \end{pmatrix} &&\quad \rightarrow \quad H^{\boldsymbol{(A)}}_{2\times 2} = 2\begin{pmatrix} \sigma_j^2 & \sigma_i \sigma_j \\ \sigma_i \sigma_j & \sigma_i^2 \end{pmatrix} \\
\boldsymbol{(B):}\quad &\frac12 H' \begin{pmatrix} \vec{r}_i \otimes \vec{v}_j  \\ -\vec{l}_i \otimes \vec{v}_j \end{pmatrix} = (\sigma_j^2+2\beta) \begin{pmatrix} \vec{r}_i \otimes \vec{v}_j  \\ -\vec{l}_i \otimes \vec{v}_j \end{pmatrix} -\sigma_i \sigma_j \begin{pmatrix} \vec{r}_j \otimes \vec{v}_i  \\ -\vec{l}_j \otimes \vec{v}_i \end{pmatrix} &&\quad \rightarrow \quad H^{\boldsymbol{(B)}}_{2\times 2} = 2\begin{pmatrix} \sigma_j^2 +2\beta & -\sigma_i \sigma_j \\ -\sigma_i \sigma_j & \sigma_i^2 + 2\beta \end{pmatrix}
\end{align*}
with the eigenvalues given in Tab.~\ref{tab:HessianSpectrum1hiddenAppendix}. Note that in case ${\boldsymbol{(B)}}$ and for $i=j$ we only get the eigenvalue $\lambda = 2\beta$.\\

\noindent \textbf{From a symmetry point of view:} we know that the the micro critical points are highly degenerate. Given a micro critical point $\vec{\theta}^*{}'$ with index set $\mathcal{S}$, the following (vectorized) microstates are critical as well ($\textcolor{orange}{\mathcal{O}}\in O(d_\text{hidden})$):
\begin{align*}
\vec{\theta}^*{}' = \sum_{i\in \mathcal{S}} \sigma_i \begin{pmatrix} \vec{r}_i \otimes \vec{v}_i  \\ \vec{l}_i \otimes \vec{v}_i \end{pmatrix} \rightarrow \vec{\theta}_{\textcolor{orange}{\mathcal{O}}}^*{}' =\sum_{i\in \mathcal{S}} \sigma_i \begin{pmatrix} \vec{r}_i \otimes \textcolor{orange}{\mathcal{O}}\vec{v}_i  \\ \vec{l}_i \otimes  \textcolor{orange}{\mathcal{O}} \vec{v}_i \end{pmatrix}.
\end{align*} 
The infinitesimal version based on a rotation angle $\epsilon \ll1$ reads:
\begin{align*}
\vec{\theta}_{\textcolor{orange}{\mathcal{O}}}^*{}' =\sum_{i\in \mathcal{S}} \sigma_i \begin{pmatrix} \vec{r}_i \otimes \textcolor{orange}{\mathcal{O}}\vec{v}_i  \\ \vec{l}_i \otimes  \textcolor{orange}{\mathcal{O}} \vec{v}_i \end{pmatrix} \approx \vec{\theta}^*{}' + \epsilon \underbrace{\sum_{i\in \mathcal{S}} \sigma_i \begin{pmatrix} \vec{r}_i \otimes A\vec{v}_i  \\ \vec{l}_i \otimes A\vec{v}_i \end{pmatrix}}_{=\delta \vec{\theta}},
\end{align*}
where $A$ is an anti-symmetric matrix (a linear combination of the anti-symmetric generators of the orthogonal group). We can directly check that $\delta \vec{\theta}$ is indeed an eigenvector of the Hessian for the eigenvalue $\lambda=0$ using \eqref{eq:FormalHessian1hidden} (and see below). Special cases are given by, e.g., $A = \vec{v}_i \vec{v}_j^T - \vec{v}_j \vec{v}_i^T$ which give rise to the explicit eigenvectors shown above in Tab.~\ref{tab:HessianSpectrum1hiddenAppendix}. \\
In case of more general (general linear) transformations $\mathcal{G} \in GL(d_\text{hidden})$ (that leave the \emph{unregularized} loss invariant), we have:
\begin{align*}
\vec{\theta}^*{}' = \sum_{i\in \mathcal{S}} \sigma_i \begin{pmatrix} \vec{r}_i \otimes \vec{v}_i  \\ \vec{l}_i \otimes \vec{v}_i \end{pmatrix} \rightarrow \vec{\theta}_{\textcolor{orange}{\mathcal{G}}}^*{}' =\sum_{i\in \mathcal{S}} \sigma_i \begin{pmatrix} \vec{r}_i \otimes \textcolor{orange}{\mathcal{G}}\vec{v}_i  \\ \vec{l}_i \otimes  \textcolor{orange}{\mathcal{G}^{-1}{}^T} \vec{v}_i \end{pmatrix} \stackrel{\text{infinitesimal}}{\approx} \vec{\theta}^*{}' + \epsilon  \underbrace{\sum_{i\in \mathcal{S}} \sigma_i \begin{pmatrix} \vec{r}_i \otimes A\vec{v}_i  \\ \vec{l}_i \otimes (-A^T)\vec{v}_i \end{pmatrix}}_{=\delta \vec{\theta}}
\end{align*} 
with $A$ being a linear combination of generators of the general linear group. The direction $\delta\vec{\theta}$ can be an eigenvector of the Hessian for two cases:
\begin{align*}
\frac12 H' \delta \vec{\theta} &= \sum_{i\in \mathcal{S}} \sigma_i \begin{pmatrix}\left[ \Sigma_{xx}\vec{r}_i \otimes W^{(2)}{}^T W^{(2)} A \vec{v}_i - \Sigma_{xx}W^{(1)}{}^T A^T \vec{v}_i \otimes W^{(2)}{}^T \vec{l}_i + \beta(\mathbb{1}\otimes A + \mathbb{1}\otimes A^T)(\vec{r}_i \otimes \vec{v}_i)\right] \\
\left[ \vec{l}_i \otimes W^{(1)} \Sigma_{xx} W^{(1)}{}^T (-A)^T \vec{v}_i - W^{(2)} A\vec{v}_i \otimes W^{(1)}\Sigma_{xx} \vec{r}_i - \beta(\mathbb{1}\otimes A + \mathbb{1}\otimes A^T)(\vec{l}_i \otimes \vec{v}_i)\right]\end{pmatrix} \\
& = \underbrace{\sum_{i \in \mathcal{S}} \begin{pmatrix} (\Sigma_{xx}W^{(1)}{}^T \otimes W^{(2)}{}^T W^{(2)} )(\mathbb{1} \otimes A - A^T \otimes \mathbb{1})(\vec{v}_i \otimes \vec{v}_i)  \\ (W^{(2)} \otimes W^{(1)} \Sigma_{xx} W^{(1)}{}^T)(A\otimes \mathbb{1} - \mathbb{1}\otimes A^T)(\vec{v}_i \otimes \vec{v}_i) \end{pmatrix} }_{=\vec{0}} + \beta \sum_{i\in \mathcal{S}} \sigma_i \begin{pmatrix} (\mathbb{1}\otimes A + \mathbb{1}\otimes A^T)(\vec{r}_i \otimes \vec{v}_i) \\ - (\mathbb{1}\otimes A + \mathbb{1}\otimes A^T)(\vec{l}_i \otimes \vec{v}_i) \end{pmatrix} \\
&= \begin{cases} \vec{0} & \text{for $A$ anti-symmetric} \\ 2\beta \delta \vec{\theta} & \text{for $A$ symmetric} \end{cases} .
\end{align*}
Therefore, the orthogonal symmetry gives rise to $\lambda=0$ eigenvalues and the explicitly broken symmetries (due to the regularizer) give rise to massive eigenvalues $\lambda=4\beta$. The massive data-dependent eigenvalues and directions (see Tab.~\ref{tab:HessianSpectrum1hiddenAppendix}) in the $\mathcal{S}$-sector are instead associated with rotations of the outer singular vectors $\textcolor{blue}{\vec{r}_i}, \textcolor{blue}{\vec{l}_i}$ (see the discussion of the Jacobian).

\subsection{2 hidden Layers}
In the following, we derive the Hessian spectrum for the 2 hidden layer case under the alignment condition and $\Sigma_{xx} = \mathbb{1}_{i,i}$ for simplicity. The rank $r$ of the critical $W^{(i)}$'s is crucial to understand the Hessian spectrum. It is convenient to first identify the (nearly) flat directions. With those (eigen-)directions at hand, we can directly construct the (low-dimensional) massive data-dependent part of the Hessian spectrum. The goal is to show that the Hessian can be decomposed into the form: $H_\text{crit} = \textcolor{orange}{H_0 + \textcolor{red!70!black}{H_{\text{null}}^{(\beta)}}+ H_{4\beta}} + \textcolor{blue}{H_{ex} + H_{r^2}}$. Starting point are the derivatives of the loss:

\begin{align}
\begin{aligned}
&\frac12\frac{\partial^2 L}{\partial^2 W_1} = \mathbb{1}_{i,i} \otimes W_2^T W_3^T W_3 W_2 + \beta \mathbb{1}_{i,i}\otimes \mathbb{1}_{h,h}, &&\frac12\frac{\partial^2 L}{\partial W_1 \partial W_2} =(\hat{N}_{\boldsymbol{W}}^T W_3 \otimes \mathbb{1}_{h,h}) \mathcal{K}_{h,h} +\Sigma_{xx}W_1^T \otimes W_2^T W_3^T W_3, \\
&\frac12\frac{\partial^2 L}{\partial^2 W_2} = W_1 W_1^T \otimes W_3^T W_3 + \beta \mathbb{1}_{h,h}\otimes \mathbb{1}_{h,h},  &&\frac12\frac{\partial^2 L}{\partial W_1 \partial W_3} = (\hat{N}_{\boldsymbol{W}}^T \otimes W_2^T)\mathcal{K}_{o,h} + \Sigma_{xx}W_1^T W_2^T \otimes W_2^TW_3^T, \\
&\frac12\frac{\partial^2 L}{\partial^2 W_3} = W_2W_1 W_1^T W_2^T \otimes \mathbb{1}_{o,o} + \beta \mathbb{1}_{h,h} \otimes \mathbb{1}_{o,o}, &&\frac12\frac{\partial^2 L}{\partial W_2 \partial W_3} = (W_1 \hat{N}_{\boldsymbol{W}}^T \otimes \mathbb{1}_{h,h})\mathcal{K}_{o,h} + W_1W_1^T W_2^T \otimes W_3^T.
\end{aligned}
\end{align}

\noindent \textbf{Motivation ($\beta$-eigenvalues and zero eigenvalues):} Those eigenvalues stem from eigendirections in which the macrostate does not change \emph{(at least) at leading order}. From a symmetry point of view, they should be associated with infinitesimal versions of symmetry transformations like:
\begin{align}
&W^{(3)} \textcolor{orange}{W^{(2)} (\mathbb{1} + \epsilon \tilde{B})}\textcolor{blue}{(\mathbb{1} + \epsilon \tilde{A})W^{(1)}} \quad \text{and} \quad \textcolor{orange}{W^{(3)} (\mathbb{1} + \epsilon \tilde{C})}\textcolor{blue}{(\mathbb{1} + \epsilon \tilde{B})W^{(2)}}W^{(1)},
\end{align}
where $\epsilon$ is a small parameter, describing the perturbation into a certain direction in parameter space. $\tilde{A},\tilde{B},\tilde{C}$ are $d_\text{hidden} \times d_\text{hidden}$ dimensional matrices. For $\tilde{B} = -\tilde{A}$ (or: $\tilde{B} = -\tilde{C}$), such perturbations leave the macrostate invariant at leading order in $\epsilon$. Such directions correspond to (e.g.) changing $W_2$ and $W_3$, while leaving $W_1$ unchanged. Therefore, it is reasonable to study the cases, where (at least) one of the matrices $W^{(i)}$ is \underline{not} changed (simplifying the calculations). We first work out the case of $\delta W_1=0$, the case $\delta W_3 =0$ follows accordingly. \\

\noindent \textbf{Consistency Condition:} For $\delta W_1=0$ being consistent with the Hessian eigenvalue equation requires:
\begin{align}
\begin{aligned}
&(\hat{N}_{\boldsymbol{W}}^T W_3 \otimes \mathbb{1}_{h,h}) \text{vec}(\delta W_2^T) + (W_1^T \otimes W_2^TW_3^TW_3) \text{vec}(\delta W_2)\\
&+ (\hat{N}_{\boldsymbol{W}}^T  \otimes W_2^T) \text{vec}(\delta W_3^T) + (W_1^TW_2^T \otimes W_2^TW_3^T) \text{vec}(\delta W_3)\stackrel{!}{=} 0.
\label{eq:deltaW1zero}
\end{aligned}
\end{align}
In the following, we identify solutions of this equation for different \emph{sectors} defined below:
\begin{itemize}
\item \textbf{In the $r$-sector:} The $r$-sector for a given micro critical point with index set $\mathcal{S}=\mathcal{S}_r$:
\begin{align}
(W^{(1)},W^{(2)},W^{(3)}) = \left( \textcolor{orange}{V_{1,\mathcal{S}}} S_{\mathcal{S}}^{(1)} \textcolor{blue}{R_{\mathcal{S}}^T}, \textcolor{orange}{V_{2,\mathcal{S}}}S_{\mathcal{S}}^{(2)}\textcolor{orange}{V_{1,\mathcal{S}}^T}, \textcolor{blue}{L_{\mathcal{S}}}S_{\mathcal{S}}^{(3)}\textcolor{orange}{V_{2,\mathcal{S}}^T}\right)
\label{eq:AppMircoCriticalPoint}
\end{align}
is defined as directions $\delta W^{(i)}$ in the subspace defined by the finite singular values (of the critical point). Formally:
\begin{align}
\boxed{\begin{aligned}
& \textbf{definition of the $r$-sector} && \\
&(\mathbb{1}-P_{W_1})\delta W_1 =0, \quad \delta W_1 (\mathbb{1}-P_{W_1^T})=0 &&\text{with:} \quad P_{W_1^T} = \textcolor{blue}{R_{\mathcal{S}}R_{\mathcal{S}}^T}, \quad P_{W_1} = \textcolor{orange}{V_{1,\mathcal{S}}V_{1,\mathcal{S}}^T}, \\
&(\mathbb{1}-P_{W_2})\delta W_2 =0, \quad \delta W_2 (\mathbb{1}-P_{W_2^T})=0 &&\text{with:} \quad P_{W_2} = \textcolor{orange}{V_{2,\mathcal{S}}V_{2,\mathcal{S}}^T}, \quad P_{W_2^T}=\textcolor{orange}{P_{W_1}}, \\
&(\mathbb{1}-P_{W_3})\delta W_3=0, \quad \delta W_3(\mathbb{1}-P_{W_3^T})=0  &&\text{with:} \quad P_{W_3} = \textcolor{blue}{L_{\mathcal{S}}L_{\mathcal{S}}^T}, P_{W_3^T}=\textcolor{orange}{P_{W_2}}.
\label{eq:DefinitionrSector}
\end{aligned}}
\end{align}

Using $\delta W_2= \tilde{B}W_2 , \delta W_3 = W_3 \tilde{C} $, \eqref{eq:deltaW1zero} becomes more symmetric:
\begin{align*}
&(\hat{N}_{\boldsymbol{W}}^T W_3 \otimes W_2^T) \text{vec}(\tilde{B}^T) + (W_1^TW_2^T \otimes W_2^TW_3^TW_3) \text{vec}(\tilde{B})\\ 
+&(\hat{N}_{\boldsymbol{W}}^T W_3 \otimes W_2^T) \text{vec}(\tilde{C}^T) + (W_1^TW_2^T \otimes W_2^TW_3^T W_3) \text{vec}(\tilde{C})\stackrel{!}{=} 0
\end{align*}
and we can directly read off that $\tilde{B}= -\tilde{C}$ fulfills this condition. This reduces the eigenvalue equations to equations for $\text{vec}_c(\tilde{B})$ only. The reduced eigenvalue equation(s) take the form:

\begin{align*}
&\left(\beta W_2^T \otimes \mathbb{1}_{h,h} +(W_1 \hat{N}_{\boldsymbol{W}}^T W_3 \otimes \mathbb{1}_{h,h})\mathcal{K}_{h,h}\right)\text{vec}(\tilde{B})=\lambda (W_2^T \otimes \mathbb{1}_{h,h}) \text{vec}(\tilde{B}).
\end{align*}
For critical points $\boldsymbol{W}_r^*$, we have: $W_1 \hat{N}_{\boldsymbol{W}} W_3 = -\beta W_2^T $ such that:

\begin{align}
\begin{aligned}
&\beta\left(W_2^T \otimes \mathbb{1}_{h,h} +(W_2^T \otimes \mathbb{1}_{h,h})\mathcal{K}_{h,h}\right)\text{vec}(\tilde{B})=\lambda (W_2^T \otimes \mathbb{1}_{h,h}) \text{vec}(\tilde{B})\\
&\Leftrightarrow \\
&\beta W_2^T \otimes \mathbb{1}_{h,h} \text{vec}(\tilde{B}) + \beta W_2^T \otimes \mathbb{1}_{h,h}\text{vec}(\tilde{B}^T)=\lambda (W_2^T \otimes \mathbb{1}_{h,h}) \text{vec}(\tilde{B}).
\end{aligned}
\end{align}

One possible set of eigenvectors $\delta \vec{\theta}$ is based on \emph{symmetric} $\tilde{B} = \tilde{B}^T$ with $(\mathbb{1}-P_{W_2})\tilde{B}=0$ and eigenvalue(s) $\lambda = 2\beta$. Correspondingly, $\tilde{C} = -\tilde{B}$ with $\tilde{C}(\mathbb{1}-P_{W_3^T})=0$ (both matrices are symmetric and effectively $r^2$-dimensional). Overall, the corresponding eigenvectors $\delta \vec{\theta} = \text{vec}_c(\delta W_1=0,\delta W_2 =\tilde{B}W_2,\delta W_3=W_3 \tilde{C})$ give rise to a $r(r+1)/2$ dimensional subspace. Similarly, we can choose $\delta W_3=0$, giving rise to another subspace of $r(r+1)/2$ dimensions (orthogonal to the other one).\\

\noindent Another possible set of eigenvectors $\delta \vec{\theta}$ is based on $\tilde{B} = \textcolor{red}{-}\tilde{B}^T$ and $\tilde{C} =-\tilde{B}$ (both matrices are anti-symmetric and effectively $r^2$-dimensional) with Hessian eigenvalues $\lambda =0$. Note that we can interpret these as the generators of orthogonal transformations (symmetries) in line with the eigenvalues $\lambda$ being zero. Overall, those eigenvectors give rise to a $r(r-1)/2$ dimensional subspace (for $\delta W_1=0$). Similarly, we can choose $\delta W_3=0$, giving rise to another subspace of $r(r-1)/2$ dimensions (orthogonal to the other one).\\

\item \textbf{In the mixing sector:} For a critical point \eqref{eq:AppMircoCriticalPoint} with an index set $\mathcal{S}$ indicating the finite singular values, the condition \eqref{eq:deltaW1zero} can also be solved for 'mixing directions' defined as:
\begin{align}
\boxed{\begin{aligned}
&\textbf{definition of mixed sector ($\delta W_1=0$)} && \\
&P_{W_2}\delta W_2 =0, \quad \delta W_2 (\mathbb{1}-P_{W_2^T})=0 &&\text{with:} \quad P_{W_2} = \textcolor{orange}{V_{2,\mathcal{S}}V_{2,\mathcal{S}}^T} = \sum_{i\in \mathcal{S}}\textcolor{orange}{\vec{v}_{2,i}}\textcolor{orange}{\vec{v}_{1,i}^T}\\
& (\mathbb{1}-P_{W_3})\delta W_3=0, \quad \delta W_3P_{W_3^T}=0  &&\text{with:} \quad P_{W_3} = \textcolor{blue}{L_{\mathcal{S}}L_{\mathcal{S}}^T} = \sum_{i\in \mathcal{S}} \textcolor{blue}{\vec{l}_{i}} \textcolor{blue}{\vec{l}^T_{i}}
\end{aligned}
\label{eq:DefinitionMixing}}
\end{align}
where the projectors are constructed from the singular vectors of $W^{(2)}{}^*, W^{(3)}{}^*$. Directions that are mixing according to the above definition \emph{and} solve \eqref{eq:deltaW1zero} are of the form $\delta W_2 = \textcolor{orange}{\vec{v}_{2,j}}\textcolor{orange}{\vec{v}_{1,i}^T}$ with $\delta W_3= \pm \textcolor{blue}{\vec{l}_{i}}\textcolor{orange}{\vec{v}_{2,j}^T}$ for $j\notin \mathcal{S}$ and $i\in \mathcal{S}$. The Hessian eigenvalue equation(s) reduce to the following set of equations (with $\delta W_1=0$):
\begin{align}
\left(W_1 W_1^T \otimes W_3^T W_3 + \textcolor{orange}{\beta \mathbb{1}\otimes \mathbb{1}}\right) \text{vec}(\delta W_2) + \left(\textcolor{orange}{(W_1 \hat{N}_{\boldsymbol{W}}^T \otimes \mathbb{1})\mathcal{K}} + W_1W_1^T W_2^T \otimes W_3^T\right) \text{vec}(\delta W_3) = \frac12\lambda \text{vec}(\delta W_2)
\label{eq:Hessian_1_zero_23_beta}
\end{align}
and the analog for $\delta W_3$, where only the \textcolor{orange}{colored} terms contribute for mixing directions \eqref{eq:DefinitionMixing}. Therefore:
\begin{align}
\begin{aligned}
&\textbf{eigenvector} && \textbf{eigenvalue} && \textbf{numerosity}\\
&\delta \vec{\theta} = (0,\textcolor{black}{\vec{v}_{1,i}}\otimes \textcolor{black}{\vec{v}_{2,j}},  \textcolor{black}{\vec{v}_{2,j}}\otimes \textcolor{black}{\vec{l}_{i}}) && \lambda =0 && r(d_\text{hidden}-r) \\ 
&\delta \vec{\theta} = (0,\textcolor{black}{\vec{v}_{1,i}}\otimes \textcolor{black}{\vec{v}_{2,j}},  -\textcolor{black}{\vec{v}_{2,j}}\otimes \textcolor{black}{\vec{l}_{i}}) && \lambda =4\beta && r(d_\text{hidden}-r)
\end{aligned}
\end{align}
with $i\in \mathcal{S}, j \notin \mathcal{S}$. A similar structure follows for $\delta W_3=0$ instead of $\delta W_1=0$.
\end{itemize}

\noindent \textbf{Null space:} Hessian directions in the null space of $(W^{(1)}{}^*,W^{(2)}{}^*,W^{(3)}{}^*)$ have to fulfill the property:
\begin{align}
\boxed{\begin{aligned}
&\textbf{definition of null sector} && \\
&P_{W_1}\delta W_1 =0, \quad \delta W_1 P_{W_1^T}=0 &&\text{with:} \quad P_{W_1^T} = \textcolor{blue}{R_{\mathcal{S}}R_{\mathcal{S}}^T}, \quad P_{W_1} = \textcolor{orange}{V_{1,\mathcal{S}}V_{1,\mathcal{S}}^T} \\
&P_{W_2}\delta W_2 =0, \quad \delta W_2 P_{W_2^T}=0 &&\text{with:} \quad P_{W_2} = \textcolor{orange}{V_{2,\mathcal{S}}V_{2,\mathcal{S}}^T}, \quad P_{W_2^T}=\textcolor{orange}{P_{W_1}} \\
&P_{W_3}\delta W_3=0, \quad \delta W_3P_{W_3^T}=0  &&\text{with:} \quad P_{W_3} = \textcolor{blue}{L_{\mathcal{S}}L_{\mathcal{S}}^T}, P_{W_3^T}=\textcolor{orange}{P_{W_2}}
\label{eq:DefinitionNullSpaceSector}
\end{aligned}}
\end{align}
In this projected subspace, only the $2\beta \mathbb{1}\otimes \mathbb{1}$ terms in the Hessian contribute, giving rise to $\lambda=2\beta$ for any $\delta \vec{\theta} = \text{vec}_c(\delta W_1, \delta W_2,\delta W_3)$ fulfilling \eqref{eq:DefinitionNullSpaceSector}. This subspace has dimension $(d_\text{in}-r)(d_\text{hidden}-r)+(d_\text{out}-r)(d_\text{hidden}-r)+(d_\text{hidden}-r)(d_\text{hidden}-r)$.\\

\noindent \textbf{Summary $\textcolor{orange}{H_0 + \textcolor{red!70!black}{H_{\text{null}}^{(\beta)}}+ H_{4\beta}}$:} So far, we have identified $H_0$ (the zero eigenvalue subspace), $H_{4\beta}$ spanned by the directions associated with (explicitly broken) symmetries of the unregularized case and $H_{\text{null}}^{(\beta)}$, stemming from perturbations in the null space. For those parts, we get $2r^2+4r(d_\text{hidden}-r) + (d_\text{in}-r)(d_\text{hidden}-r)+(d_\text{out}-r)(d_\text{hidden}-r)+(d_\text{hidden}-r)(d_\text{hidden}-r)$ directions with eigenvalues between $0$ and $4\beta$ (referred to as $\beta$-sector and the zero sector) for a rank-$r$ critical point. This leaves  $r^2+r(d_\text{in}-r)+r(d_\text{out}-r)$ massive \emph{data-dependent} directions. From those, $r^2$ are associated with $\textcolor{blue}{H_{r^2}}$ and the rest with $\textcolor{blue}{H_{ex}}$.\\
\hrule 
\vspace{1cm}

\noindent \textbf{Massive data-dependent directions $\textcolor{blue}{H_{ex} + H_{r^2}}$:} The relevant data-dependent perturbation directions come in two flavors: (i) diagonal perturbations (changes in the singular values) and (ii) de-alignment. Both can be understood as perturbations of the form:
\begin{align}
\begin{aligned}
&W^{(1)} \to W^{(1)}(\mathbb{1} + \epsilon \tilde{A}) \quad \text{with: $\delta W_1 = W^{(1)} \tilde{A}$}, \\
&W^{(2)} \to W^{(2)}  + \epsilon\delta W_2, \\
&W^{(3)} \to (\mathbb{1}+\epsilon \tilde{C})W^{(3)} \quad \text{with: $\delta W_3 = \tilde{C} W^{(3)}$},
\end{aligned}
\end{align}
which result in changes of the macrostate.

\begin{itemize}
\item \textbf{$r$-sector $\textcolor{blue}{H_{r^2}}$:} In this sector, the overall idea is that the zero- and $\beta$-directions span a subspace of dimension $2r^2$, leaving only $r^2$ (potentially) massive directions. By constructing an (orthonormal) basis for the $r^2$-dim. complement, we obtain a reduced Hessian matrix acting only in this subspace. By diagonalizing this reduced matrix, we obtain the $r^2$ \emph{massiv data-dependent} eigenvalues. This procedure is not limited to $2$ hidden layers $(n=3)$, but we will outline the steps in detail for this case.

Basis vectors $\vec{v}$ spanning the complement to the subspace spanned by the zero and $\beta$ directions can be constructed by collecting the known eigendirections in a matrix $U$ and finding a basis for the null space of $U^T$. In the given case, this can be done by inspection and the following orthonormal vectors have this property (for $i,j \le r$):

\begin{align}
\begin{aligned}
&\text{diagonal perturbations:} \quad \vec{v}^{(ii)} =  \frac{1}{\sqrt{3}}\begin{pmatrix} \textcolor{blue}{\vec{r}_i} \otimes \vec{v}^{(1)}_i \\ \vec{v}^{(1)}_i \otimes \vec{v}^{(2)}_i \\\vec{v}^{(2)}_i \otimes \textcolor{blue}{\vec{l}_i}\end{pmatrix},\\
& \text{off-diagonal perturbations:} \quad \vec{v}^{(i\neq j)} = \frac{1}{\sqrt{\sigma_i^4+\sigma_j^4 + \sigma_i^2 \sigma_j^2}}\begin{pmatrix} (\textcolor{blue}{\mathbb{1}} \otimes W_2^2) (\textcolor{blue}{\vec{r}_j} \otimes \vec{v}^{(1)}_i )\\ (W_2 \otimes W_2)(\vec{v}^{(1)}_j \otimes \vec{v}^{(2)}_i) \\ (W^2 \otimes \textcolor{blue}{\mathbb{1}}) (\vec{v}^{(2)}_j \otimes \textcolor{blue}{\vec{l}_i})\end{pmatrix}. 
\label{eq:orthogonalbasismassivesector2hidden}
\end{aligned}
\end{align}
The part $\textcolor{blue}{H_{r^2}}$ of the Hessian only acts nontrivially in the $r^2$-dim. subspace spanned by $\{\vec{v}^{(ii)}, \vec{v}^{(i\neq j)}\}$. Using bra-ket notation, the matrix elements of $\textcolor{blue}{H_{r^2}}$ are given as: $\left(\textcolor{blue}{H_{r^2}}\right)_{ij,kl} = \langle v^{(ij)} | H |v^{(kl)}\rangle$. Due to the simplicity of the critical solutions, $\textcolor{blue}{H_{r^2}}$ in this subspace can be represented in the following form 
\begin{align*}
\textcolor{blue}{H_{r^2}}^\text{($v$ subspace)} = 2\begin{pmatrix} \textcolor{orange}{3\sigma_1^4-\beta} & & & & & &  \\ & \textcolor{orange}{3\sigma_2^4-\beta} & & & & & \\ & & \ddots & & & & \\  & & & \textcolor{orange}{3\sigma_r^4-\beta} & & & \\ & & & & \begin{pmatrix} \gamma_{12} & \delta_{12} \\ \delta_{12} & \gamma_{12} \end{pmatrix}& & \\ & & & & & \begin{pmatrix} \gamma_{13} & \delta_{13} \\ \delta_{13} & \gamma_{13} \end{pmatrix} & \\ & & & & & & \ddots \end{pmatrix}
\end{align*}
for the basis ordering $\{\textcolor{orange}{\vec{v}^{(11)}, \dots , \vec{v}^{(rr)}}, \vec{v}^{(12)}, \vec{v}^{(13)}, \dots \}$. The \textcolor{orange}{diagonal sector} decouples from the off-diagonal one. For the off-diagonal sector, we obtain in each $(i,j)$-pairing a $2\times2$-matrix $M^{(ij|r)}$. Defining  $z_{ij} := \sigma_i^4+\sigma_j^4+\sigma_i^2\sigma_j^2$, the matrix reads:
\begin{align*}
M^{(ij|r)} = 2\begin{pmatrix} z_{ij} + \beta & -\beta \cdot \frac{(\sigma_i^2 +\sigma_j^2)}{\sigma_i \sigma_j} \\ -\beta \cdot\frac{(\sigma_i^2 +\sigma_j^2)}{\sigma_i \sigma_j} & z_{ij} + \beta \end{pmatrix}:= \begin{pmatrix} \gamma_{ij} & \delta_{ij} \\ \delta_{ij} & \gamma_{ij} \end{pmatrix}.
\end{align*}
The eigenvectors in this $2\times2$ subspace are $\vec{v}_\pm = (1,\pm1)^T$ with eigenvalues 
\begin{align}
\begin{aligned}
&\frac12\lambda_+^{(ij)} = z_{ij} + \beta -\beta \cdot \frac{(\sigma_i^2 +\sigma_j^2)}{\sigma_i \sigma_j},\\
&\frac12\lambda_-^{(ij)} = z_{ij} + \beta +\beta \cdot \frac{(\sigma_i^2 +\sigma_j^2)}{\sigma_i \sigma_j}. 
\label{eq:DefLambdaplusminus2hidden}
\end{aligned}
\end{align}
Note again that formally, this subspace is spanned by $\{\vec{v}^{(ij)}, \vec{v}^{(ji)}\}$ and the eigenvectors correspond to a symmetric or anti-symmetric combination of these $\vec{v}$-vectors. In summary, we get:
\begin{align}
\begin{aligned}
&\textbf{eigenvector} && \textbf{eigenvalue} && \textbf{numerosity} \\
&\begin{pmatrix} \textcolor{blue}{\vec{r}_i} \otimes \vec{v}^{(1)}_i \\ \vec{v}^{(1)}_i \otimes \vec{v}^{(2)}_i \\\vec{v}^{(2)}_i \otimes \textcolor{blue}{\vec{l}_i}\end{pmatrix} && \lambda^{(i)} = 6\sigma_i^4-2\beta && r \\
&\begin{pmatrix} (\textcolor{blue}{\mathbb{1}} \otimes W_2^2) (\textcolor{blue}{\vec{r}_j} \otimes \vec{v}^{(1)}_i \pm \textcolor{blue}{\vec{r}_i} \otimes \vec{v}^{(1)}_j )\\ (W_2 \otimes W_2)(\vec{v}^{(1)}_j \otimes \vec{v}^{(2)}_i \pm \vec{v}^{(1)}_i \otimes \vec{v}^{(2)}_j) \\ (W^2 \otimes \textcolor{blue}{\mathbb{1}}) (\vec{v}^{(2)}_j \otimes \textcolor{blue}{\vec{l}_i} \pm \vec{v}^{(2)}_i \otimes \textcolor{blue}{\vec{l}_j})\end{pmatrix}  && \lambda^{(ij)}_\pm = z_{ij} + \beta \mp \beta \cdot \frac{(\sigma_i^2 +\sigma_j^2)}{\sigma_i \sigma_j} && r(r-1)
\end{aligned}
\end{align}
giving rise to $r^2$ eigendirections. Together with the zero and $\beta$ directions in the $r$-subspace we therefore have $3r^2$ directions in accordance with the dimensionality of the $r$-sector of the Hessian vector space. \\

\item \textbf{exchange sector $\textcolor{blue}{H_{ex}}$:} The following direction(s) can be seen as exchanging a finite singular value direction (in $\vec{\theta}^*$) with a direction where the singular value is zero - 'exchanging' the singular directions. In the mixed sector for $\delta W_2=0$, the combination $\delta W_1 = \textcolor{orange}{\vec{v}_{1,a}} \textcolor{blue}{\vec{r}_b^T}$, $\delta W_3 = \pm\textcolor{blue}{\vec{l}_b} \textcolor{orange}{\vec{v}_{2,a}^T}$ with $a\in \mathcal{S} $, $b\notin \mathcal{S}$ give rise to eigenvectors with eigenvalues $\lambda_{ab} = 2\sigma_a(\eta_a\pm \eta_b)$:
\begin{align}
\begin{aligned}
&\textbf{eigenvector} && \textbf{eigenvalue} && \textbf{numerosity} \\
&\delta \vec{\theta} = \begin{pmatrix} \textcolor{blue}{\vec{r}_b}\otimes \textcolor{orange}{\vec{v}_{1,a}} \\ \vec{0} \otimes \vec{0} \\ \pm \textcolor{orange}{\vec{v}_{2,a}} \otimes  \textcolor{blue}{\vec{l}_b^T} \end{pmatrix} && \lambda_{ab} = 2\sigma_a(\eta_a\pm \eta_b) && 2r(d_\text{min}-r)
\end{aligned}
\end{align}
with $d_\text{min}=d_\text{out}$ in our case. For the singular values being ordered $\mathcal{S} = [[1,r]]$, $\eta_b < \eta_a$ and therefore the eigenvalue is positive. In the unordered case, $\mathcal{S} \neq [[1,r]]$, there is at least one $\eta_b$ such that $\eta_b > \eta_a$ for at least one $a \in \mathcal{S}$, resulting in a negative direction. Therefore, the unordered states are strict saddles and \underline{not} stable (local) minima.

\end{itemize}

\noindent \textbf{Overall Summary (2 hidden layers):} The Hessian spectrum can be split into three main categories: (i) massive data-dependent eigenvalues; (ii) (only) $\beta$-dependent eigenvalues and (iii) zero eigenvalues.

\begin{itemize}
\item $\textcolor{blue}{H_{r^2}}, \quad \textcolor{blue}{H_{ex}}$: The massive data-dependent directions are associated with changes in the microstate $(W^{(1)},\dots, W^{(n)})$ that also change the macrostate. Practically speaking, they belong to (a1) changing the singular values; (a2) de-aligning the $\boldsymbol{W}$ from $\boldsymbol{W}^*$ as well as (b) swapping a finite with a zero singular value.

\item $\textcolor{orange}{H_0}$ (symmetries): The zero eigenvalues correspond to those rotations that lead to different microstates belonging to the same macrostate.

\item $\textcolor{orange}{H_{4\beta},\quad \textcolor{red!70!black}{H_{\text{null}}^{(\beta)}}}$: The $\beta$-directions belong to (a) the explicitly broken part of the GL inner symmetries (forming a basin for the $0$-balanced solutions) and (b) perturbations in the Null space of the solution. Both perturbations leave the macrostate invariant \emph{at leading order}.

\end{itemize}

\subsection{Deeper Networks: $n>2$ Layers}
The strategy for $n=3$ ($2$ hidden layers) can be extended to arbitrary depth. The main point is that the flat directions and $\beta$-directions do not depend on the model details as they are associated with (broken) symmetry transformations. In the general case, we have:

\begin{align}
\begin{aligned}
&\text{for $j<i$:} \quad \frac12 \frac{\partial^2 \mathcal{L}}{\partial W^{(i)} \partial W^{(j)}} &&= \left((W^{(j+1)}{}^T \cdots W^{(i-1)}{}^T \otimes W^{(i+1)}{}^T\cdots W^{(n)}{}^T \hat{N} W^{(1)}{}^T \cdots W^{(j-1)}{}^T\right) \mathcal{K} \\
&&&+ W^{(j-1)} \cdots W^{(1)} \Sigma_{xx} W^{(1)}{}^T \cdots W^{(i-1)}{}^T \otimes W^{(i+1)}{}^T \cdots W^{(n)}{}^T W^{(n)} \cdots W^{(j+1)}, \\
&\text{for $j=i$:} \quad \frac12 \frac{\partial^2 \mathcal{L}}{\partial W^{(j)} \partial W^{(j)}} &&= W^{(j-1)}\cdots W^{(1)} \Sigma_{xx} W^{(1)}{}^T \cdots W^{(j-1)}{}^T \otimes W^{(j+1)}{}^T \cdots W^{(n)}{}^T W^{(n)} \cdots W^{(j+1)} + \beta \mathbb{1}\otimes \mathbb{1}, \\
&\text{for $j>i$:} \quad \frac12 \frac{\partial^2 \mathcal{L}}{\partial W^{(i)} \partial W^{(j)}} &&= \left(\left(W^{(j+1)}{}^T \cdots W^{(n)}{}^T \hat{N} W^{(1)}{}^T \cdots W^{(i-1)}{}^T\right)^T \otimes W^{(i+1)}{}^T \cdots W^{(j-1)}{}^T \right) \mathcal{K}\\
&&& + \left(W^{(j-1)} \cdots W^{(1)} \Sigma_{xx} W^{(1)}{}^T \cdots W^{(i-1)}{}^T \right)^T \otimes W^{(i+1)}{}^T \cdots W^{(n)}{}^T W^{(n)} \cdots W^{(j+1)}.
\label{eq:secondderivativesarbitrarydepth}
\end{aligned}
\end{align}
The spectrum of the Hessian at the critical points is summarized in Tab.~\ref{tab:HessianSpectrumgeneral}.\\ \\

\noindent \textbf{Symmetry-based} $\textcolor{orange}{H_0}$, $\textcolor{orange}{H_{4\beta}}$: The flat eigendirections and $\beta$-directions follow the same derivation and (symmetry-based) logic as before: between any adjacent weight matrices $W^{(i+1)}$ and $W^{(i)}$ perturbations of the form:
\begin{align}
W^{(n)} \cdots W^{(i+1)}(\mathbb{1}+\epsilon \tilde{B})(\mathbb{1}+\epsilon \tilde{A})W^{(i)}\cdots W^{(1)}
\end{align} 
give rise to either flat directions (associated with orthogonal rotations) or $4\beta$ directions (associated with explicitly symmetry-broken transformations). Further flat or $4\beta$ directions stem from mixing perturbations between two adjacent weight matrices with eigenvectors constructed from ($a \in \mathcal{S}, b \notin \mathcal{S}$):
\begin{align}
&(i) &&\delta W_1 = \textcolor{orange}{\vec{v}_b^{(1)}} \textcolor{blue}{\vec{r}_a^T}, \quad \delta W_{2} = \pm \textcolor{orange}{\vec{v}_a^{(2)} \vec{v}_b^{(1)}{}^T}, \quad \delta W_{l}=0 \quad l\neq 1,2, \\
&(ii) &&\delta W_j = \textcolor{orange}{\vec{v}_b^{(j)} \vec{v}_a^{(j-1)}{}^T}, \quad \delta W_{j+1} = \pm \textcolor{orange}{\vec{v}_a^{(j+1)} \vec{v}_b^{(j)}{}^T}, \quad \delta W_{l}=0 \quad l\neq j,j+1, \\
&(iii)&&\delta W_{n-1} = \textcolor{orange}{\vec{v}_b^{(n-1)} \vec{v}_a^{(n-2)}{}^T}, \quad \delta W_{n} = \pm \textcolor{blue}{\vec{l}_a} \textcolor{orange}{\vec{v}_b^{(n-1)}{}^T}, \quad \delta W_{l}=0 \quad l\neq j,j+1,
\end{align}
giving rise to $(n-1)\cdot r(d_\text{hidden}-r)$ eigenvectors for $\lambda=0$ and the same number for $\lambda=4\beta$. The proof follows by inspection of \eqref{eq:secondderivativesarbitrarydepth} and is based on the identity for $i=j+1$ (valid for $W_i$ being critical points):
\begin{align}
&(\mathbb{1} \otimes W_{i+1}^T \cdots W_n^T \hat{N}_{\boldsymbol{W}} W_1^T \cdots W_{i-2}^T)\mathcal{K} (\vec{v}_a^{(i-2)} \otimes \vec{v}_b^{(i-1)}) = -\beta (\vec{v}_b^{(i-1)} \otimes \vec{v}_a^{(i)}), \\
&(\vec{v}_b^{(i-1)} \otimes \vec{v}_a^{(i)}) = \text{vec}_c(\delta W_i).
\end{align}
\textbf{Null sector $\textcolor{red!70!black}{H_{\text{null}}^{(\beta)}}$:} Qualitatively speaking, only for $j=i\pm1$ the $\mathcal{K}$-terms in \eqref{eq:secondderivativesarbitrarydepth} feature a $\mathbb{1}$-term which does \emph{not} act as a projector onto the finite singular value directions. As for 2 hidden layers, any set of matrices $\delta W_i$ with the property $P_{W_i}\delta W_i=0$ and $\delta W_iP_{W_i^T}=0$ corresponds - after vectorization - to eigenvectors of $H$ with eigenvalue $\lambda=2\beta$. \\

\noindent \textbf{Massive data-dependent sector $\textcolor{blue}{H_{r^2}}$:} The remaining subspace is spanned by basis vectors that are orthogonal to all eigenvectors of the $0-$ and $\beta-$sectors. The following set of normalized vectors with $i,j \in \mathcal{S}$ forms a basis for the 'massive' $r^2$-dim. subspace that is orthogonal to the 'flat' and $\beta$-subspaces in direct generalization of \eqref{eq:orthogonalbasismassivesector2hidden}:

\begin{align}
\vec{a}_{lm} = \frac{1}{\sum_{k=1}^n \sigma_l^{2(k-1)} \sigma_m^{2(n-k)}}\begin{pmatrix} \sigma_m^{n-1} \textcolor{blue}{\vec{r}_l} \otimes \textcolor{orange}{\vec{v}^{(1)}_m} \\ \sigma_m^{n-1} \textcolor{orange}{\vec{v}^{(1)}_l} \otimes \textcolor{orange}{\vec{v}^{(2)}_m} \\ \vdots \\  \sigma_m^{n-j} \sigma_l^{j-1} \textcolor{orange}{\vec{v}_l^{(j)}} \otimes \textcolor{orange}{\vec{v}_m^{(j+1)}} \\ \vdots \\ \sigma_l^{n-1} \textcolor{orange}{\vec{v}_l^{(n)}} \otimes \textcolor{blue}{\vec{l}_m} \end{pmatrix} \quad \stackrel{\text{rotated frame}}{\rightarrow} \quad \frac{1}{\sum_{k=1}^n \sigma_l^{2(k-1)} \sigma_m^{2(n-k)}}\begin{pmatrix} \sigma_m^{n-1} \vec{e}_l \otimes \vec{e}_m \\ \vdots \\  \sigma_m^{n-j} \sigma_l^{j-1} \vec{e}_l \otimes \vec{e}_m \\ \vdots \\ \sigma_l^{n-1} \vec{e}_l \otimes \vec{e}_m \end{pmatrix}.
\end{align}
The Hessian in this $r^2$-dim. subspace can again be decomposed into $1$ and $2$-dim. subspaces (here assuming $\Sigma_{xx}=\mathbb{1}$), as the sets $\{\vec{a}_{ij} , \vec{a}_{ji}\}$ are invariant under $H$. For $l=m$, we have:
\begin{align}
H\vec{a}_{ll} = 2(n \sigma_l^{2n-2}+(2-n)\beta)\vec{a}_{ll}, \quad \lambda_l = 2(n \sigma_l^{2n-2}+(2-n)\beta).
\end{align}
In the subspace $\{\vec{a}_{i\neq j} , \vec{a}_{j \neq i}\}$ the Hessian takes the form $M^{(ij|r)}$ with:
\begin{align}
M^{(ij|r)} = 2\begin{pmatrix} \frac{\sigma_i^{2n}-\sigma_j^{2n}}{\sigma_i^2 -\sigma_j^2} + \beta & -\beta \frac{\sigma_j^{2-n}\sigma_i^n - \sigma_i^{2-n}\sigma_j^2}{\sigma_i^2 -\sigma_j^2} \\  -\beta \frac{\sigma_j^{2-n}\sigma_i^n - \sigma_i^{2-n}\sigma_j^2}{\sigma_i^2 -\sigma_j^2} & \frac{\sigma_i^{2n}-\sigma_j^{2n}}{\sigma_i^2 -\sigma_j^2} +\beta \end{pmatrix}.
\label{eq:HessianDeepNetwork2x2general}
\end{align} 
To show this, we have to evaluate (using $(\hat{N}_{\boldsymbol{W}} \textcolor{blue}{\vec{r}_i}) \otimes \vec{v}_j = -\beta \sigma_i^{2-n}\textcolor{blue}{\vec{l}_i} \otimes \vec{v}_j$ for $i\in \mathcal{S}$ at the critical points):
\begin{align}
\frac12H\vec{a}_{lm} = \left(\sum_{j=1}^n \sigma_l^{2(j-1)} \sigma_m^{2(n-j)} + \beta \right)\vec{a}_{lm} -\beta \left(\sum_{j=1}^{i-1}(\sigma_m^{i-j-1}\sigma_l^{j-i+1}) \sigma_m^{n-j}\sigma_l^{j-1} + \sum_{j=i+1}^n (\sigma_m^{i-j+1}\sigma_l^{j-i-1}) \sigma_m^{n-j}\sigma_l^{j-1}\right) \begin{pmatrix} \vdots \\ \vec{e}_m \otimes \vec{e}_l \\ \vdots \end{pmatrix},
\end{align}
where the second contribution stems from the terms with $\mathcal{K}$ in the Hessian (note that we are working in the rotated (`covariance diagonalized') frame now). In the first term, we have
\begin{align}
\sum_{j=1}^n \sigma_l^{2(j-1)} \sigma_m^{2(n-j)} +\beta = \sum_{j=0}^{n-1} \sigma_l^{2j} \sigma_m^{2(n-j-1)} +\beta = \sigma_m^{2(n-1)} \left(\frac{1- (\sigma_l^2 \sigma_m^{-2})^n}{1-\sigma_l^2 \sigma_m^{-2}}\right) +\beta = \frac{\sigma_m^{2n} -\sigma_l^{2n}}{\sigma_m^2 -\sigma_l^2}+\beta,
\end{align}
giving rise to the diagonal terms in $M^{(ij|r)}$. For the off-diagonal terms, we have to calculate multiple geometric series with 
\begin{align}
&\begin{aligned}
\textbf{(i)} \quad \sum_{j=1}^{i-1}(\sigma_m^{i-j-1}\sigma_l^{j-i+1}) \sigma_m^{n-j}\sigma_l^{j-1} &= \sigma_m^{i-1} \sigma_l^{n-i} \sigma_m^n \sigma_l^{-n} \sum_{j=1}^{i-1} (\sigma_m^{-2}\sigma_l^2)^j =  \sigma_m^{i-1} \sigma_l^{n-i} \sigma_m^{n-2} \sigma_l^{2-n} \sum_{j=0}^{i-2} (\sigma_m^{-2}\sigma_l^2)^j \\
&=  \sigma_m^{i-1} \sigma_l^{n-i} \left(\frac{1- (\sigma_m^{-2}\sigma_l^2)^{i-1}}{1-\sigma_m^{-2}\sigma_l^2}\right) =  \textcolor{blue}{\sigma_m^{i-1} \sigma_l^{n-i}} \left( \frac{\sigma_m^n\sigma_l^{2-n} \textcolor{red}{- \sigma_m^{n-2i+2}\sigma_l^{2i-n}}}{\sigma_m^2-\sigma_l^2}\right),
\end{aligned}\\
&\begin{aligned}
\textbf{(ii)} \quad \sum_{j=i+1}^n (\sigma_m^{i-j+1}\sigma_l^{j-i-1}) \sigma_m^{n-j}\sigma_l^{j-1} &= \sigma_m^{i-1} \sigma_l^{n-i} \sum_{j'=0}^{n-(i+1)} \sigma_m^{n-2(j'+i+1)+2}\sigma_l^{2(j'+i+1)-n-2} \\
& = \sigma_m^{i-1}\sigma_l^{n-i}\sigma_m^{n-2i}\sigma_l^{2i-n} \sum_{j'=0} (\sigma_m^{-2}\sigma_l^2)^{j'} = \sigma_m^{i-1}\sigma_l^{n-i}\sigma_m^{n-2i}\sigma_l^{2i-n} \left( \frac{1- \sigma_m^{-2}\sigma_l^2)^{n-i}}{1-\sigma_m^{-2}\sigma_l^2}\right)\\
& = \textcolor{blue}{\sigma_m^{i-1}\sigma_l^{n-i}} \frac{\textcolor{red}{\sigma_m^{n-2i+2}\sigma_l^{2i-n}}-\sigma_m^{2-n}\sigma_l^n}{\sigma_m^2-\sigma_l^2}.
\end{aligned}
\end{align}
The red-colored terms drop out of the overall sum, leaving us in the rotated frame with:
\begin{align}
\frac12H\vec{a}_{lm} = \left(\frac{\sigma_m^{2n} -\sigma_l^{2n}}{\sigma_m^2 -\sigma_l^2} +\beta \right)\vec{a}_{lm} -\beta \frac{\sigma_m^{2-n}\sigma_l^n - \sigma_l^{2-n}\sigma_j^m}{\sigma_m^2 -\sigma_l^2}  \underbrace{\begin{pmatrix} \vdots \\ \textcolor{blue}{\sigma_l^{n-i}\sigma_m^{i-1}}\vec{e}_m \otimes \vec{e}_l \\ \vdots \end{pmatrix}}_{=\vec{a}_{ml}},
\end{align}
which shows that the subspace $\{\vec{a}_{ij} , \vec{a}_{ji}\}$ is indeed invariant under $H$ with $M^{(ij|r)}$ in \eqref{eq:HessianDeepNetwork2x2general} describing the action of $H$ on this subspace. The corresponding eigenvalues $\lambda^{(ij)}_\pm$ of the Hessian are the eigenvalues of \eqref{eq:HessianDeepNetwork2x2general}.\\

\noindent \textbf{Exchange Sector $\textcolor{blue}{H_{ex}}$:} The exchange sector is based on 'exchanging' one learned singular direction (of the covariance matrix) with another (not learned) one. The corresponding direction in parameter space is given by an (infinitesimal) change of the \textcolor{blue}{outer singular vectors} (here for $\Sigma_{xx} = \mathbb{1}$):

\begin{align}
\delta \vec{\theta}_{lm}^{(ex)} = \begin{pmatrix} \textcolor{blue}{\vec{r}_l} \otimes \vec{v}_m^{(1)} \\ 0 \\\vdots \\ 0 \\ \pm \vec{v}_m^{(n-1)} \otimes \textcolor{blue}{\vec{l}_l} \end{pmatrix}
\end{align}
with $l \notin \mathcal{S}$ and $m\in \mathcal{S}$ such that:
\begin{align}
H \delta \vec{\theta}_{lm}^{(ex)} = \textcolor{red!70!black}{(\sigma_m^{2(n-1)} + \beta \mp \eta_l\sigma_m^{n-2})}\delta \vec{\theta}_{lm}^{(ex)}=\textcolor{red!70!black}{2\sigma_m^{n-2}(\eta_m \pm \eta_l)} \delta \vec{\theta}_{lm}^{(ex)}.
\end{align}
For more general $\Sigma_{xx}$ (that are still compatible with $\Sigma_{yx}$), the two finite entries of the eigenvector have to be rescaled and the eigenvalues do change as well. In case of $d_\text{in} > d_\text{out}$, we additionally have the eigenvectors for $l>d_{\text{in}}$ and $m \in \mathcal{S}$ with massive data-dependent eigenvalues:
\begin{align}
\delta \vec{\theta}_{lm}^{(ex)} = \begin{pmatrix} \textcolor{blue}{\vec{r}_l} \otimes \vec{v}_m^{(1)} \\ 0 \\\vdots \\ 0 \\ 0 \end{pmatrix}, \quad \lambda = 2\sigma_m^{n-2}\eta_m.
\end{align}

\noindent \textbf{Overall Summary:}

\begin{table}[h]
\centering
\begin{tabular}{l | l | l| l | l}
\textbf{perturbation} & \textbf{eigenvalue} & \textbf{number of eigenvalues} & total \\ \hline
 \multicolumn{3}{l}{\textbf{ \textsf{massive sector}}} & $r^2+r(d_\text{in}-r)+r(d_\text{out}-r)$ \\ \hline
diagonal & $2(n \sigma_l^{2n-2}+(2-n)\beta)$ & $r$  \\
off-diagonal & $\lambda^{(ij)}_+$, $\lambda^{(ij)}_-$ & $ r(r-1)$  \\
exchange & $2\sigma_a^{n-2}(\eta_a \pm \eta_l)$ & $2r(d_\text{min}-r)$ \\ 
exchange & $2\sigma_a^{n-2}\eta_a$ & $r(d_\text{in}-d_\text{out})$ \\\hline
 \multicolumn{3}{l}{\textbf{\textsf{$\beta$-sector (null space)}}} & \\ \hline
null sector & $2\beta$ & $(d_\text{in}+d_\text{out}+(n-2)\cdot d_\text{hidden}-n\cdot r)(d_\text{hidden}-r)$&  \\ \hline
 \multicolumn{3}{l}{\textbf{\textsf{$\beta$-sector (explicitly broken symmetry)}}} &  $\frac{(n-1)}{2}\cdot r(r+1) + (n-1)\cdot r(d_\text{hidden}-r)$ \\ \hline
$r$-sector & $4\beta$ & $(n-1)\cdot\frac12 r(r+1)$ \\
mixing & $4\beta$ & $(n-1)\cdot r(d_\text{hidden}-r)$ \\ \hline
\multicolumn{3}{l}{\textbf{\textsf{zero eigenvalues (from symmetries)}}} &  $\frac{(n-1)}{2} \cdot r(r-1) + (n-1)\cdot r(d_\text{hidden}-r)$ \\ \hline
$r$-sector & $0$ & $(n-1)\cdot\frac12 r(r-1)$ \\
mixing & $0$ & $(n-1)\cdot r(d_\text{hidden}-r)$
\end{tabular}
\caption{Overview of the Hessian spectrum at a critical point with rank $r$ for a $n>2$ layer networks for $\Sigma_{xx}=\mathbb{1}$. The eigenvectors are defined in the text.}
\label{tab:HessianSpectrumgeneral}
\end{table}

\section{Macroperspective in the $0$-balance sector}
The macro dynamics can be described as a purely macro theory in the $0$-balanced subspace \eqref{eq:MacroDynamicsAppendix}. From the macro-dynamics, we can as well extract certain properties of the microscopic Hessian based on the macroscopic Jacobian: $\mathcal{J} \delta \bar{\boldsymbol{W}} = \lambda^{(\mathcal{J})}\delta \bar{\boldsymbol{W}}$ (see \eqref{eq:JacobianDerivatives}).

\subsection{Jacobian} 
Due to the restriction to this subspace, we will not be able to extract all information about, e.g., the Hessian from this dynamics. However, we should get access to some of the eigenvectors and eigenvalues (namely those that do not break the $0$-balance condition). The Jacobian is constructed from the following matrix derivatives from \eqref{eq:MacroDynamicsAppendix}:

\begin{align}
\mathcal{J} = 2\begin{pmatrix}
\frac{\partial \vec{g}_{\textcolor{blue_L}{\mathfrak{W}_L}}}{\partial \textcolor{blue_L}{\textcolor{blue_L}{\mathfrak{W}_L}}} & \frac{\partial \vec{g}_{\textcolor{blue_L}{\mathfrak{W}_L}}}{\partial \boldsymbol{W}} & \frac{\partial \vec{g}_{\textcolor{blue_L}{\mathfrak{W}_L}}}{\partial \textcolor{blue_R}{\textcolor{blue_R}{\mathfrak{W}_R}}} \\
 \frac{\partial \vec{g}_{\boldsymbol{W}}}{\partial \textcolor{blue_L}{\textcolor{blue_L}{\mathfrak{W}_L}}} & \frac{\partial \vec{g}_{\boldsymbol{W}}}{\partial \boldsymbol{W}} & \frac{\partial \vec{g}_{\boldsymbol{W}}}{\partial \textcolor{blue_R}{\textcolor{blue_R}{\mathfrak{W}_R}}} \\
 \frac{\partial \vec{g}_{\textcolor{blue_R}{\mathfrak{W}_R}}}{\partial \textcolor{blue_L}{\textcolor{blue_L}{\mathfrak{W}_L}}} & \frac{\partial \vec{g}_{\textcolor{blue_R}{\mathfrak{W}_R}}}{\partial \boldsymbol{W}} & \frac{\partial \vec{g}_{\textcolor{blue_R}{\mathfrak{W}_R}}}{\partial \textcolor{blue_R}{\textcolor{blue_R}{\mathfrak{W}_R}}}
\end{pmatrix}  \begin{cases} &\begin{aligned}
&- \frac{\partial \vec{g}_{\boldsymbol{W}}}{\partial \textcolor{blue_R}{\mathfrak{W}_R}} = 
\sum_{l=0}^{n-2} \sum_{k=0}^{n-2-l} (\textcolor{blue_R}{\mathfrak{W}_R^T})^k \otimes \textcolor{blue_L}{\mathfrak{W}_L}^l\hat{N} (\textcolor{blue_R}{\mathfrak{W}_R})^{n-2-l-k}, \\
&-\frac{\partial \vec{g}_{\boldsymbol{W}}}{\partial \boldsymbol{W}} = n\beta \mathbb{1}_{i,i} \otimes \mathbb{1}_{o,o} + \sum_{k=0}^{n-1} (\Sigma_{xx} \textcolor{blue_R}{\mathfrak{W}_R}^k)^T \otimes (\textcolor{blue_L}{\mathfrak{W}_L})^{n-1-k},\\
&-\frac{\partial \vec{g}_{\boldsymbol{W}}}{\partial \textcolor{blue_L}{\textcolor{blue_L}{\mathfrak{W}_L}}} = \sum_{l=0}^{n-2} \sum_{k=0}^{n-2-l} (\textcolor{blue_L}{\mathfrak{W}_L}^{n-2-l-k} \hat{N}\textcolor{blue_R}{\mathfrak{W}_R}^l) \otimes (\textcolor{blue_L}{\mathfrak{W}_L})^K,
\end{aligned}\\
&\begin{aligned}\\
&-\frac{\partial \vec{g}_{\textcolor{blue_R}{\mathfrak{W}_R}}}{\partial \textcolor{blue_R}{\mathfrak{W}_R}} = 2\beta \mathbb{1}_{o,o} \otimes \mathbb{1}_{o,o}, \\
&-\frac{\partial \vec{g}_{\textcolor{blue_R}{\mathfrak{W}_R}}}{\partial \boldsymbol{W}} = (\boldsymbol{W}^T \otimes \Sigma_{xx})\mathcal{K}_{o,i} + \Sigma_{xx} \otimes \boldsymbol{W}^T  + \mathbb{1}_{o,o} \otimes \hat{N}^T + (\hat{N}^T \otimes \mathbb{1}_{i,i})\mathcal{K}_{o,i},\\
&-\frac{\partial \vec{g}_{\textcolor{blue_R}{\mathfrak{W}_R}}}{\partial \textcolor{blue_L}{\mathfrak{W}_L}} = 0,
\end{aligned} \\
&\begin{aligned}\\
&-\frac{\partial \vec{g}_{\textcolor{blue_L}{\mathfrak{W}_L}}}{\partial \textcolor{blue_R}{\mathfrak{W}_R}} = 0, \\
&-\frac{\partial \vec{g}_{\textcolor{blue_L}{\mathfrak{W}_L}}}{\partial \boldsymbol{W}} = \boldsymbol{W}\Sigma_{xx}\otimes \mathbb{1}_{i,i} +(\mathbb{1}_{o,o} \otimes \hat{N})\mathcal{K}_{o,i} + \hat{N}\otimes \mathbb{1} +(\mathbb{1}_{o,o} \otimes \boldsymbol{W} \Sigma_{xx})\mathcal{K}_{o,i}, \\
&-\frac{\partial \vec{g}_{\textcolor{blue_L}{\mathfrak{W}_L}}}{\partial \textcolor{blue_L}{\mathfrak{W}_L}} = 2\beta \mathbb{1}_{o,o} \otimes \mathbb{1}_{o,o}.
\end{aligned} \end{cases}
\label{eq:JacobianDerivatives}
\end{align}

\subsection{1 hidden Layer} 
In principle, \eqref{eq:JacobianDerivatives} can be further evaluated for arbitrary depth. The special case of 1 hidden layer ($n=2$) instead can also be approached differently by making use of the Riccati equation. For $d_\text{in}=d_\text{out}=d_\text{hidden}=d$ and $\Sigma_{xx} = \mathbb{1}$, the macro dynamics can be written in terms of a matrix Riccati differential equation \eqref{eq:RiccatiEquation} (for the derivation see App.~\ref{App:AlgebraicScaling}) for $M=QQ^T$:
\begin{align}
 \frac{d}{dt} M(t) = \mathcal{G}(M(t)):= 2F M(t) + 2M(t) F -2M^2(t), 
\end{align}
where $F$ is defined in \eqref{eq:RiccatiEquation}. Working with the macrovariable $M(t)$ directly (collecting $\textcolor{blue_L}{\mathfrak{W}_L},\textcolor{blue_R}{\mathfrak{W}_R}, \boldsymbol{W},\boldsymbol{W}^T$), we can work out the Jacobian for $M$: $\mathcal{J}'=\mathcal{J}'(M)$ according to:
\begin{align}
\mathcal{J}'(M)= \frac{\partial \mathcal{G}}{\partial M} = 2\left(\mathbb{1} \otimes F + F \otimes \mathbb{1} - M \otimes \mathbb{1} -\mathbb{1} \otimes M \right).
\end{align}
Making use of symmetries and the alignment condition, $M$ takes the following form at a critical point in the frame of reference where $\Sigma_{yx}$ is diagonal:
\begin{align*}
M^* = \begin{pmatrix} \textcolor{blue_R}{\mathfrak{W}_R}{}^* & \boldsymbol{W}^T{}^* \\  \boldsymbol{W}^* & \textcolor{blue_L}{\mathfrak{W}_L}{}^* \end{pmatrix} = \begin{pmatrix}  \hat{P}_\mathcal{S} \tilde{\Sigma} \hat{P}_\mathcal{S} &  \hat{P}_\mathcal{S} \tilde{\Sigma} \hat{P}_\mathcal{S} \\  \hat{P}_\mathcal{S} \tilde{\Sigma} \hat{P}_\mathcal{S} &  \hat{P}_\mathcal{S} \tilde{\Sigma} \hat{P}_\mathcal{S} \end{pmatrix}, \quad \tilde{\Sigma} = \sum_{i\in \mathcal{S}} \sigma_i^2 \vec{e}_i \vec{e}_i^T ,
\end{align*}
where $\hat{P}_\mathcal{S} = \sum_{i\in \mathcal{S}} \vec{e}_i \vec{e}_i^T $ projects onto the subspace defined by $\mathcal{S}$. Therefore, the Jacobian at $M^*$ reads (with $\hat{P}_\perp = \mathbb{1}-\hat{P}_{\mathcal{S}}$):
\begin{align*}
 \mathcal{J}'(M^*) = 2\left(\mathbb{1} \otimes \begin{pmatrix} - \hat{P}_{\mathcal{S}} \tilde{\Sigma} \hat{P}_{\mathcal{S}} & \hat{P}_\perp \tilde{\Sigma} \hat{P}_\perp \\ \hat{P}_\perp \tilde{\Sigma} \hat{P}_\perp & - \hat{P}_{\mathcal{S}} \tilde{\Sigma} \hat{P}_{\mathcal{S}} \end{pmatrix} + \begin{pmatrix} - \hat{P}_{\mathcal{S}} \tilde{\Sigma} \hat{P}_{\mathcal{S}} & \hat{P}_\perp \tilde{\Sigma} \hat{P}_\perp \\ \hat{P}_\perp \tilde{\Sigma} \hat{P}_\perp & - \hat{P}_{\mathcal{S}} \tilde{\Sigma} \hat{P}_{\mathcal{S}} \end{pmatrix} \otimes \mathbb{1}\right).
\end{align*}

Since this is a Kronecker sum, the eigenvectors of the Jacobian are simply given by combinations of:
\begin{align*}
\left\{ \hat{P}_{\mathcal{S}} \begin{pmatrix} \vec{e}_j \\ \vec{e}_j \end{pmatrix}, \hat{P}_\perp \begin{pmatrix} \vec{e}_j \\ \vec{e}_j \end{pmatrix}, \hat{P}_\perp \begin{pmatrix} \vec{e}_j \\ -\vec{e}_j \end{pmatrix}\right\} \otimes \left\{ \hat{P}_{\mathcal{S}} \begin{pmatrix} \vec{e}_l \\ \vec{e}_l \end{pmatrix}, \hat{P}_\perp \begin{pmatrix} \vec{e}_l \\ \vec{e}_l \end{pmatrix}, \hat{P}_\perp \begin{pmatrix} \vec{e}_l \\ -\vec{e}_l \end{pmatrix}\right\}
\end{align*}
with $j,l \in [1,d]$. But note that $\hat{P}_\perp (\vec{e}_j ,-\vec{e}_j)^T$ is not compatible with the $0$-balanced condition. Therefore we do not include this term for the analysis on the $0$-balanced subspace. For the remaining combinations we get the following eigenvalues $\lambda^{(\mathcal{J})}$ for $\mathcal{S}=[[1,r]]$:
\begin{itemize}
\item \textbf{$r$-sector} ($j,l \in \mathcal{S}$): $\frac12\lambda^{(\mathcal{J})} = -(\eta_j+\eta_l-2\beta) = -(\sigma_j^2+\sigma_l^2) \,\ (<0)$,
\item \textbf{mixing-sector} ($j\in\mathcal{S}$ and $l\notin \mathcal{S}$): $\frac12\lambda^{(\mathcal{J})} = \eta_j-\eta_l \,\ (<0)$,
\item \textbf{zero-sector} ($j,l\notin \mathcal{S}$): $\frac12\lambda^{(\mathcal{J})} = \eta_j+\eta_l-2\beta \,\ (<0)$.
\end{itemize}
In the $r$-sector as well as the mixing sector we obtain exactly the eigenvalues of the microscopic Hessian (for the $0$-balance preserving perturbations), see Tab.~\ref{tab:HessianSpectrum1hidden}. \\

\subsection{2 hidden Layers ($n=3$) \label{App:Jacobian2Hidden}} 
In case of two hidden layers, the Jacobian is constructed from the derivatives, evaluated at the critical points:
\begin{align}
&\begin{aligned}
&- \frac{\partial \vec{g}_{\boldsymbol{W}}}{\partial \textcolor{blue_R}{\mathfrak{W}_R}} = \textcolor{blue_R}{\mathfrak{W}_R}^T \otimes \hat{N} + \mathbb{1}_{i,i} \otimes \hat{N}\textcolor{blue_R}{\mathfrak{W}_R} + \mathbb{1}_{i,i} \otimes \textcolor{blue_L}{\mathfrak{W}_L}\hat{N} 
= R S_{i,i}^2 R^T \otimes (L S_{o,i}^3 R^T - \Sigma_{yx})-2\beta  \mathbb{1}_{i,i} \otimes L S_{o,i}^2 R, \\
&= -\beta RS^2 R^T \otimes L S^{-1} R - RS^2 R^T \otimes \hat{P}_{o,\perp} \Sigma_{yx}  \hat{P}_{i,\perp} - 2\beta \mathbb{1}_{i,i} \otimes L S_{o,i}^2 R, \\
&-\frac{\partial \vec{g}_{\boldsymbol{W}}}{\partial \boldsymbol{W}} = (\Sigma_{xx} \textcolor{blue_R}{\mathfrak{W}_R}^2)^T \otimes \mathbb{1}_{o,o} + (\Sigma_{xx} \textcolor{blue_R}{\mathfrak{W}_R})^T \otimes \textcolor{blue_L}{\mathfrak{W}_L} + \Sigma_{xx} \otimes \textcolor{blue_L}{\mathfrak{W}_L}^2 + 3\beta \mathbb{1}_{i,i} \otimes \mathbb{1}_{o,o}\\ 
& = R S_{r,r}^4 R \otimes \mathbb{1}_{o,o} + R S_{i,i}^2 R^T \otimes L S_{o,o}^2 L^T + \mathbb{1}_{i,i} \otimes L S_{o,o}^4 L^T + 3\beta \mathbb{1}_{i,i} \otimes \mathbb{1}_{o,o}, \\
&-\frac{\partial \vec{g}_{\boldsymbol{W}}}{\partial \textcolor{blue_L}{\textcolor{blue_L}{\mathfrak{W}_L}}} = (\hat{N}\textcolor{blue_R}{\mathfrak{W}_R})^T \otimes \mathbb{1} + (\textcolor{blue_L}{\mathfrak{W}_L}\hat{N})^T \otimes \mathbb{1}+ \hat{N}^T \otimes \textcolor{blue_L}{\mathfrak{W}_L} \\ 
& = -2\beta R S_{i,o}L^T \otimes \mathbb{1}_{o,o}  - \beta R  \hat{P}_{i} S^{-1} \hat{P}_{o}  L^T \otimes L S_{o,o}^2 L^T    -\hat{P}_{o,\perp} \Sigma_{yx}^T  \hat{P}_{i,\perp} \otimes L S_{o,o}^2 L^T,
\end{aligned} \\
\hline
&\begin{aligned}
&-\frac{\partial \vec{g}_{\textcolor{blue_R}{\mathfrak{W}_R}}}{\partial \textcolor{blue_R}{\mathfrak{W}_R}} = 2\beta \mathbb{1}_{o,o} \otimes \mathbb{1}_{o,o}, \\
&-\frac{\partial \vec{g}_{\textcolor{blue_R}{\mathfrak{W}_R}}}{\partial \boldsymbol{W}} = (\boldsymbol{W}^T \otimes \Sigma_{xx})\mathcal{K}_{o,i} + \Sigma_{xx} \otimes \boldsymbol{W}^T  + \mathbb{1}_{i,i} \otimes \hat{N}^T + (\hat{N}^T \otimes \mathbb{1}_{i,i})\mathcal{K}_{o,i} \\
& = (R S_{i,o}^3 L^T \otimes \mathbb{1}_{i,i}) \mathcal{K}_{o,i} + \mathbb{1}_{i,i} \otimes R S_{i,o}^3 L^T - \beta  \mathbb{1}_{i,i} \otimes R_r S^{-1} L_r^T - \mathbb{1}_{i,i} \otimes \hat{P}_{o,\perp} \Sigma_{yx}^T  \hat{P}_{i,\perp} \\
&+ ( - \beta  \mathbb{1}_{i,i} \otimes R_r S^{-1} L_r^T \otimes \mathbb{1}_{i,i} - \hat{P}_{i,\perp} \Sigma_{yx}^T  \hat{P}_{o,\perp} \otimes \mathbb{1}_{i,i}) \mathcal{K}_{o,i}, \\
&-\frac{\partial \vec{g}_{\textcolor{blue_R}{\mathfrak{W}_R}}}{\partial \textcolor{blue_L}{\mathfrak{W}_L}} = 0,
\end{aligned} \\
\hline
&\begin{aligned}
&-\frac{\partial \vec{g}_{\textcolor{blue_L}{\mathfrak{W}_L}}}{\partial \textcolor{blue_R}{\mathfrak{W}_R}} = 0, \\
&-\frac{\partial \vec{g}_{\textcolor{blue_L}{\mathfrak{W}_L}}}{\partial \boldsymbol{W}} = \boldsymbol{W}\Sigma_{xx}\otimes \mathbb{1}_{i,i} +(\mathbb{1}_{o,o} \otimes \hat{N})\mathcal{K}_{o,i} + \hat{N}\otimes \mathbb{1} +(\mathbb{1}_{o,o} \otimes \boldsymbol{W} \Sigma_{xx})\mathcal{K}_{o,i} \\
& = L S_{o,i}^3 R^T \otimes \mathbb{1}_{i,i} + (-\beta \mathbb{1}_{o,o} \otimes L_r S^{-1} R_r^T - \mathbb{1}_{o,o} \otimes \hat{P}_{o,\perp} \Sigma_{yx} \hat{P}_{i,\perp})) \mathcal{K}_{o,i} \\
&+ (-\beta L_r S^{-1} R_r^T \otimes \mathbb{1}_{i,i} - \hat{P}_{o,\perp} \Sigma_{yx} \hat{P}_{i,\perp} \otimes \mathbb{1}_{i,i}) + (\mathbb{1}_{o,o} \otimes L S_{o,i}^3 R^T)\mathcal{K}_{o,i},\\
&-\frac{\partial \vec{g}_{\textcolor{blue_L}{\mathfrak{W}_L}}}{\partial \textcolor{blue_L}{\mathfrak{W}_L}} = 2\beta \mathbb{1}_{o,o} \otimes \mathbb{1}_{o,o}.
\end{aligned}
\end{align}

\noindent Based on our assumption on the data (and working in a rotated frame where $L$ and $R$ are the identity), the critical points are given by $\boldsymbol{W}^* =S_r = \sum_{i\in \mathcal{S}} \sigma_i^3 \vec{e}_i \vec{e}_i^T, \mathfrak{W}_R^* = \mathfrak{W}_L^* = (\boldsymbol{W}^*)^{2/3}$. At the critical point, the Jacobian can be split into three parts $\mathcal{J}^* = \mathcal{J}_{r}^* + \mathcal{J}_\perp^* + \mathcal{J}_{\text{mixed}}^*$ associated with different invariant subspaces. The invariant subspace relevant for us is the one composed of the vectors associated with the $r$-sector of the matrices $\boldsymbol{W}^*, \mathfrak{W}_R^*, \mathfrak{W}_L^*$:

\begin{align}
\left\{ \begin{pmatrix} \hat{P}_{\mathcal{S}}^{(o|o)} \vec{v}  \\ \hat{P}_{\mathcal{S}}^{(o|i)} \vec{w} \\ \hat{P}_{\mathcal{S}}^{(i|i)}  \vec{u} \end{pmatrix} | \vec{v} \in \mathbb{R}^{d_\text{out}^2}, \vec{w} \in \mathbb{R}^{d_\text{out}\cdot d_\text{in}}, \vec{u} \in \mathbb{R}^{d_\text{in}^2}\right\}.
\end{align}

On this subspace, the Jacobian acts as $\mathcal{J}_r^*$:

\begin{align}
\frac12 \mathcal{J}_r^* = \begin{pmatrix} 0 & (S_r^3-\beta S_r^{-1}) \otimes \mathbb{1})\mathcal{K} + \mathbb{1} \otimes (S_r^3-\beta S_r^{-1}) & 0 \\
-\beta S_r^2 \otimes S_r^{-1}  & S_r^4 \otimes \mathbb{1}+S_r^2 \otimes S_r^2 + \mathbb{1} \otimes S_r^4 + \beta \mathbb{1} & -\beta S_r \otimes \mathbb{1} - \beta S_r^{-1} \otimes S_r^2 \\
0 & (S_r^3-\beta S_r^{-1})\otimes \mathbb{1} +(\mathbb{1} \otimes (S_r^3-\beta S_r^{-1}))\mathcal{K} & 0 \end{pmatrix} +2\beta \mathbb{1}.
\end{align}
As in the Hessian case of the microscopic parameter space, we are expecting that perturbing the singular values as well as de-aligning should be massive perturbations. Infinitesimal versions of those rotations $\mathcal{O}$ can be of the form:
\begin{align*}
\begin{pmatrix} \textcolor{blue_L}{\mathfrak{W}_L} \\ \boldsymbol{W} \\ \textcolor{blue_R}{\mathfrak{W}_R} \end{pmatrix} \rightarrow  \begin{pmatrix} \mathcal{O}_o \textcolor{blue_L}{\mathfrak{W}_L} \mathcal{O}_o^T \\ \mathcal{O}_o\boldsymbol{W} \mathcal{O}_i^T\\ \mathcal{O}_i\textcolor{blue_R}{\mathfrak{W}_R} \mathcal{O}_i^T \end{pmatrix} \approx \begin{pmatrix} \textcolor{blue_L}{\mathfrak{W}_L}^* \\ \boldsymbol{W}^* \\ \textcolor{blue_R}{\mathfrak{W}_R}^* \end{pmatrix} + \epsilon \underbrace{\begin{pmatrix} B \textcolor{blue_L}{\mathfrak{W}_L}^* + \textcolor{blue_L}{\mathfrak{W}_L}^* B^T \\ B \boldsymbol{W}^* + \boldsymbol{W}^* A \\ A^T \textcolor{blue_R}{\mathfrak{W}_R}^* +\textcolor{blue_R}{\mathfrak{W}_R}^*A \end{pmatrix}}_{=\delta \bar{\boldsymbol{W}}},
\end{align*}
where $A$ and $B$ are anti-symmetric matrices. Due to the diagonal structures in $\mathcal{J}_r^*$, $A = e_{ij}-e_{ji}$, $B = e_{ij}-e_{ji}$ with $i,j \in \mathcal{S}$ are reasonable try-out states, corresponding to 
\begin{align}
\text{unrotated frame:} \quad A = \textcolor{blue}{\vec{r}_i \vec{r}_j^T} - \textcolor{blue}{\vec{r}_j \vec{r}_i^T}, \quad B = \textcolor{blue}{\vec{l}_i \vec{l}_j^T} - \textcolor{blue}{\vec{l}_j \vec{l}_i^T}.
\end{align}
The corresponding Ansatz for eigenstates $\delta \bar{\boldsymbol{W}}$ of $\mathcal{J}_r^*$ are (with $\vec{s}_{ij}^{(k|m)}:= \vec{e}_i \otimes \vec{e}_j + \vec{e}_j \otimes \vec{e}_i \in \mathbb{R}^{k\cdot m}$ with $k$ and $m$ denoting the respective dimensions):
\begin{align}
\text{vec}_c\left(\begin{pmatrix} B \textcolor{blue_L}{\mathfrak{W}_L}^* + \textcolor{blue_L}{\mathfrak{W}_L}^* B^T \\ B \boldsymbol{W}^* + \boldsymbol{W}^* A \\ A^T \textcolor{blue_R}{\mathfrak{W}_R}^* +\textcolor{blue_R}{\mathfrak{W}_R}^*A \end{pmatrix}\right) = \begin{pmatrix} (\sigma_i^2 - \sigma_j^2) \vec{s}_{ij}^{(o|o)} \\ (\sigma_i^3-\sigma_j^3)\vec{s}_{ij}^{(o|i)} \\(\sigma_i^2 - \sigma_j^2) \vec{s}_{ij}^{(i|i)} \end{pmatrix} \propto \begin{pmatrix} \frac{\sigma_i^2-\sigma_j^2}{\sigma_i^3-\sigma_j^3} \vec{s}_{ij}^{(o|o)} \\ \vec{s}_{ij}^{(o|i)} \\\frac{\sigma_i^2-\sigma_j^2}{\sigma_i^3-\sigma_j^3} \vec{s}_{ij}^{(i|i)} \end{pmatrix} \stackrel{\text{unrotated frame}}{\rightarrow} \begin{pmatrix} \frac{\sigma_i^2-\sigma_j^2}{\sigma_i^3-\sigma_j^3} \left(\textcolor{blue}{\vec{r}_j \otimes \vec{r}_i} + \textcolor{blue}{\vec{r}_i \otimes  \vec{r}_j}\right) \\ \left(\textcolor{blue}{\vec{r}_i \otimes \vec{l}_j} + \textcolor{blue}{\vec{r}_j \otimes \vec{l}_i} \right)  \\\frac{\sigma_i^2-\sigma_j^2}{\sigma_i^3-\sigma_j^3} \left(\textcolor{blue}{\vec{l}_i \otimes \vec{l}_j} + \textcolor{blue}{\vec{l}_j \otimes \vec{l}_i} \right) \end{pmatrix}.
\end{align}

We can directly check that those are eigenvectors of $\mathcal{J}_r^*$:
\begin{align}
\textbf{rot. perturb.:} \quad &\text{vec}_c(\delta \bar{\boldsymbol{W}}_{ij}) =\begin{pmatrix} \frac{\sigma_i^2-\sigma_j^2}{\sigma_i^3-\sigma_j^3} \vec{s}_{ij}^{(o|o)} \\ \vec{s}_{ij}^{(o|i)} \\ \frac{\sigma_i^2-\sigma_j^2}{\sigma_i^3-\sigma_j^3} \vec{s}_{ij}^{(i|i)} \end{pmatrix} \quad \text{eigenvalues:  } \frac12\lambda_{ij} = \sigma_i^4+ \sigma_j^4 + \sigma_i^2 \sigma_j^2 + \beta - \beta \left(\frac{\sigma_i^2+\sigma_j^2}{\sigma_i \sigma_j}\right),\\
\textbf{diag. perturb.:} \quad &\text{vec}_c(\delta \bar{\boldsymbol{W}}_{jj}) = \begin{pmatrix} \frac{2}{3} \vec{e}_j \otimes \vec{e}_j \\ \vec{e}_j \otimes \vec{e}_j \\ \frac{2}{3} \vec{e}_j \otimes \vec{e}_j \end{pmatrix} \quad \text{eigenvalues:  } \frac12\lambda_{jj} =3\sigma_j^4 -\beta,
\end{align}
with $i,j \in \mathcal{S}$. The eigenvalues are the same as the ones in the massive data-dependent sector of the microscopic Hessian (which correspond to perturbations that are compatible with the $0$-balanced condition), see Tab.~\ref{tab:HessianSpectrum2hidden}. The diagonal perturbation can be seen as the limiting case of the rotation case for $\sigma_i \to \sigma_j$. The eigenvectors and corresponding matrices are indeed (i) rescaled versions of the \emph{generators} of rotations of the stationary state and (ii) diagonal shifts of the singular values. Note that those perturbations/directions are in line with the $0$-balanced condition and therefore carry useful information.\\

\subsection{Generalization to deeper Networks}
For networks with arbitrary depth $n>2$, the Jacobian can be analyzed in a fashion similar to the $n=3$ case. For a finite rank $r$ critical point, the Jacobian in the $r$-sector reads (for $\Sigma_{xx} = \mathbb{1}$ and $\Sigma_{yx}$ diagonalized from the onset):

\begin{align}
&\frac12 \mathcal{J}_r = {\tiny\begin{pmatrix} 2\beta \mathbb{1} & S_r^n \otimes \mathbb{1} - \beta (\mathbb{1} \otimes S_r^{(2-n)})\mathcal{K} - \beta (S_r^{2-n} \otimes \mathbb{1} )+ (\mathbb{1}\otimes S_r^n) \mathcal{K} & 0 \\
- \beta \sum_{l=0}^{n-2} \sum_{k=0}^{n-2-l} S_r^{n-2-2k} \otimes S_r^{2k} & n\beta \mathbb{1} \otimes \mathbb{1} + \sum_{k=0}^{n-1} S^{2k} \otimes S^{2(n-1-k)} & -\beta \sum_{l=0}^{n-2} \sum_{k=0}^{n-2-l} S_r^{2k} \otimes S_r^{n-2-2k} \\
0 & \mathbb{1} \otimes S^n - \beta (S^{2-n}\otimes \mathbb{1})\mathcal{K} - \beta(\mathbb{1} \otimes S^{2-n}) + (S^n \otimes \mathbb{1}) \mathcal{K} & 2\beta\mathbb{1} \end{pmatrix}}.
\end{align}
Generalizing the result from $n=3$ layers, we check that the following are eigenvectors (based on the symmetry-based intuition):
\begin{align}
&\text{vec}_c(\delta \bar{\boldsymbol{W}}_{ij}) =\begin{pmatrix} \frac{\sigma_i^2-\sigma_j^2}{\sigma_i^n-\sigma_j^n} \vec{s}_{ij}^{(o|o)} \\ \vec{s}_{ij}^{(o|i)} \\ \frac{\sigma_i^2-\sigma_j^2}{\sigma_i^n-\sigma_j^n} \vec{s}_{ij}^{(i|i)} \end{pmatrix} &&\quad \text{eigenvalues:  } \frac12\lambda_{ij} = \left( \frac{\sigma_i^{2n}-\sigma_j^{2n}}{\sigma_i^2-\sigma_j^2}\right) + \beta\left( 2-(\sigma_i^{2-n}+\sigma_j^{2-n})\left( \frac{\sigma_i^n -\sigma_j^n}{\sigma_i^2 -\sigma_j^2}\right)\right), \\
&\text{vec}_c(\delta \bar{\boldsymbol{W}}_{i=j}) =\begin{pmatrix} \frac{2}{n} \vec{e}_i \otimes \vec{e}_i \\ \vec{e}_i \otimes \vec{e}_i \\ \frac{2}{n} \vec{e}_i \otimes \vec{e}_i \end{pmatrix} &&\quad \text{eigenvalues:  } \frac12\lambda_{ii} =  n\sigma_i^{2(n-1)}+\beta(2-n).
\end{align}
Based on the $\mathcal{J}_r$, this is an eigenstate once:
\begin{align}
\begin{aligned}
&\frac12 \mathcal{J}_r \text{vec}_c(\delta \bar{\boldsymbol{W}}_{ij}) = \\
&\begin{pmatrix} 
\left[\textcolor{orange}{\beta \left(2\frac{\sigma_i^2-\sigma_j^2}{\sigma_i^n-\sigma_j^n} -(\sigma_i^{2-n} + \sigma_j^{2-n})\right)} + (\sigma_i^n+\sigma_j^n)\right]\vec{s}_{ij} \\
\left[\sum_{k=0}^{n-1} (\sigma_i^2)^k (\sigma_j^2)^{n-1-k} + \textcolor{orange}{\beta\left(n - \sum_{l=0}^{n-2} \sum_{k=0}^{n-2-l} ( \sigma_i^{n-2-2k}\sigma_j^{2k} + \sigma_j^{n-2-2k}\sigma_i^{2k}) \frac{\sigma_i^2-\sigma_j^2}{\sigma_i^n-\sigma_j^n}\right)} \right] \vec{s}_{ij}\\ \left[\textcolor{orange}{\beta \left(2\frac{\sigma_i^2-\sigma_j^2}{\sigma_i^n-\sigma_j^n} -(\sigma_i^{2-n} + \sigma_j^{2-n})\right)} + (\sigma_i^n+\sigma_j^n)\right]\vec{s}_{ij}  \end{pmatrix} \stackrel{!}{=} \lambda_{ij} \text{vec}_c(\delta \bar{\boldsymbol{W}}_{ij}).
\end{aligned}
\end{align}
Using the geometric series, we have $\sum_{k=0}^{n-1} (\sigma_i^2)^k (\sigma_j^2)^{n-1-k} = \frac{\sigma_i^{2n}-\sigma_j^{2n}}{\sigma_i^2-\sigma_j^2}$ such that we only need to check the equivalence of the orange-marked terms according to:
\begin{align*}
n - \sum_{l=0}^{n-2} \sum_{k=0}^{n-2-l} \left( \sigma_i^{n-2-2k}\sigma_j^{2k} + \sigma_j^{n-2-2k}\sigma_i^{2k} \right) \frac{\sigma_i^2-\sigma_j^2}{\sigma_i^n-\sigma_j^n} \stackrel{!}{=} 2 - \frac{1}{\sigma_i^2-\sigma_j^2}\left( (\sigma_i^n -\sigma_j^n)(\sigma_i^{2-n} + \sigma_j^{2-n})\right).
\end{align*}
To show that these expressions are the same, we start from the left hand side and use multiple times the geometric series in the form of:
\begin{align*}
&\sum_{k=0}^{\frac{p}{2} -1} x^{p-2-2k}y^{2k} = \frac{x^p-y^p}{x^2-y^2}: \quad  \sum_{l=0}^{n-2} \sum_{k=0}^{n-2-l}\sigma_i^{n-2-2k}\sigma_j^{2k}= \sum_{l=0}^{n-2} \sigma_i^{-n+2(l+1)} \frac{(\sigma_i^{2(n-l-1)} - \sigma_j^{2(n-l-1)})}{\sigma_i^2-\sigma_j^2},
\end{align*}
where we identified $p=2(n-l-1)$. Using the geometric series again, we get:
\begin{align*}
&\sum_{l=0}^{n-2} \sum_{k=0}^{n-2-l}\sigma_i^{n-2-2k}\sigma_j^{2k}= \frac{1}{\sigma_i^2 -\sigma_j^2} \left[(n-1)\sigma_i^n- \frac{1}{\sigma_j^2 -\sigma_i^2}\left(\sigma_i^{-n+2}\sigma_j^{2n}-\sigma_i^n \sigma_j^2 \right)\right]
\end{align*}
and the other term follows in a similar fashion. Therefore, we have:

\begin{align*}
&n - \sum_{l=0}^{n-2} \sum_{k=0}^{n-2-l} ( \sigma_i^{n-2-2k}\sigma_j^{2k} + \sigma_j^{n-2-2k}\sigma_i^{2k}) \frac{\sigma_i^2-\sigma_j^2}{\sigma_i^n-\sigma_j^n} = 1 - \frac{1}{\sigma_i^2 -\sigma_j^2} \frac{1}{\sigma_i^n -\sigma_j^n} \underbrace{\left( \sigma_i^{-n+2}\sigma_j^{2n}-\sigma_i^n \sigma_j^2 + \sigma_j^{-n+2}\sigma_i^{2n} - \sigma_j^n \sigma_i^2\right)}_{= (\sigma_i^n-\sigma_j^n)((\sigma_j^2-\sigma_i^2) + (\sigma_i^n-\sigma_j^n)(\sigma_i^{2-n}+\sigma_j^{2-n}))} \\
&= 1- \frac{1}{\sigma_i^2 -\sigma_j^2}\left[ (\sigma_j^2-\sigma_i^2) + (\sigma_i^n-\sigma_j^n)(\sigma_i^{2-n}+\sigma_j^{2-n})\right] = 2- \frac{1}{\sigma_i^2 -\sigma_j^2}\left[(\sigma_i^n-\sigma_j^n)(\sigma_i^{2-n}+\sigma_j^{2-n})\right].
\end{align*}
This implies that $\text{vec}_c(\delta \bar{\boldsymbol{W}}_{ij})$ is indeed an eigenvector with the eigenvalue $\lambda_{ij}$ given above. In summary, the \emph{massive data-dependent eigenvalues} of the microscopic Hessian (at a critical point), see Tab.~\ref{tab:HessianSpectrumgeneral}, are as well the eigenvalues of the Jacobian of the macro dynamics for those directions compatible with the $0$-balanced condition.

\section{Exact dynamical Solution \& Algebraic Scaling at the Second Order Transition \label{App:AlgebraicScaling}}
For special cases, the gradient flow dynamics for 1 hidden layer networks with regularization can be solved exactly. In particular, at the phase transitions \emph{algebraic} behaviour in the form of critical slowing down becomes observable. For 1 hidden layer, the macro dynamics (for $\Sigma_{xx}=\mathbb{1}$) can be cast in the form of a matrix Riccati differential equation, adapting \cite{Fukumizu1998,Braun2022}. Starting point are the equations of motion:

\begin{align}
&-\frac12 \frac{d}{dt} \mathfrak{W}_L = \underbrace{2 \boldsymbol{W}\boldsymbol{W}^T}_{=\mathfrak{W}_L^2 + \boldsymbol{W}\boldsymbol{W}^T} - \Sigma_{yx} \boldsymbol{W}^T-\boldsymbol{W} \Sigma_{yx}+2\beta \mathfrak{W}_L, \\
&- \frac12 \frac{d}{dt} \boldsymbol{W} = \boldsymbol{W} \mathfrak{W}_R + \mathfrak{W}_L \boldsymbol{W} - \Sigma_{yx} \mathfrak{W}_R - \mathfrak{W}_L \Sigma_{yx} + 2\beta \boldsymbol{W}, \\
&-\frac12 \frac{d}{dt} \mathfrak{W}_R = \underbrace{2\boldsymbol{W}^T \boldsymbol{W}}_{\mathfrak{W}_R^2+\boldsymbol{W}^T \boldsymbol{W}} - \Sigma_{yx}^T - \boldsymbol{W}^T \Sigma_{yx} +2\beta \mathfrak{W}_R,
\end{align}
such that we get (for $\tau = \frac12$):
\begin{align}
\begin{aligned}
&\tau\frac{d}{dt} \underbrace{\begin{pmatrix} \mathfrak{W}_R & \boldsymbol{W}^T \\ \boldsymbol{W} & \mathfrak{W}_L \end{pmatrix}}_{:=M(t)} = \begin{pmatrix} 0 & \Sigma_{yx}^T \\  \Sigma_{yx} & 0 \end{pmatrix} M(t) + M(t) \begin{pmatrix} 0 & \Sigma_{yx}^T \\  \Sigma_{yx} & 0 \end{pmatrix} -2\beta M(t)  - M^2(t) =: \boldsymbol{F} M(t) + M(t)\boldsymbol{F} -M^2(t) \\
&\text{with:} \quad \boldsymbol{F}:= \begin{pmatrix} -\beta \mathbb{1} & \Sigma_{yx}^T \\  \Sigma_{yx} & -\beta \mathbb{1} \end{pmatrix},
\label{eq:RiccatiEquation}
\end{aligned}
\end{align}
which is a matrix Riccati differential equation. Solutions of this kind of differential equations are well-known (see, e.g., \cite{Sasagawa1982}). For convenience, we reproduce the solution step by step. The Riccati equation can be solved exactly by mapping to a (quadratic) Hamiltonian system:
\begin{align}
\frac{d}{dt}\begin{pmatrix} A(t) \\ B(t) \end{pmatrix} = \frac{1}{\tau}\underbrace{\begin{pmatrix} -\boldsymbol{F} & \mathbb{1} \\ 0 & \boldsymbol{F} \end{pmatrix}}_{:=\boldsymbol{H}_{AB}}\begin{pmatrix} A(t) \\ B(t) \end{pmatrix}  \Rightarrow \boxed{M(t) = B(t)A^{-1}(t)} \quad \text{initial cond.:} \quad \begin{pmatrix} A(0) \\ B(0) \end{pmatrix} = \begin{pmatrix} \mathbb{1} \\ M_0 \end{pmatrix} = \begin{pmatrix} \mathbb{1} \\ Q_0Q_0^T \end{pmatrix}
\end{align}
with $\boldsymbol{F}^T = \boldsymbol{F}$ in our case and $M_0=M(0)$. The general solution of these coupled linear differential equations is given by applying the time evolution operator $\exp(\boldsymbol{H}_{AB}\frac{t}{\tau})$ to the initial state. The matrix exponential can be evaluated explicitly:
\begin{align}
\exp\left(\boldsymbol{H}_{AB}\frac{t}{\tau}\right) = \begin{pmatrix} \sum_{n=0}^\infty \frac{1}{n!}(-\boldsymbol{F})^n (\frac{t}{\tau})^n & \boldsymbol{F}^{-1}\sum_{n=0}^\infty \frac{1}{(2n+1)!} (-\boldsymbol{F})^{2n+1}(\frac{t}{\tau})^{2n+1} \\ 0 & \sum_{n=0}^\infty \frac{1}{n!} (\boldsymbol{F})^n (\frac{t}{\tau})^n \end{pmatrix} = \begin{pmatrix} \exp(-\boldsymbol{F}\frac{t}{\tau}) & \boldsymbol{F}^{-1} \sinh(-\boldsymbol{F}\frac{t}{\tau}) \\ 0 & \exp(\boldsymbol{F}\frac{t}{\tau}) \end{pmatrix},
\end{align}
valid as long as $\boldsymbol{F}$ is invertible. Therefore, the exact solutions reads (using the matrix identity $(*): CC^T(\mathbb{1}+DCC^T)^{-1}= C(\mathbb{1}+C^TDC)^{-1}C^T$):
\begin{align}
M(t) &= B(t)A^{-1}(t) = \exp\left(\boldsymbol{F}\frac{t}{\tau}\right)M_0 \left( \exp\left(\boldsymbol{-F}\frac{t}{\tau}\right) + \boldsymbol{F}^{-1} \sinh\left(-\boldsymbol{F} \frac{t}{\tau}\right) M_0 \right)^{-1}  \\
& = \exp\left(\boldsymbol{F}\frac{t}{\tau}\right) \underbrace{M_0}_{=Q_0Q_0^T} \left( \mathbb{1} - \frac12 \left(\boldsymbol{F}^{-1} - \exp\left(\boldsymbol{F}\frac{t}{\tau}\right)\boldsymbol{F}^{-1} \exp\left(\boldsymbol{F}\frac{t}{\tau}\right) \right)M_0 \right)^{-1} \exp\left(\boldsymbol{F}\frac{t}{\tau}\right) \\
& \stackrel{(*)}{=} \exp\left(\boldsymbol{F}\frac{t}{\tau}\right)Q_0 \left( \mathbb{1} - \frac12 Q_0^T\left(\boldsymbol{F}^{-1} - \exp\left(\boldsymbol{F}\frac{t}{\tau}\right)\boldsymbol{F}^{-1} \exp\left(\boldsymbol{F}\frac{t}{\tau}\right) \right)Q_0 \right)^{-1} Q_0^T\exp\left(\boldsymbol{F}\frac{t}{\tau}\right). \label{eq:explicitfulltimeevolution1hidden}
\end{align}

\noindent \textbf{Long-term dynamics:} 
The dynamics at long times is determined by the positive (and vanishingly small/zero) eigenvalues of $\boldsymbol{F}$. For $\beta = \beta^*_j = \eta_j$, the matrix $\boldsymbol{F}$ features a zero-eigenvalue, which results in \emph{algebraic} decay at long times in the corresponding subspace. The goal is to show that the dynamics at large times is roughly given by $M(t) \approx M^* + \frac{1}{2\frac{t}{\tau}}M_j$ with $M_j$ corresponding to a matrix in this subspace. For simplicity, we 'diagonalize' $\Sigma_{yx} = L S_{yx}R^T$ such that only the diagonal matrix $S_{yx}$ of (non-negative) singular values in descending order remains. Afterwards, we use an orthogonal rotation to bring $\boldsymbol{F}$ into diagonal form:
\begin{align}
\boldsymbol{F} \quad \stackrel{U = \frac{1}{\sqrt{2}} \begin{pmatrix} \mathbb{1} & R^TL \\ LR^T & -\mathbb{1} \end{pmatrix}}{\rightarrow} \quad  U\boldsymbol{F}U^T = \tilde{\boldsymbol{F}} = \begin{pmatrix}S_{yx}- \beta \mathbb{1} & \boldsymbol{0} \\ \boldsymbol{0} & -S_{yx} -\beta \mathbb{1} \end{pmatrix}=: \begin{pmatrix} \tilde{F}_+ & 0 \\ 0 & \tilde{F}_- \end{pmatrix},
\end{align}
where $\tilde{F}_+$ contains $r_\text{max}(\beta)$ positive eigenvalues $\lambda_i$ (with the corresponding eigenvectors $|\lambda_i\rangle$ of $\boldsymbol{F}$). $\tilde{F}_-$ instead only contains negative eigenvalues. Using the same transformation, we are working with the transformed weight matrices:
\begin{align}
Q_0=\begin{pmatrix} W_1^T(t=0) \\ W_2(t=0) \end{pmatrix} \quad \rightarrow \quad UQ := \begin{pmatrix} \tilde{W}_1^T \\ \tilde{W}_2 \end{pmatrix}, \quad \tilde{M}_0 = U M_0 U^T = \begin{pmatrix} \tilde{W}_1^T \\ \tilde{W}_2 \end{pmatrix} \begin{pmatrix} \tilde{W}_1 & \tilde{W}_2^T \end{pmatrix}.
\end{align}
The full time evolution is given by \eqref{eq:explicitfulltimeevolution1hidden}. The main task is to evaluate the inverse for long times:
\begin{align}
\left( \mathbb{1} - \frac12 \left(\boldsymbol{F}^{-1} - \exp\left(\boldsymbol{F}\frac{t}{\tau}\right)\boldsymbol{F}^{-1} \exp\left(\boldsymbol{F}\frac{t}{\tau}\right) \right)M_0 \right)^{-1}.
\end{align}
The positive eigenvalues of $\boldsymbol{F}$ will dominate the expression for long times, whereas the negative eigenvalues lead to an exponential decay. To formally distinguish the different subspaces (positive, negative, zero eigenvalues), we define suitable projection operators
\begin{align}
P_< = \sum_{\lambda_i <0} |\lambda_i \rangle \langle \lambda_i|, \quad P_> = \sum_{\lambda_i >0} |\lambda_i \rangle \langle \lambda_i|, \quad P^* = |0 \rangle \langle 0| , \quad \tilde{F}_> = \sum_{\lambda_i >0}\lambda_i |\lambda_i \rangle \langle \lambda_i|.
\end{align}
The zero eigenvalue(s) (and sector) have to be considered with care: strictly speaking the inverse does not exist right at a transition. However, for an infinitesimal $\beta$-distance to the transition that goes to zero, we have [adapting \cite{Braun2022}]: $\lim_{\epsilon\to 0} [ e^{\epsilon \frac{t}{\tau}} \epsilon^{-1} e^{\epsilon \frac{t}{\tau}}- \epsilon^{-1}] = 2\frac{t}{\tau}$, resulting in:
\begin{align}
-\frac12 \left(\boldsymbol{F}^{-1} - \exp\left(\boldsymbol{F}\frac{t}{\tau}\right)\boldsymbol{F}^{-1} \exp\left(\boldsymbol{F}\frac{t}{\tau}\right)\right)P^* = \frac{t}{\tau}P^*.
\end{align}
For long times, we can approximate (using the $+$-index to indicate that an operator only acts on the $+$-subspace):
\begin{align}
&\mathbb{1} - \frac12 \left(\tilde{\boldsymbol{F}}^{-1} - \exp\left(\tilde{\boldsymbol{F}}\frac{t}{\tau}\right)\tilde{\boldsymbol{F}}^{-1} \exp\left(\tilde{\boldsymbol{F}}\frac{t}{\tau}\right) \right)\textcolor{orange}{(P_>+P_<+P^*)}M_0 \\
&\approx \mathbb{1} + \left(-\frac12 \tilde{\boldsymbol{F}}^{-1}P_< + \frac12\tilde{\boldsymbol{F}}^{-1} \exp\left(\tilde{\boldsymbol{F}}2\frac{t}{\tau}\right) P_>\right) \begin{pmatrix} \tilde{W}_1^T \\ \tilde{W}_2 \end{pmatrix} \begin{pmatrix} \tilde{W}_1 & \tilde{W}_2^T \end{pmatrix} + \frac{t}{\tau} P^* \begin{pmatrix} \tilde{W}_1^T \\ \tilde{W}_2 \end{pmatrix} \begin{pmatrix} \tilde{W}_1 & \tilde{W}_2^T \end{pmatrix} \\
&=\begin{pmatrix} (\mathbb{1} - \frac12\tilde{F}_{+,<}^{-1}P_{+,<} + \frac12\tilde{F}_{+,>}^{-1} e^{\tilde{F}_{+,>} 2\frac{t}{\tau}}P_{+,>} + \frac{t}{\tau} P_+^*) \tilde{W}_1^T \tilde{W}_1 & \underbrace{(-\frac12\tilde{F}_{+,<}^{-1}P_{+,<} +\tilde{F}_{+,>}^{-1} e^{\tilde{F}_{+,>} 2\frac{t}{\tau}}P_{+,>} + \frac{t}{\tau} P_+^*)}_{:=\mathcal{F}_+(\frac{t}{\tau})} \tilde{W}_1^T \tilde{W}_2^T \\ \boldsymbol{0} & \mathbb{1} \end{pmatrix}, \label{eq:longtime1hiddenbeforeinverse}
\end{align}
which holds for sufficiently large times such that the terms coupled to negative eigenvalues of $\boldsymbol{F}$ have already decayed. To check for the asymptotics of the exact solution, we have to evaluate the inverse of  \eqref{eq:longtime1hiddenbeforeinverse} using 
\begin{align}
\begin{pmatrix} A & B \\ 0 & \mathbb{1} \end{pmatrix}^{-1} = \begin{pmatrix} A^{-1} & - A^{-1} B\\ 0 & \mathbb{1}\end{pmatrix},
\end{align}
valid for $A$ being an invertible matrix. In our case:
\begin{align}
&A^{-1} = \left(\left(\mathbb{1} - \frac12\tilde{F}_{+,<}^{-1}P_{+,<} +  \frac12\tilde{F}_{+,>}^{-1} e^{\tilde{F}_{+,>} 2\frac{t}{\tau}}P_{+,>} + 2t P_+^*\right) \tilde{W}_1^T \tilde{W}_1\right)^{-1} \\
& \quad \quad = \left(\tilde{W}_1^T \tilde{W}_1\right)^{-1} \underbrace{\left(\mathbb{1} - \frac12\tilde{F}_{+,<}^{-1}P_{+,<} + \frac12\tilde{F}_{+,>}^{-1} e^{\tilde{F}_{+,>} 2\frac{t}{\tau}}P_{+,>} + \frac{t}{\tau} P_+^*\right)^{-1}}_{\text{diagonal - easy to invert}},\\
&-A^{-1}B = -\left(\tilde{W}_1^T \tilde{W}_1\right)^{-1} \left(\mathbb{1}+ \frac12\mathcal{F}_+\left(\frac{t}{\tau}\right)\right)^{-1} \mathcal{F}_+\left(\frac{t}{\tau}\right) \tilde{W}_1^T \tilde{W}_2^T,
\end{align}
where we assumed that $\tilde{W}_1^T \tilde{W}_1$ is invertible. For the full evolution in the long time limit, we get (in the frame of reference, where $\boldsymbol{F}$ was diagonalized):
\begin{align}
\tilde{M}(t) \approx &e^{\tilde{F}\frac{t}{\tau}} \begin{pmatrix} \tilde{W}_1^T \\ \tilde{W}_2 \end{pmatrix} \begin{pmatrix} \tilde{W}_1 & \tilde{W}_2^T \end{pmatrix} \begin{pmatrix} \left(\tilde{W}_1^T \tilde{W}_1\right)^{-1} \left(\mathbb{1} + \mathcal{F}_+(\frac{t}{\tau})\right)^{-1} & -\left(\tilde{W}_1^T \tilde{W}_1\right)^{-1} (\mathbb{1}+ \mathcal{F}_+(\frac{t}{\tau}))^{-1} \mathcal{F}_+(\frac{t}{\tau}) \tilde{W}_1^T \tilde{W}_2^T \\ \boldsymbol{0} & \mathbb{1} \end{pmatrix}e^{\tilde{F}\frac{t}{\tau}}\\
&= e^{\tilde{F}\frac{t}{\tau}} \begin{pmatrix} \tilde{W}_1^T \\ \tilde{W}_2 \end{pmatrix}  \begin{pmatrix} \tilde{W}_1 \left(\tilde{W}_1^T \tilde{W}_1\right)^{-1} \left(\mathbb{1} +  \mathcal{F}_+(\frac{t}{\tau})\right)^{-1} & -\tilde{W}_1 \left(\tilde{W}_1^T \tilde{W}_1\right)^{-1} (\mathbb{1}+ \mathcal{F}_+(t))^{-1} \mathcal{F}_+(\frac{t}{\tau}) \tilde{W}_1^T \tilde{W}_2^T + \tilde{W}_2^T \end{pmatrix} e^{\tilde{F}\frac{t}{\tau}} \\
&= \begin{pmatrix} e^{\tilde{F}_+\frac{t}{\tau}}\tilde{W}_1^T \\ e^{\tilde{F}_-\frac{t}{\tau}}\tilde{W}_2 \end{pmatrix}  \begin{pmatrix} \tilde{W}_1 \left(\tilde{W}_1^T \tilde{W}_1\right)^{-1} \left(\mathbb{1} +  \mathcal{F}_+(\frac{t}{\tau})\right)^{-1} e^{\tilde{F}_+ \frac{t}{\tau}} & \underbrace{\left(-\tilde{W}_1 \left(\tilde{W}_1^T \tilde{W}_1\right)^{-1} (\mathbb{1}+ \mathcal{F}_+(\frac{t}{\tau}))^{-1} \mathcal{F}_+(\frac{t}{\tau}) \tilde{W}_1^T \tilde{W}_2^T + \tilde{W}_2^T\right) e^{\tilde{F}_- \frac{t}{\tau}}}_{\text{exp. suppressed}} \end{pmatrix}\\
&\approx \begin{pmatrix} e^{\tilde{F}_+ \frac{t}{\tau}}  \tilde{W}_1^T \tilde{W}_1\left(\tilde{W}_1^T \tilde{W}_1\right)^{-1} \left(\mathbb{1} +  \mathcal{F}_+(\frac{t}{\tau})\right)^{-1} e^{\tilde{F}_+ \frac{t}{\tau}}& \boldsymbol{0} \\ \boldsymbol{0} & \mathbb{1} \end{pmatrix} \\
&\approx \begin{pmatrix} 2\tilde{F}_{+,>} & \boldsymbol{0} \\ \boldsymbol{0} & \boldsymbol{0} \end{pmatrix} + \frac{1}{1+\frac{t}{\tau}}|0\rangle \langle 0|.
\end{align}
In the original frame of reference, we therefore have
\begin{align}
M(t) = \begin{pmatrix} \mathfrak{W}_R & \boldsymbol{W}^T \\ \boldsymbol{W} & \mathfrak{W}_L \end{pmatrix} \stackrel{\text{long times}}{\rightarrow} \begin{pmatrix} R\tilde{F}_{+,>}R^T & R\tilde{F}_{+,>}L^T \\ L\tilde{F}_{+,>}R^T & L\tilde{F}_{+,>}L^T \end{pmatrix} \textcolor{orange}{+ \frac{1}{1+\frac{t}{\tau}} \frac12 \begin{pmatrix} RP_{+}^*R^T & RP_{+}^*L^T \\ LP_{+}^*R^T & LP_{+}^* L^T \end{pmatrix}},
\end{align}
where the second contribution describes the slow algebraic relaxation towards the minimum at a transition point $\beta^* = \eta_j$. A special (and simpler) case arises once we have an initial state of the form $\boldsymbol{W} = \sigma_0^2 L e_{jj} R^T$ for $\sigma_0^2 \in \mathbb{R}_+$. In this case, $\tilde{W}_1^T \tilde{W}_1$ would not be invertible, but we can solve the dynamics directly:
\begin{align}
M_0 = \sigma_0^2 \begin{pmatrix} Re_{jj}R^T & Re_{jj}L^T \\ Le_{jj}R^T & Le_{jj}L^T \end{pmatrix}: \quad M(t) \approx \frac{\sigma_0^2}{1+2\frac{t}{\tau}\sigma_0^2} \begin{pmatrix} Re_{jj}R^T & Re_{jj}L^T \\ Le_{jj}R^T & Le_{jj}L^T \end{pmatrix},
\end{align}
which describes the $t^{-1}$ scaling for long times, corresponding to the slow relaxation of the singular value $\sigma_j(t) \to 0$.

\end{widetext}

\end{document}